\documentclass[aps,prapplied,%
  superscriptaddress,%
  reprint,%
  floatfix%
]{revtex4-2}

\usepackage[utf8]{inputenc}
\usepackage[T1]{fontenc}
\usepackage[english]{babel}
\usepackage[caption=false,labelformat=simple]{subfig}
\usepackage{mathptmx,upgreek,graphicx,amsmath,amssymb,dcolumn,contour,siunitx,chemmacros,xspace,physics,todonotes,lipsum}
\usepackage[hidelinks]{hyperref}
\usepackage[normalem]{ulem}
\usepackage{xcolor}
\usepackage{changes}

\makeatletter
\newcommand\thefontsize{The current font size is: \f@size pt}
\makeatother

\definecolor{color_out_of_plane}{RGB}{112,173,71}
\definecolor{color_in_plane}{RGB}{127,127,127}
\definecolor{color_torsional}{RGB}{0,0,153}

\newcommand\labo[1][]{{\bf\color{color_out_of_plane}o#1}\xspace}
\newcommand\labi[1][]{{\bf\color{color_in_plane}i#1}\xspace}
\newcommand\labt[1][]{{\bf\color{color_torsional}t#1}\xspace}

\makeatletter
\def\@email#1#2{%
 \endgroup
 \patchcmd{\titleblock@produce}
  {\frontmatter@RRAPformat}
  {\frontmatter@RRAPformat{\produce@RRAP{*#1\href{mailto:#2}{#2}}}\frontmatter@RRAPformat}
  {}{}
}%
\makeatother

\renewcommand{\selectlanguage}[1]{}

\setuptodonotes{inline,color=tabgreen!50}

\newcommand{\addDocumentEditor}[2]{
  \expandafter\newcommand\csname todo#1\endcsname[1]{\todo[color=#2!50]{#1: ##1}}

  \definechangesauthor[name={#1}, color=#2!50!black]{#1}
  \expandafter\newcommand\csname added#1\endcsname[1]{\added[id=#1]{##1}}
  \expandafter\newcommand\csname deleted#1\endcsname[1]{\deleted[id=#1]{##1}}
  \expandafter\newcommand\csname replaced#1\endcsname[2]{\replaced[id=#1]{##1}{##2}}
}
\addDocumentEditor{AH}{tabcyan}
\addDocumentEditor{EW}{taborange}
\addDocumentEditor{FD}{tabred}
\addDocumentEditor{HW}{tabblue}
\addDocumentEditor{PB}{tabpurple}
\addDocumentEditor{BD}{tabolive}

\usepackage{etoolbox}
\robustify{\subref}

\graphicspath{{./img/}}

\newcommand\Figref[1]{Figure~\ref{#1}}   
\newcommand\figref[1]{Fig.~\ref{#1}}     
\newcommand\eqnref[1]{Eq.~\ref{#1}}      
\newcommand\tabref[1]{Table~\ref{#1}}    
\newcommand\secref[1]{Section~\ref{#1}}  
\newcommand\appref[1]{Appendix~\ref{#1}} 

\definecolor{tabblue}{HTML}{1f77b4}
\definecolor{taborange}{HTML}{ff7f0e}
\definecolor{tabred}{HTML}{d62728}
\definecolor{tabgreen}{HTML}{2ca02c}
\definecolor{tabcyan}{HTML}{17becf}
\definecolor{tabolive}{HTML}{bcbd22}
\definecolor{tabpurple}{HTML}{9467bd}

\newcommand\emx[1]{\ensuremath{#1}\xspace}

\newcommand\timevar{\emx{t}}
\newcommand\freq{\emx{f}}
\newcommand\freqeig{\emx{f_0}}
\newcommand\om{\emx{\omega}}
\newcommand\omeig{\emx{\omega_0}}
\newcommand\omdrive{\emx{\omega_d}}
\newcommand\amp{\emx{A}}
\newcommand\ampdrive{\emx{A_d}}
\newcommand\damping{\emx{\Gamma}}
\newcommand\Q{\emx{Q}}
\newcommand\Qmax{\emx{Q_\mathrm{max}}}
\newcommand\Qintr{\emx{Q_\mathrm{intr}}}
\newcommand\Qvol{\emx{Q_\mathrm{vol}}}
\newcommand\Qsurf{\emx{\beta}}
\newcommand\Qted{\emx{Q_\mathrm{ted}}}
\newcommand\Qconst{\emx{Q_\mathrm{d}}}

\newcommand\Qfprod{\emx{Q\cdot f}}  

\newcommand\stress{\emx{\sigma}}
\newcommand\modeno{\emx{n}}
\newcommand\youngs{\emx{E}}
\newcommand\thick{\emx{h}}
\newcommand\width{\emx{w}}
\newcommand\length{\emx{l}}
\newcommand\dens{\emx{\rho}}
\newcommand\poisson{\emx{\nu}}

\newcommand\Vsi{\emx{\mathrm{V}_\mathrm{Si}}}

\newcommand\posvar{\emx{x}}
\newcommand\deflvar{\emx{u}}

\newcommand\clampangle{\emx{\theta}}

\newcommand\youngsNumVal{437(90)}
\newcommand\youngsVal{\SI{\youngsNumVal}{\GPa}}
\newcommand\QconstVal{\num{1.38(9)e5}}
\newcommand\ourQvolBound{\num{1.4e5}}
\newcommand\ourQsurfBoundNum{\num{2.4e4}}
\newcommand\ourQsurfBound{\ourQsurfBoundNum/\SI{100}{\nm}}

\newcommand\figscale{0.78}

\begin{document}


\title{Monolithic 4H-SiC nanomechanical resonators with high intrinsic quality factors}

\author{A. Hochreiter}
\thanks{These authors contributed equally to this work.}
  \affiliation{Chair for Applied Physics, Friedrich-Alexander-Universität Erlangen-Nürnberg, 91058 Erlangen, Germany}
\author {P. Bredol}
\thanks{These authors contributed equally to this work.}
\affiliation{School of Computation, Information and Technology, Technical University of Munich, Germany}
\author{F. David}
\affiliation{School of Computation, Information and Technology, Technical University of Munich, Germany}
\author {B. Demiralp}
\affiliation{School of Computation, Information and Technology, Technical University of Munich, Germany}
\author{H. B. Weber}%
  \affiliation{Chair for Applied Physics, Friedrich-Alexander-Universität Erlangen-Nürnberg, 91058 Erlangen, Germany}
\author{E. M. Weig}
\thanks{eva.weig@tum.de}
\affiliation{School of Computation, Information and Technology, Technical University of Munich, Germany}
\affiliation{Munich Center for Quantum Science and Technology (MCQST), 80799 Munich, Germany}
\affiliation{TUM Center for Quantum Engineering (ZQE), 85748 Garching, Germany}

\date{\today}



\begin{abstract}
We present an extensive study of 4H-SiC nanomechanical resonators electrochemically etched out of a monocrystalline wafer. Combining piezo-driven interferometric determination of the mechanical spectra with scanning-laser-Doppler vibrometry, an unambiguous assignment of resonance peaks to flexural and torsional modes is achieved.
The investigation of multiple harmonic eigenmodes of singly and doubly clamped resonators with varying geometry allows for a comprehensive characterization. Excellent intrinsic mechanical quality factors up to \SI{2e5} are found at room temperature, approaching the thermoelastic limit at eigenfrequencies exceeding \SI{10}{\MHz}. Mechanical stress is essentially absent. Young's modulus in agreement with literature.
These findings are robust under post-processing treatments, in particular atomic layer etching and high-temperature thermal annealing. The resulting on-chip high-quality mechanical resonators represent a valuable technological element for a broad range of applications. In particular, the monolithic architecture meets the requirements of spin-based photonic quantum technologies on the upcoming SiC platform.
\end{abstract}

\maketitle

\section{Introduction}

Silicon Carbide (SiC), in particular its 4H-SiC polytype, has a unique potential as a technology platform for monolithically integrating classical technologies and quantum applications on the very same chip. Being a prime candidate for high-power electronic applications, there are today excellent SiC single-crystal materials on the wafer scale commercially available as well as a mature process technology. Moreover, SiC provides excellent spin carrying color centers, for which coherent operations with long coherence time have been reported \cite{Lukin_Color_center_SiC_review_2020}. Photonic waveguides and resonators have been fabricated using SiC-on-insulator technologies \cite{Lukin_Purcell_enhancement_SiC_SiCOI-2020, Yang_SiCOI_photonics_2023}. 
The interest of photonic waveguides is motivated on the one hand by gaining photonic access to the spin, but also by an unusual combination of non-linearity parameters: both ${\chi^{(2)}}$ and ${\chi^{(3)}}$ coefficients are significant, providing a rare opportunity for non-linear photonics on the SiC-platform \cite{Sato_Chi2_2009,DeLeonardis_Chi3_2017}.
Beyond these unique research opportunities for charge, spin and photons, we focus here on mechanical degrees of freedom, i.e. high-quality nanomechanical resonators. Mechanical oscillations are interacting with all of the above quantities. In particular, their interplay with spin (e.g. in Silicon vacancies \Vsi)  is expected to lead to a hybrid spin-mechanical state, which further enhances the quantum-classical portfolio of SiC \cite{Dietz_V_Si_spin-acoustic_coupling-2023, Nagy_V_Si-2019}. As the ground state of the silicon vacancy is sensitive to strain~\cite{Udvarhelyi_strain_V_Si-2020}, color-center-based quantum technologies benefit from a strain-free crystal structure.
To realize the full potential of SiC spin-mechanics, the C3 symmetry of the silicon vacancy \Vsi should remain unperturbed.

In consequence, neither of the tensile-strain related strategies to enhance the mechanical quality factor~\cite{sementilli_review_2021,engelsen_review_2024}, including dissipation dilution, soft clamping or strain engineering, can be exploited. It is thus essential for the performance of the hybrid spin-mechanical device to achieve benchmark values of the \emph{intrinsic} mechanical quality factor. This may be facilitated by another favorable property of SiC, which also stands out as a unique mechanical material. It has been pointed out that the Akhiezer damping in SiC is significantly lower than in any other material as a result of its remarkably low Grüneisen parameter~\cite{Ghaffari_Akhiezer_limit_damping-2013}. This damping mechanism related to phonon scattering processes sets the ultimate limit of dissipation, and is referred to as the quantum limit of dissipation~\cite{Ghaffari_Akhiezer_limit_damping-2013}. Unlike in other materials, quality factors reaching the Akhiezer limit are yet to be accomplished in SiC~\cite{Ghaffari_Akhiezer_limit_damping-2013, Hamelin_Akhiezer_SiC_limit-2019}. However, following the successful suppression of all other dissipation mechanisms, unprecedented intrinsic quality factors can be expected.

This work describes the first important step towards this goal. In order to access the regime in which dissipation is governed by fundamental phonon-related processes including thermoelastic damping or the Akhiezer effect, extrinsic damping from environmental effects and intrinsic mechanisms related to material imperfections have to be addressed~\cite{imboden_review_dissipation_2014}. The former includes damping from the surrounding medium, which can be mitigated by conducting the experiment in vacuum at a sufficiently low pressure, as well as clamping loss, which is targeted by appropriate resonator layouts. The latter involve defects residing both in the bulk and on the surface of the resonator material. Depending on the thickness of the resonator, one or a combination of these often limit the performance. This guides our attention to the correlation of process technology and the quality factor.

For the targeted hybrid spin-mechanical device architectures, we opt for a monolithic fabrication scheme \cite{hochreiter_electrochemical_etching_2023}. Only with a monolithic scheme, the best crystal quality and the full portfolio of high-temperature SiC process technology is available. This includes high-temperature annealing to remove crystal imperfections and undesired point defects, allowing for accurate control of the defect budget, which is of utmost importance for quantum technologies \cite{Ruehl_controlled_color_center_generation_2018}. Even more, the growth of epitaxial graphene on the \ch{SiC} (0001)-surface at temperatures well beyond \SI{1300}{\celsius} allows for accurately terminated surfaces and entails novel functionalities~\cite{emtsev_graphene_growth_Argon_6H-SiC_2009}. All in all, our approach is in stark contrast to the SiC-on-insulator technology that has been developed for photonic applications~\cite{Lukin_Purcell_enhancement_SiC_SiCOI-2020}. 

On the first glance, however it is challenging to fabricate 3D-structures out of a monocrystalline and nearly inert wafer. This is why we have developed an isotropic electrochemical etching scheme that can tailor the geometry by accurate control of the doping contrasts. In combination with the anisotropic dry etching, a rich variety of 3D devices like singly clamped, doubly clamped resonators, membranes and disk-resonators can be monolithically manufactured out of a single wafer~\cite{hochreiter_electrochemical_etching_2023}.

Here, we present an extensive characterization of 4H-SiC mechanical resonators and demonstrate remarkably high intrinsic quality factors. We will identify the evolution of quality factors, Young's modulus, and pre-stress conditions in the course of subsequent processing steps pointing out strategies for further improvement.

\section{Methods}

\subsection{Sample fabrication and mechanical characterization}

\begin{figure*}
  \centering
  \includegraphics[width=1\linewidth]{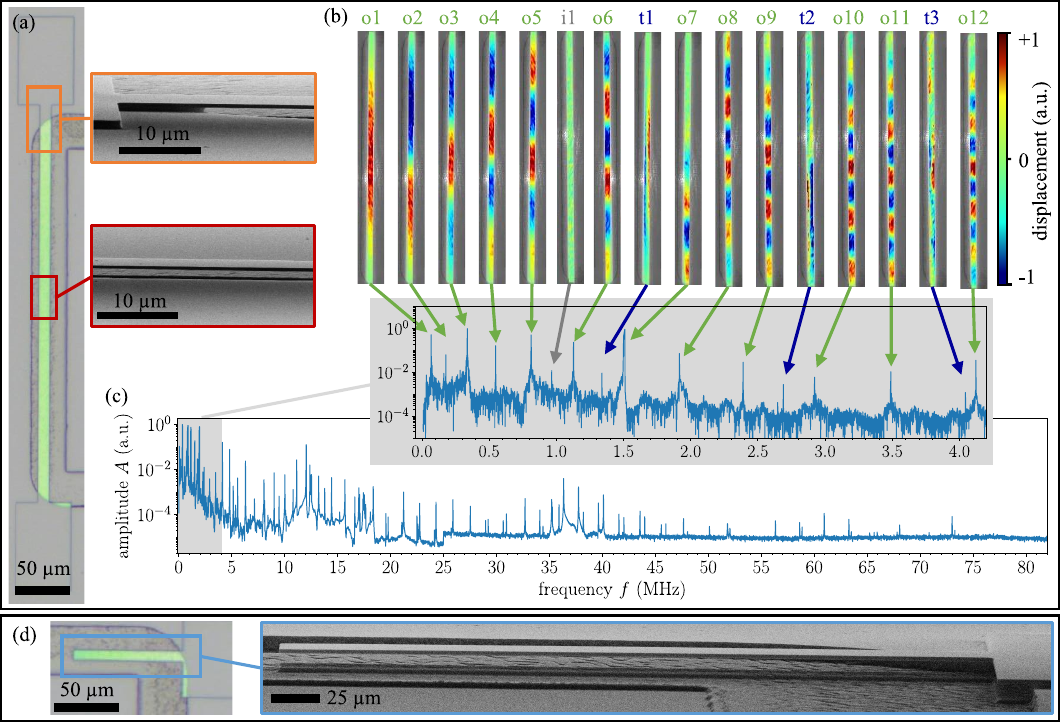}
  \caption{Examples of investigated devices on sample B. Top (a)-(c) and bottom (d) feature a ${\SI{344}{\um}\times\SI{10}{\um}}$ beam and a  ${\SI{70}{\um}\times\SI{8}{\um}}$ cantilever, respectively. (a)~Optical and SEM micrographs of the beam, the green color indicates underetched areas. 
  (b)~Mode shapes obtained as Laser-Doppler vibrometry maps (red and blue indicate opposite directions of the relative amplitude with respect to a reference point; 594 scan points (6 along width, 99 along length) form a 2D surface representation with interpolated data). Out-of-plane modes are labelled with \labo, in-plane modes with \labi and torsional modes with \labt.
  (c)~Spectrum from \SI{0}{\MHz} to about \SI{80}{\MHz} recorded with interferometric read-out and a close-up into the low-frequency range. Comparison of (b) and (c) enables a unique mode assignment, which we further confirm by COMSOL simulations, \figref{fig:PN-14_LDV_spectrum}.
  (d)~Optical and SEM micrographs of a ${\SI{70}{\um}\times\SI{6}{\um}}$ cantilever.}
  \label{fig:Panel_SEM_optical_modes}
\end{figure*}

Two chips (sample A and sample B) have been prepared from monolithic 4H-SiC utilizing the electrochemical etching process previously described~\cite{hochreiter_electrochemical_etching_2023}. 
A total of eight mechanical resonators, four from each sample, have been selected for detailed analysis and will be discussed in the following. They comprise both singly- and doubly-clamped resonators (subsequently denoted as cantilevers and beams, respectively). Their length ranges from $15$ to \SI{344}{\um}, examples are depicted in \figref{fig:Panel_SEM_optical_modes}.

For the characterization of surface quality and geometry, we employ scanning electron microscopy (SEM) and atomic force microscopy (AFM). SEM and AFM images are recorded with a Zeiss Supra 40, and a Park Systems NX10 AFM in non-contact mode at ambient conditions with a nominal resolution of \SI{10}{\nm}, respectively.

For the analysis of the mechanical mode shapes, we use a scanning-laser-Doppler vibrometer (LDV, Polytec MSA-600) with the sample in vacuum (\SI{\approx 10}{\milli\bar}). No external excitation was applied, the measurement solely relies on thermomechanical motion. A \SI{532}{\nm} wavelength CW laser with \SI{\leq2}{\mW} is used and focused to about \SI{2.3}{\um} with a plan-corrective objective.
The vibrational map was generated by a point-wise recording of mechanical spectra with a phase-reference near the center of the resonator. The scanning step size was adapted to the sample size with typical values between $2$ to \SI{3}{\um}. Integration times were chosen significantly large such that the frequency resolution of our results were not affected.

In order to precisely determine the remaining mechanical properties, specifically the quality factor, the Young's modulus and eventual residual stress, we measure mechanical resonances also in a different, interferometric setup both by spectral analysis and ringdown measurements under vacuum conditions sufficient to suppress gas damping (${p<\SI[parse-numbers=false]{10^{-3}}{\milli\bar}}$). The samples were excited with a piezo shaker. The measurement is performed using optical interferometric read-out as described in more detail in \figref{fig:Setup} of the Appendix, using a Koheras BasiK E15 \SI{1550}{\nm} laser with \SI{100}{\uW} optical power on the device under investigation. The laser power of \SI{100}{\uW} is large enough to attain a high signal-to-noise ratio and much smaller than the power at which laser heating effects become visible (see \appref{app:laser-power}).
Spectral analysis is performed by sweeping the drive frequency across each resonance, and by probing the response to the drive at each drive frequency using a vector network analyzer. The drive power is adjusted for each mode to not exceed the linear response regime, and set close to the upper edge of the dynamic range to maximize the signal-to-noise ratio.
The eigenfrequency \omeig and damping \damping are extracted from the resulting amplitude spectra by fitting to the linear response curve (see \appref{app:conventions}).

Ringdown measurements are obtained by coherently driving the mode exactly on resonance, before switching off the drive power abruptly. The resulting transient decay of the intensity is recorded with a spectrum analyzer and fit with an exponential function to extract \damping via the energy decay rate (see \appref{app:conventions}).

All measurements discussed in this work are performed at room temperature.

\subsection{Sample post-processing and surface optimizations}

Sample A was exposed to several thermal annealing (TA) steps, already described in Ref.~\onlinecite{hochreiter_electrochemical_etching_2023}. Here, we characterize the mechanical properties of the sample in its final state. The investigation of sample B followed an overall strategy of alternating sample processing steps and characterization, in order to understand the correlation between surface modifications and mechanical properties (i.e. quality factor). In total, sample B was exposed to a sequence of four processes: two atomic layer etching (ALE) steps followed by two thermal annealing (TA) steps.

The TA-process was performed in a vertical cold-wall furnace with $900$\,mbar Argon pressure (purity 5.0) \cite{emtsev_graphene_growth_Argon_6H-SiC_2009} for $15$ to \SI{30}{min}. 
The ALE process was performed at room temperature in an Oxford Instruments PlasmaPro 100 Cobra System under \SI{25}{mTorr} with consecutive chlorine and argon steps.

\section{Results}

\subsection{Mode identification}

Figure \ref{fig:Panel_SEM_optical_modes} gives an example of sample geometries, the resulting vibrational mode maps and the recorded spectra. With this extended set of data, a clear assignment of the mechanical resonances can be performed. In the optical image on the left of \figref{fig:Panel_SEM_optical_modes}(a), a ${\SI{344}{\um}\times\SI{10}{\um}}$ SiC-beam is displayed, the underetched part of which appears in green. The closeups depict two SEM images that illustrate the free-standing geometry of the long beam resonator. The mechanical spectra have been recorded using two complementary techniques. Spatial mode maps of the observed mechanical resonances recorded using the LDV are depicted as false color images in \ref{fig:Panel_SEM_optical_modes}(b) (red/blue = positive/negative displacement). Out-of-plane (\labo), in-plane (\labi) and torsional modes (\labt) are labelled in green, gray and blue, respectively, where the numbers indicate the harmonic mode index. A wide spectrum from \SI{10}{\kHz} to \SI{75}{\MHz} recorded with the optical interferometer is displayed in \figref{fig:Panel_SEM_optical_modes}(c), along with a closeup of the low-frequency range. The measured eigenfrequencies of the modes observed in Figs.~\ref{fig:Panel_SEM_optical_modes}(b) and (c) do not completely coincide, as the high laser power employed in the LDV measurement induces heating of the sample and thus slightly lowers the eigenfrequencies (see Appendix \figref{fig:PN-14_LDV_spectrum}). However, an unambiguous assignment of each vibrational mode map to a spectral peak can be obtained. The modes that are indicated with green and blue arrows are identified as out-of-plane flexural and torsional modes, respectively. Due to the imperfect, slightly oblique clamping, which can be discerned in the optical micrograph in Fig.~\ref{fig:Panel_SEM_optical_modes}(a), a certain admixture between torsional and flexural modes can in principle appear beyond the first torsional mode \labt[1]. 
However, any signs of hybridization are absent in the mode maps shown in Fig.~\ref{fig:Panel_SEM_optical_modes}(b). The lowest hybridized torsional-flexural mode pair we observed in this device is found at \SI{>6}{\MHz} and entails \labo[15] and \labt[5] (see Fig.~\ref{fig:torsional_admixture}).
For the first six out-of-plane modes labeled \labo[1] to \labo[6], lying below the first torsional mode \labt[1], the admixture of torsional components can be neglected. 
Interestingly, the mode close to \SI{1}{\MHz} labeled by a gray arrow is not observed by the LDV and only appears in the interferometric data. We interpret this resonance as the fundamental in-plane mode \labi[1]. 
The reason for the peak's absence in the LDV data may be the small thermal vibration amplitude due to the beam's large width. The interferometric detection involves a piezo drive, whereas the LDV measurements in this work rely on Brownian motion. Both techniques are predominantly sensitive to out-of-plane motion, but alignment or device geometry imperfections may allow for the detection of in-plane modes.
The eigenfrequency of \labi[1] is consistent with the expected in-plane fundamental eigenfrequency, which is obtained by scaling the \labo[1] frequency by the ratio of the area moments of inertia of the beam in in-plane and out-of-plane directions.

For the entity of resonances, by combining interferometric read-out data with the LDV data as well as Euler-Bernoulli theory, we obtain a convincing assignment scheme that continues up to about \SI{10}{\MHz}. For higher frequencies, we interferometrically observe further resonances to up to \SI{75}{\MHz}, which can be quantitatively analyzed Euler-Bernoulli theory. The complementary mode-mapping information is not available in this frequency regime as a result of an insufficient signal-to-noise ratio in the LDV due to the absence of external excitation.

\subsection{Intrinsic mechanical quality factors exceeding ${\mathbf{10^5}}$}
\label{sec:qintr}

\begin{figure}
  \centering
  %
  %
  \includegraphics[scale=\figscale]{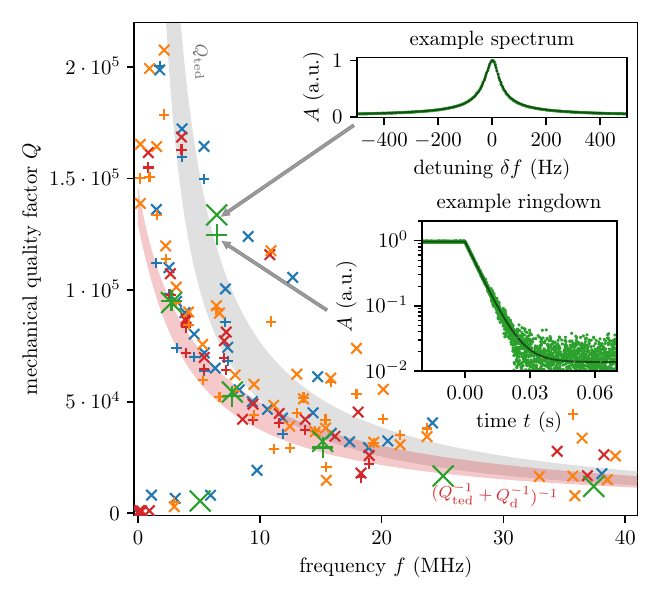}\\[-.6cm]
  ~\raisebox{.65cm}{sample~A:}\includegraphics[scale=\figscale]{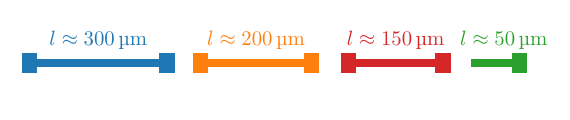}\\[-.6cm]
  \caption{
    Mechanical quality factors of sample A as a function of frequency. The dataset comprises three beams and one cantilever, indicated by different colors in the legend, and combining data obtained from spectral~(\textsf{x}) and ringdown~(\textsf{+}) measurements. The insets display example traces (green dots) for spectral and ringdown measurements with the respective fits (black solid line). The expected \Q of pure thermoelastic damping \Qted and the combined ${(\Qted^{-1}+\Qconst^{-1})^{-1}}$ are indicated by the gray and red shaded areas, respectively. The width of the shaded areas
    arises from the uncertainty in the determination of \youngs (see \secref{sec:youngs}).
  }
  \label{fig:quality-factors}
\end{figure}

The interferometric measurement setup is employed to analyze the mechanical quality factor $\Q=\om/\damping$. \Figref{fig:quality-factors} provides an overview of the results for sample A, which features a cantilever of length \SI{50}{\um} and beams of length $150$, $200$ and \SI{300}{\um}, encoded by color. It includes all detected modes up to \SI{40}{\MHz}. Quality factors obtained from spectral (linewidth) and ringdown measurements are denoted by different symbols (\textsf{x}, \textsf{+}) and illustrated in the upper and lower inset of \figref{fig:quality-factors}, respectively.  
We observe high mechanical quality factors exceeding $10^5$ for frequencies up to approx. \SI{10}{\MHz}.

These quality factors are denoted intrinsic quality factors in Fig.~\ref{fig:quality-factors}, as the investigated samples exhibit only marginal residual stresses (see \secref{sec:youngs}).
Dissipation dilution, by definition the only effect that results in measured \Q exceeding \Qintr, therefore only minimally affects the quality factor.
Given that we cannot exclude a certain extrinsic contribution (\appref{app:extrinsic}), the measured \Q should be seen as a lower bound of \Qintr.
The highest quality factors around \num{2e5} are observed for the \SI{200}{\um} and \SI{300}{\um}-long beams at frequencies between ${0.9}$ and \SI{2.2}{\MHz} (blue and orange data in \figref{fig:quality-factors}). 
Note that the \Q value of each of these modes is obtained from averaging the fit results of at least 8 response curve or ringdown measurements, respectively.
Spectral and ringdown analysis methods yield consistent \Q results.

The highest observed quality factors for a cantilever device with a length of \SI{50}{\um} are highlighted by slightly larger symbols and displayed in the inset of \figref{fig:quality-factors}. The singly-clamped geometry of the cantilever guarantees a stress-free resonator, such that even marginal dissipation dilution effects can be excluded. We emphasize the excellent data quality. Again, the quality factors extracted from spectral (${\Q=\num{1.3e5}}$) and ringdown data (${\Q=\num{1.2e5}}$) are in very good agreement.

\begin{table}
  \centering
  \begin{ruledtabular}
    \begin{tabular}{ccccc}
      \multicolumn{1}{c}{mat.} &
      \multicolumn{1}{c}{thickness (\unit{\um})} &
      \multicolumn{1}{c}{\Qintr} &
      \multicolumn{1}{c}{\Qvol} &
      \multicolumn{1}{c}{${\Qsurf\cdot\SI{100}{\nm}}$}
      \\\hline\\[-3mm]
    a-\ch{SiN} \cite{villanueva_evidence_surface_2014} & $0.03-2$ & \num{2.8e4} 
        & \num{2.8e4} & \num{6e3} 
        \\
    3C-\ch{SiC} \cite{romero_engineering_dissipation_2020} & $0.07-0.34$ & \num{8e3} 
        & \num{8e3} & \num{1.2e4} 
        \\
    a-\ch{SiC} \cite{xu_highstrength_amorphous_2024} & $0.07-0.14$ & \num{7.3e3} 
        & ${\num{7.3e3}~{^\mathrm{(*)}}}$ & \num{5.2e3} 
        \\
    \ch{InGaP} \cite{manjeshwar_highq_trampoline_2023} & $0.073$ & \num{8e3} 
        & ${\num{8e3}\,{^\mathrm{(*)}}}$ 
        & ${\num{1e4}~{^\mathrm{(*)}}}$ 
        \\
    \ch{AlN} \cite{ciers_AlN_2024} & $0.29$ & \num{8e3} 
        & ${\num{8e3}\,{^\mathrm{(*)}}}$ 
        & ${\num{3e3}~{^\mathrm{(*)}}}$ 
        \\
    \ch{GaAs} \cite{cadeddu_timeresolved_nonlinear_2016}
        & \num{.1} & \num{1.2e4} 
        & ${\num{1.2e4}\,{^\mathrm{(*)}}}$ 
        & ${\num{6e3}~{^\mathrm{(*)}}}$ 
     \\\hline\\[-3mm]
    \ch{Si} \cite{tao_singlecrystal_diamond_2014} & $0.07 ~\&~ 1.3$ & \num{3.8e5} 
        & ${-~^\mathrm{(**)}}$ 
        & ${\num{1e4}\,{^\mathrm{(*)}}}$
        \\
    \ch{Si} \cite{metcalf_thermoelastic_damping_2009} & \num{1.5} & \num{1.4e6} 
        & ${\num{1.4e6}\,{^\mathrm{(*)}}}$ 
        & ${\num{9.3e4}\,{^\mathrm{(*)}}}$
        \\
    diam. \cite{tao_singlecrystal_diamond_2014} & $0.1-1$ & \num{1.8e6} 
        & ${-~^\mathrm{(**)}}$ & \num{1e5}
        \\
    4H-\ch{SiC} \cite{adachi_singlecrystalline_4hsic_2013} & 0.9 & \num{2.3e5}
        & ${\num{2.3e5}\,{^\mathrm{(*)}}}$ 
        & ${\num{2.6e4}~{^\mathrm{(*)}}}$ 
        \\
    4H-\ch{SiC} \cite{sementilli_lowdissipation_nanomechanical_2025} & $0.12-0.5$ & \num{4e4} 
        & \num{1.5e5} & \num{1.2e4}
        \\
    4H-\ch{SiC}$^\mathrm{\,here}$\hspace{-2mm} & 0.58 & \num{1.4e5} 
        & ${\ourQvolBound\,{^\mathrm{(*)}}}$ 
        & ${\ourQsurfBoundNum~{^\mathrm{(*)}}}$ 
        \\
    \end{tabular}
  \end{ruledtabular}
  \caption{
    Overview of reported values for the highest measured intrinsic \Qintr and the highest reported volume and surface loss contributions \Qvol and \Qsurf. The first five results as well as 4H-\ch{SiC} \cite{sementilli_lowdissipation_nanomechanical_2025}
    have been obtained from pre-stressed devices by mathematically excluding the effects of dissipation dilution. The quality factors for \ch{Si}~\cite{metcalf_thermoelastic_damping_2009} were measured after in-situ thermal annealing. The prefix `a' denotes amorphous materials, otherwise crystalline resonator materials are assumed. 
    For results marked with ${^\mathrm{(*)}}$ a conclusive analysis of \Qvol and \Qsurf is not feasible, because not enough device thicknesses have been investigated. In this case, we give lower bounds deduced from the highest reported \Qintr. For the results marked with ${^\mathrm{(**)}}$, there is no sign of saturation to a \Qvol even for the thickest cantilevers.
  }
  \label{tab:qintr-qvol-qsurf}
\end{table}

The observation of intrinsic room temperature quality factors in excess of ${\Qintr>10^5}$ is a remarkable finding. Most materials considered so far for nanomechanical resonators with sub-micron thicknesses show intrinsic quality factors ${\Qintr<\num{5e4}}$ for flexural modes at room temperature (\tabref{tab:qintr-qvol-qsurf}). 
Three materials are known to exceed this range significantly: Diamond~\cite{tao_singlecrystal_diamond_2014}, Si~\cite{tao_singlecrystal_diamond_2014,metcalf_thermoelastic_damping_2009} and monolithically etched 4H-SiC~\cite{adachi_singlecrystalline_4hsic_2013,sato_fabrication_electrostatically_2014}, for which intrinsic quality factors up to ${\Qintr\approx\num{1.8e6}}$, ${\Qintr\approx\num{1.4e6}}$ and ${\Qintr\approx\num{2.3e5}}$ have been reported, respectively. Recent measurements of 4H-SiC-on-insulator cantilevers and bridges \cite{sementilli_lowdissipation_nanomechanical_2025} were consistent with a volumetric \Qintr of \num{1.5e5}, but were limited to ${\Qintr\approx\num{4e4}}$ by surface losses.
Even higher intrinsic quality factors at room temperature have only been reported for sapphire~\cite{locke_sapphire_2000} and quartz~\cite{galliou_quartz_2011} macroscopic bulk acoustic resonators.

The analysis of sample B yields results qualitatively similar to those of sample A, which had already undergone the post-treatment described in Ref.~\onlinecite{hochreiter_electrochemical_etching_2023}. All measured quality factors of sample A and the pristine sample B before post-processing are compiled in Fig.~\ref{fig:full-q-vs-f-plot}. However, the $Q$ values observed in sample B are slightly lower than those found in sample A, with a maximum \Q of \num{1.3e5} for doubly-clamped beams and \num{6e4} for cantilevers.
%
The overall trend in Figs.~\ref{fig:quality-factors} and~\ref{fig:full-q-vs-f-plot} suggests a $1/f$ scaling of the quality factor, irrespective of the structural length.
Notice that the modes on the lower left corner of both figures exhibiting significantly lower quality factors despite displaying a good signal-to-noise ratio are likely attributed to experimental artifacts caused by the piezo drive.

\begin{figure} 
  \centering
  %
  %
  %
  \includegraphics[scale=\figscale]{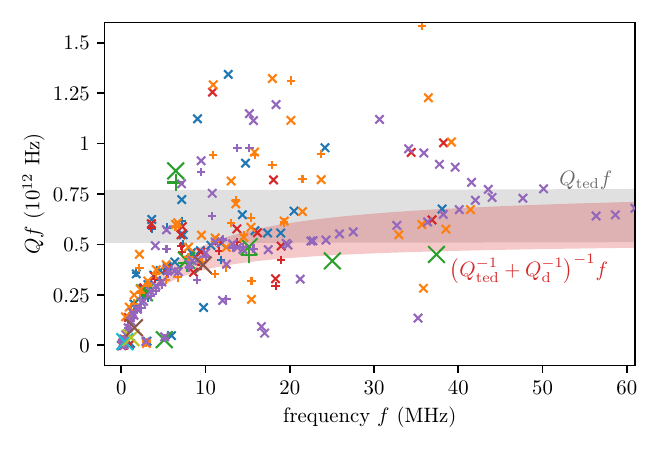}\\[-.3cm]
  ~\raisebox{.65cm}{sample~A:}\includegraphics[scale=\figscale]{pn25-legend}\\[-.6cm]
  ~\raisebox{.65cm}{sample~B:}\includegraphics[scale=\figscale]{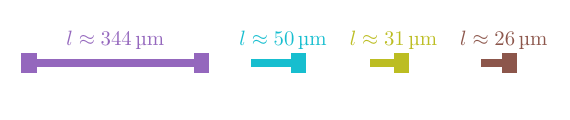}\\[-.6cm]
  \caption{
    \Qfprod products of samples A and the pristine sample B. The dataset comprises three beams and one cantilever on sample A and one beam and three cantilevers on sample B, indicated by different colors in the legend, and combining data obtained from spectral~(\textsf{x}) and ringdown~(\textsf{+}) measurements. Larger symbols were chosen for resonators with only a few data points. The gray and red lines trace the \Qfprod products for \Qted and ${(\Qted^{-1}+\Qconst^{-1})^{-1}}$ with a frequency-independent ${\Qconst=\QconstVal}$, respectively, analogous to \figref{fig:quality-factors}.
    The width of the shaded areas arises from the uncertainty in the determination of \youngs (see \secref{sec:youngs}).
   }
  \label{fig:ted}
\end{figure}

In Fig.~\ref{fig:ted} 
we show that the $1/f$ behavior can be accounted for by thermoelastic damping. 
The gray shaded area represents the theoretical low-frequency limit for the thermoelastic damping of flexural out-of-plane 4H-SiC resonators (orders of magnitude below the Debye frequency). 
In this limit, $\Qted\cdot\freq$ has a constant value for flexural modes of beams which is entirely defined by bulk material parameters and the thickness in the direction of vibration.
It thus sets an upper bound for $\Qfprod$ values of flexural modes. 
Note that the gray shaded area includes a comparatively large error. This is a consequence of the large error on Young's modulus E, which we determine from cantilever frequency measurements (see \secref{sec:youngs}).
The result we obtain for \youngs has an uncertainty of \SI{\approx20}{\percent} due to geometric imperfections of our devices.
It is thus important to propagate this \youngs uncertainty throughout the analysis to recognize potential error amplification.

When further assuming frequency independent intrinsic damping represented by \Qconst, a common hypothesis for defect-mediated losses~\cite{gonzalez_brownian_1994, kajima_1999,bernardini_1999,fedorov_structural-damping_2017,klass_unpublished},
the resulting expectation is 
\begin{align}
  \Qfprod=(\Qconst^{-1} + \Qted^{-1})^{-1}\cdot\freq. 
  \label{eqn:qconst-qted}
\end{align}
We apply a least-squares fit to all data with ${\Q>10^4}$ to estimate the value of \Qconst.

One may be concerned about the feasibility of this fit, because some measured \Q lie below the common trend or exceed \Qted.
%
As the theoretical model underlying \Qted is only valid for flexural modes and thermoelastic damping is negligible in torsional modes, we
hypothesize the outliers exceeding \Qted to be torsional modes (see \figref{fig:Panel_SEM_optical_modes}) or of torsional admixture (\figref{fig:torsional_admixture}), supported by LDV mode shape analysis. The majority of the data however follows the trend implied by \eqnref{eqn:qconst-qted}, suggesting a predominant flexural character of these modes. 
Lacking an unambiguous criterion to classify the observed modes, we include all data points in the fit, and only omit those with ${\Q<10^4}$ attributed to resonances damped by local contamination or to drive piezo artifacts.

Fortunately, the large dataset enables a robust fit despite the presence of these outliers and is shown in Figs.~\ref{fig:quality-factors},~\ref{fig:ted}~and~\ref{fig:full-q-vs-f-plot} by the area shaded in red, yielding ${\Qconst=\QconstVal}$. The good agreement between the fit and the majority of data confirms that the outliers did not significantly distort the fit result.
The width of the shaded areas in Figs.~\ref{fig:quality-factors},~\ref{fig:ted}~and~\ref{fig:full-q-vs-f-plot} represent the propagated effect of the \youngs uncertainty on the \Qted and \Qconst analysis. 
Being obtained by a fit, our \Qconst result may be affected by small extrinsic contributions in the data (\appref{app:extrinsic}). 

At about \SI{10}{\MHz}, both contributions, \Qted and \Qconst have equal weights. For lower frequencies, the value of \Qconst limits the \Qfprod product, while for higher frequencies, the \Qted contribution dominates.
It is remarkable to find quality factors close to the thermoelastic limit for resonators with a thickness of only \SI{500}{\nm}. In this regime, a pronounced influence of surface-limited dissipation~\cite{villanueva_evidence_surface_2014,metcalf_thermoelastic_damping_2009} is expected, which in our case clearly does not constitute the dominant contribution to the quality factor.
From the resulting ${\Qconst=\QconstVal}$ of the above analysis, we deduce lower bounds for the entailed surface and volume contributions using
\begin{align}
  \Qconst^{-1} = \Qvol^{-1} + \qty(\Qsurf\thick)^{-1},
  \label{eqn:qconst}
\end{align}
with the volume quality factor \Qvol and the surface quality factor per thickness \Qsurf.
We exploit the fact that \eqnref{eqn:qconst} implies lower bounds for \Qvol and ${\Qsurf\thick}$ when \Qconst is known.
Therefore, we conclude ${\Qvol\gtrsim\ourQvolBound}$ and ${\Qsurf\gtrsim\ourQsurfBound}$.
This bound for \Qvol is comparable to a recent work on 4H-\ch{SiC}-on-insulator \cite{sementilli_lowdissipation_nanomechanical_2025} and \SI{\approx30}{\percent} lower than the reported \Q in a previous work on monolithic 4H-SiC~\cite{adachi_singlecrystalline_4hsic_2013}, which reports on a \SI{50}{\percent} thicker device, however (\tabref{tab:qintr-qvol-qsurf}).
The surface quality factor bound ${\Qsurf\gtrsim\ourQsurfBound}$  that we find is comparable to the previous monolithic 4H-SiC results~\cite{adachi_singlecrystalline_4hsic_2013} and exceeds the \Qsurf found for 4H-\ch{SiC}-on-insulator devices \cite{sementilli_lowdissipation_nanomechanical_2025} by a factor of two.

\subsection{Young's modulus and residual stress}
\label{sec:youngs}

\begin{figure*}
    \centering
    \begin{tikzpicture}
      \node[anchor=south west] at (0,0) {\includegraphics[scale=\figscale]{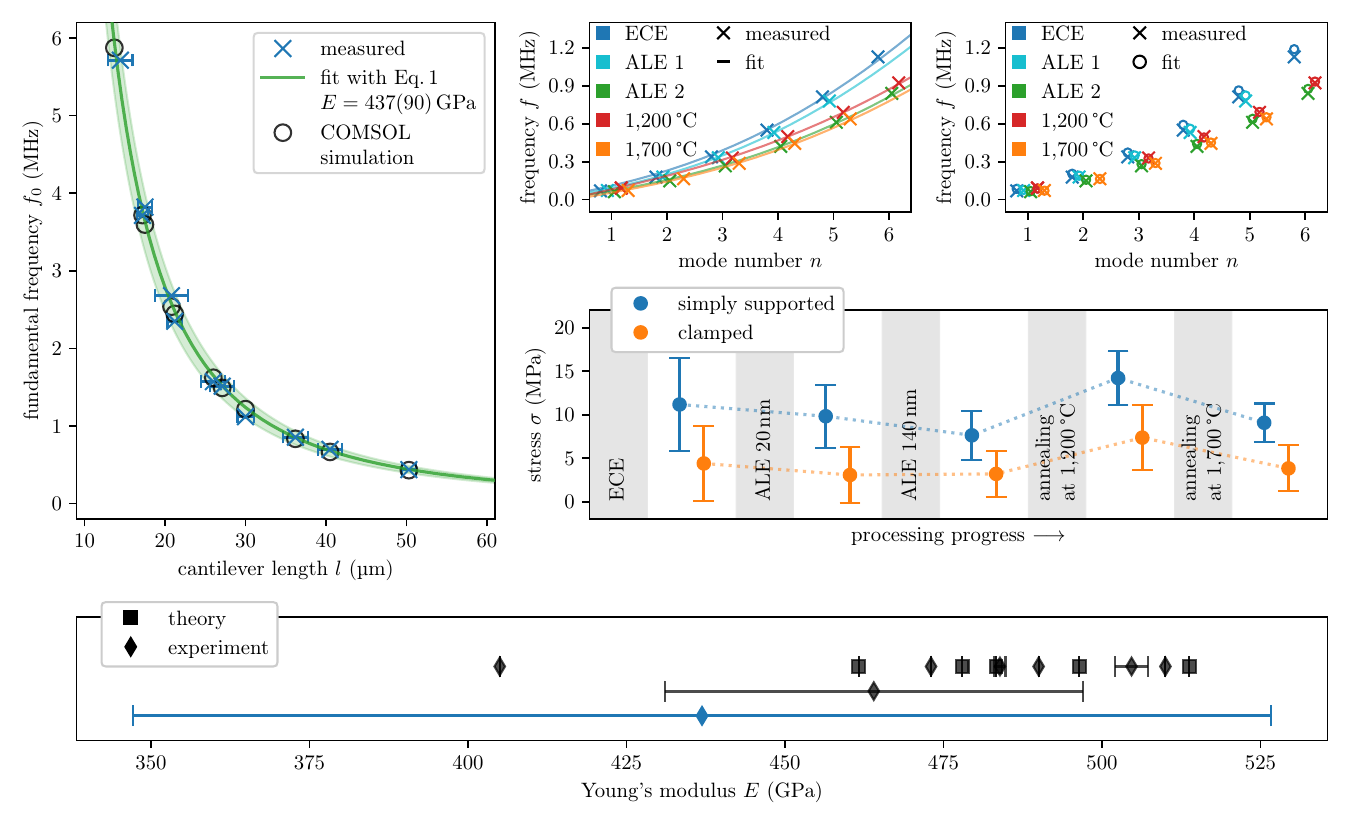}};
      %
      \begin{scope}[shift={(0,.6)}, x=1cm, y=1.04cm]
        \node at (.5,   9.9) {\subfloat[]{~\label{subfig:youngs-determination}}};
        \node at (7,   10.1) {\subfloat[]{~\label{subfig:stress-fits-simply}}};
        \node at (12.4,10.1) {\subfloat[]{~\label{subfig:stress-fits-clamped}}};
        \node at (7,    6.3) {\subfloat[]{~\label{subfig:stress-fits-results}}};
        \node at (.5,   2.5) {\subfloat[]{~\label{subfig:youngs-literature}}};
      \end{scope}
      %
      \begin{scope}[shift={(1.15,1.7)}, x=16.55cm, y=0.8cm]
        \node[anchor=center] at (0.33657,1) {\scriptsize\cite{messaoud_investigation_youngs_2019}};
        \node[anchor=center] at (0.62212,1) {\scriptsize\cite{xu_hightemperature_mechanical_2018}};
        \node[anchor=center] at (0.64424,.6) {\scriptsize\cite{islam_singlecrystal_sic_2012}};
        \node[anchor=center] at (0.67980,1) {\scriptsize\cite{karmann_piezoelectric_properties_1989}};
        \node[anchor=center] at (0.70467,1) {\scriptsize\cite{pizzagalli_stability_mobility_2014}};
        \node[anchor=center] at (0.73141,1.25) {\scriptsize\cite{yahagi_deformation_behavior}};
        \node[anchor=center] at (0.73141,.95) {\scriptsize\cite{nuruzzaman_structural_elastic_2015}};
        \node[anchor=center] at (0.76541,1) {\scriptsize\cite{yang_investigating_elastic_2020}};
        \node[anchor=center] at (0.79988,1) {\scriptsize\cite{sakakima_development_method_2018}};
        \node[anchor=center] at (0.83637,1) {\scriptsize\cite{kamitani_elastic_constants_1997}};
        \node[anchor=center] at (0.86331,1) {\scriptsize\cite{chen_study_elastic_2019}};
        \node[anchor=center] at (0.89144,1) {\scriptsize\cite{iuga_abinitio_simulation_2007}};
        \node[anchor=center] at (0.50200,-.42) {\color{tabblue}\scriptsize(this work)};
      \end{scope}
    \end{tikzpicture}
    \vspace{-.8cm}
    \caption{Characterization of Young's modulus and tensile pre-stress.
      \subref*{subfig:youngs-determination}~Cantilever data for determination of Young's modulus. Fundamental resonance frequency (blue symbols) as a function of length
      and fit with ~\eqnref{eq:Eulerplate} (green line) for eleven cantilevers on sample A, yielding a Young's modulus ${\youngs=\youngsVal}$. Eigenfrequencies simulated using COMSOL Multiphysics using this value as an input parameter are indicated by black circles to illustrate eventual deviations from the Euler-Bernoulli assumptions. The uncertainty of \youngs obtained from propagating the uncertainties of the cantilever lengths (see error bars, thickness ${\thick=\SI{574\pm28}{\nm}}$, density ${\dens=\SI{3.2(1)e3}{\kg\per\cubic\m}}$ and Poisson ratio ${\poisson=\num{0.25\pm0.1}}$ is indicated by a green shading.
      \subref*{subfig:stress-fits-simply} and~\subref*{subfig:stress-fits-clamped}~Beam data for determination of tensile pre-stress. Eigenfrequencies of the first 6 modes of the \SI{344}{\um} beam of sample B (crosses), and fit with simply supported (line) and clamped boundary conditions (open circles), respectively. State of post-processing discussed in Sec.~\ref{post-processing} indicated by color.
      \subref*{subfig:stress-fits-results} Tensile pre-stress obtained using both boundary conditions after each processing step. 
      \subref*{subfig:youngs-literature} Comparison of determined Young's modulus (blue) with literature values of 4H-SiC in the (0001)-plane (black). Squares and diamonds correspond to theoretical results obtained with ab initio calculation and experimentally determined values using various methods, respectively. For references specifying only elastic constants, we calculate the Young's modulus (see \appref{app:youngs-from-cij}). The value reported in Ref.~\cite{islam_singlecrystal_sic_2012} was determined through calculations based on the resonant frequencies provided in the original reference. The result of Ref.~\cite{iuga_abinitio_simulation_2007} was not calculated for 4H-\ch{SiC} in particular, but for hexagonal \ch{SiC} in general.}
    \label{fig:Euler-Bernoulli-fit}
\end{figure*}

The detailed mode analysis of sample A allows to quantitatively extract the Young's modulus \youngs of 4H-SiC from the experimental data \cite{chen_study_elastic_2019}. Due to the crystal direction of our samples, we measure the Young's modulus in the SiC-(0001) plane. We investigate eleven cantilevers with lengths \length ranging from \SI{14}{\um} to \SI{50}{\um} with a common thickness ${\thick=\SI{574\pm28}{\nm}}$. A twelfth cantilever exhibited an irregular eigenfrequency and was excluded from the analysis. For the determination of Young's modulus, we consider the Euler-Bernoulli framework of a vibrating plate, yielding 
\begin{equation}
    f_\mathrm{res} = \lambda_\modeno^2/\length^2\sqrt{\youngs\thick^2/[{12(1-\poisson^2)\dens}]}
    \label{eq:Eulerplate}
\end{equation}
with  mode number \modeno, density \dens, Poisson ratio \poisson and ${\lambda_{1}=1.8751}$. This formula is valid under the assumption of a long beam geometry~\cite{funda-of-nanom-reson}, exhibiting ${\length/\thick\gg1}$ and ${\width/\thick>5}$. In our dataset, all cantilevers fulfill this with ${\length/\thick>25}$, and ${\width/\thick>12}$.
To further ensure that the geometry ratios are sufficiently large, we performed COMSOL simulations (\appref{app:comsol}). We find that the frequency deviations between \eqnref{eq:Eulerplate} and the COMSOL model are below \SI{2.5}{\percent} for all investigated cantilevers.
The measured eigenfrequencies as a function of length are displayed in Fig.~\ref{fig:Euler-Bernoulli-fit}(a).
The green curve is a least-square fit of Eq.~\ref{eq:Eulerplate} to the data yielding ${\youngs=\youngsVal}$. The uncertainty of the obtained Young's modulus is dominated by the geometric uncertainty.

The determination of Young's modulus allows for a more detailed analysis of the doubly-clamped, \SI{344}{\um} beam on sample B, revealing an eventual fabrication-induced pre-stress which may lead to dissipation dilution effects. To this end, we fit the frequencies of the observed out-of-plane modes with increasing mode number (${\modeno = 1 \dots 6}$) with the Euler-Bernoulli prediction for a stressed plate with simply supported boundary conditions (c.f. Eq.~\ref{eq:ebb-analytic} in Appendix~\ref{app:stress-fit}) using stress as the only free parameter.
For comparison, we numerically solve the corresponding Euler-Bernoulli equation with clamped boundary conditions, which is more accurately suited to our monolithic device geometry (see \appref{app:stress-fit}).
The best fits are displayed in Fig.~\ref{fig:Euler-Bernoulli-fit}(b) and (c). 
We find a tensile pre-stress below \SI{10}{\MPa} (${<\num{2.5e-5}}$ strain), as shown in Fig.~\ref{fig:Euler-Bernoulli-fit}(d).
Notice that the simply supported boundary condition only approximately describes the clamping conditions and slightly overestimates the stress.

This finding confirms our initial assumption that the monolithic fabrication process results in nearly unstressed devices, even for doubly-clamped geometries.
The effect of the post-processing steps discussed in Sec.~\ref{post-processing} is also illustrated in Figs.~\ref{fig:Euler-Bernoulli-fit}(b), (c) and (d).

Finally, we discuss the obtained value of ${\youngs=\youngsVal}$ in comparison with experimental values and theoretical predictions of Young's modulus for the same crystal direction, see Fig.~\ref{fig:Euler-Bernoulli-fit}(e). 
This value is in good agreement with previous literature, where the squares and diamonds indicate theoretical and experimental results, respectively.  

\subsection{Influence of post-processing protocols on mechanical parameters}
\label{post-processing}

\begin{figure}
  \centering
  %
  %
  \captionsetup[subfigure]{position=top,
                           singlelinecheck=off,
                           justification=raggedright}
  \subfloat[]{\includegraphics[scale=\figscale,
                trim={.2cm .4cm .2cm .7cm}]{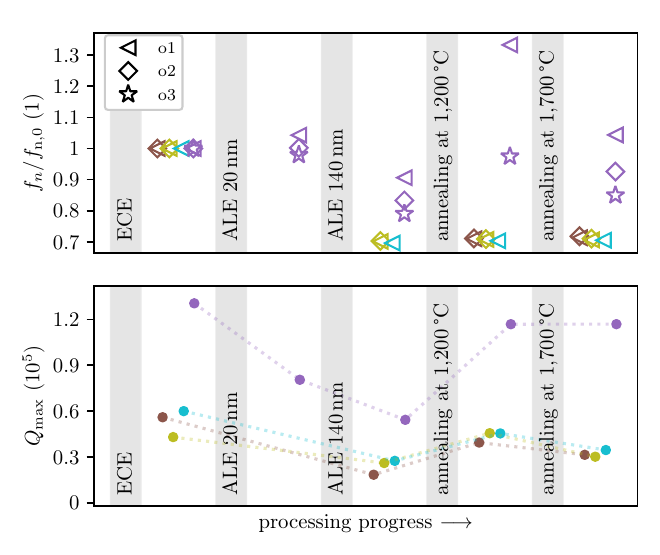}
              \label{subfig:processing-freq}}\\[-3.8cm]
  \subfloat[]{\hspace{\linewidth}
              \label{subfig:processing-quality}}\\[3.3cm]
  ~\raisebox{.65cm}{sample~B:}\includegraphics[scale=\figscale]{pn14-legend}\\[-.6cm]
  \caption{
    Evolution of frequencies and quality factors of sample B after step-wise ALE and thermal annealing. Colors indicate device and symbols indicate mode index. \subref*{subfig:processing-freq}~Normalized frequency shifts of the first three out-of-plane modes \labo[1] -- \labo[3]. \subref*{subfig:processing-quality}~Highest mechanical quality factor observed for each device.
  }
  \label{fig:processing}
\end{figure}

The properties of on-chip mechanical devices critically depend on process protocols. 
In particular, surface conditions may have critical influence on both quality factors and strain.
In the following, we use sample B to shed light on the impact of post-processing routines on the mechanical quality and to obtain insights into the role of the underlying surface modifications. The data of sample B discussed in the context of Figs.~\ref{fig:Panel_SEM_optical_modes},~\ref{fig:ted} and~\ref{fig:full-q-vs-f-plot} has been obtained on the pristine sample, i.e. right after the electrochemical etching (ECE) protocol~\cite{hochreiter_electrochemical_etching_2023} used to realize the freely-suspended devices, where $Q=\num{1.3e5}$ was found for the \SI{344}{\um} long beam and ${\Q\approx\num{5e4}}$ was found for the three cantilevers with lengths $26$, $31$ and \SI{50}{um}.
Here we employ this complete characterization as the starting point of a further analysis. Sample B is exposed to a series of surface treatments, consisting of atomic layer etching (ALE) and thermal annealing. After each step, the sample is thoroughly re-characterized. This iterative scheme allows to map out the effect of the individual procedures, and to identify the optimum post-processing protocol. Figure~\ref{fig:processing} traces the effect of the post-processing steps on the eigenfrequency and the quality factor. A more detailed summary of the quantitative analysis of all sample parameters monitored during the post-processing is included in Table~\ref{tab:Sample_parameters}.

As a first step, sample B is exposed to an ALE process consisting of \num{98} cycles resulting in a mild and accurately-controlled \SI{20}{\nm}-deep etch of the SiC surface.  
According to the AFM images depicted in Fig.~\ref{fig:AFM_scans_fabrication_processes}(a) and (b), this procedure has removed nanometer-sized point-like contaminations and has further smoothened the step edges on the SiC(0001) surface (\SI{4}{\degree}-off-axis). With its low plasma power ALE is expected to introduce only minor damage to the bulk material \cite{Khan-RIE_ICP_damage_Cl_F_chemistry-2002,Michaels_SiC_ALE-2023}.
However, the apparent improvement of the sample surface quality does not favorably affect the mechanical performance as shown in Fig.~\ref{fig:processing}. The quality factor of the \SI{344}{\um} long beam drops to approx. $\num{8e4}$. Given the small reduction of the thickness, the resonance frequency remains essentially unchanged.
This result is further confirmed by a second ALE process consisting of $340$ cycles to remove approx. \SI{140}{\nm} of \ch{SiC}. The AFM characterization suggests a further improvement of the surface morphology with low and homogeneously distributed terraces as apparent from Fig.~\ref{fig:AFM_scans_fabrication_processes}(c). However, Fig.~\ref{fig:processing} shows that the mechanical quality factor is further reduced to approx. \num{6e4}, while the resonance frequency slightly decreases as expected for the diminished thickness of the device.
The data on cantilevers follows a similar trend.
While a mild quality factor reduction due to the increased contribution of surface loss is expected for the thinned devices, the observed \Q reduction clearly exceeds the reduction in thickness of \SI{\approx10}{\percent} (\tabref{tab:Sample_parameters}). This indicates that ALE induces additional mechanical damping.
We conclude that ALE is well suited to smoothen the sample surface and to adjust the thickness of the resonator and hence the resonance frequency in a very controlled fashion. However, the quality factor deteriorates significantly, such that ALE is not a suitable post-processing technique to optimize the mechanical performance.

As a next step, we perform high temperature annealing on sample B. SiC on-chip devices have proven to withstand thermal annealing up \SI{1550}{\celsius} without a reconfiguration of the device geometry \cite{hochreiter_electrochemical_etching_2023}. Initially, the sample is annealed at \SI{1200}{\celsius} under \SI{900}{\milli\bar} Ar atmosphere. As shown in Fig.~\ref{fig:processing}, this significantly increases the quality factor, approximately re-establishing the quality of the pristine device.
The increase in the quality factor goes along with a further reduction of the surface roughness according to Fig.~\ref{fig:AFM_scans_fabrication_processes}(d). 
In a subsequent step, sample B is annealed at \SI{1700}{\celsius} under \ch{Ar} atmosphere, with minor effects of the quality factor. Under these conditions, epitaxial graphene growth is observed, along with significant reconfigurations of the surface terraces, see Fig.~\ref{fig:AFM_scans_fabrication_processes}(e). For the freely-suspended devices under investigation, this even leads to minor modifications of the device shape \cite{hochreiter_electrochemical_etching_2023}.

All in all, the post-processing procedures are interpreted as follows: While ALE leads to a reduction of the quality factor, thermal annealing at \SI{1200}{\celsius} effectuates a significant improvement. We hypothesize that thermal annealing of the pristine sample immediately following the ECE procedure will increase the mechanical quality beyond the values reported in this work. As ALE allows to fine-tune the device thickness, a combination of ALE and thermal annealing can be employed to adjust the eigenfrequencies without compromising the quality factor.
There is no significant change of Young's modulus and the tensile stress (c.f.~\ref{fig:Euler-Bernoulli-fit}(b)-(d)).


\begin{table*}
  \centering
  \begin{ruledtabular}
    \newcolumntype{d}{D{+}{\,\pm\,}{-1}}
    \begin{tabular}{rdcddd}
      \multicolumn{1}{r}{last process step} &
      \multicolumn{1}{c}{\stress (\si{\MPa})} &
      \multicolumn{1}{c}{\Qmax} &
      \multicolumn{1}{c}{\Qintr} &
      \multicolumn{1}{c}{thickness (\si{nm})} &
      \multicolumn{1}{c}{RMS roughness (\si{pm})}
      \\\hline\\[-3mm]
    %
    %
      ECE                                 & \num{4.4} + \num{4.3}  &   \num{1.2e5}     &   \num{5e4} + \num{2e4}   &   585 + 21   &   322 + 12 \\
      ALE \SI{20}{\nm}                    & \num{3.1} + \num{3.2}  &   \num{8.0e4}     &   \num{4e4} + \num{2e4}   &   565 + 21    &   283 + 15 \\
      ALE \SI{140}{\nm}                   & \num{3.2} + \num{2.6}  &   \num{5.0e4}     &   \num{2e4} + \num{1e4}   &   425 +  8    &   185 +  8 \\
      annealing at \SI{1200}{\celsius}    & \num{7.4} + \num{3.8}  &   \num{1.2e5}     &   \num{4e4} + \num{1e4}   &   425 +  8    &   160 +  10 \\
      annealing at \SI{1700}{\celsius}    & \num{3.8} + \num{2.6}  &  \num{1.2e5}                &      \num{3e4} + \num{1e4}                &   427+16      & 1880+120
    \end{tabular}
  \end{ruledtabular}
    \caption{Sample B parameters after each processing step: stress \stress, 
    highest observed mechanical quality factor \Qmax, intrinsic quality factor \Qintr, thickness of the devices \thick and RMS surface roughness.}
    \label{tab:Sample_parameters}
\end{table*}

\section{Summary}
We explore the vibrational properties of mechanical resonators monolithically fabricated from crystalline 4H-SiC by means of electrochemical etching. Our modal analysis of singly-clamped cantilevers and doubly-clamped beams expose excellent intrinsic quality factors. 
While the explored cantilevers are guaranteed stress-free, the beams are shown to exhibit a negligible fabrication-induced tensile stress, with an upper bound of approximately \SI{10}{\MPa}. 
Hence effects of dissipation dilution are imperceptible in all of the explored structures, such that the observed quality factors are interpreted as intrinsic values.
The out-of-plane flexural modes feature quality factors up to ${10^5}$, and are shown to approach the fundamental limit of thermoelastic damping. Even more, we find higher-order modes of admixed character involving torsional
components that feature even higher quality factors approaching \num{2e5}. This corresponds to a \Qfprod-product exceeding \SI[parse-numbers=false]{10^{12}}{\Hz}.

Furthermore, we study the effect of post-processing treatments such as atomic layer etching and thermal annealing on the mechanical quality. While atomic layer etching leads to a decrease of the quality factor, it is shown to be restored under high temperature annealing at \SI{1200}{\celsius} and \SI{1700}{\celsius}, pointing out a potential pathway toward further improvement of the mechanical quality.
The value found for the Young's modulus is in agreement with the literature.  

The outstanding quality factors observed in monolithically etched crystalline 4H-SiC are in line with previous work~\cite{adachi_singlecrystalline_4hsic_2013,sato_fabrication_electrostatically_2014}. A recent study on 4H-SiC-on-insulator nanomechanical resonators reports quality factors up to \num{2e7}~\cite{sementilli_lowdissipation_nanomechanical_2025}, albeit under the influence of a more significant fabrication-induced tensile pre-stress leading to dissipation-dilution. Comparing the underlying surface quality factors, SiC-on-insulator falls short compared with monolithic architectures.
This is likely a consequence of surface imperfections induced in the wafer bonding and polishing procedure employed for the creation of the 4H-SiC-on-insulator hybrids.

The unparalleled quality of 4H-SiC nanomechanical resonators suggests 4H-SiC as a promising material for applications targeting highly coherent vibrations, e.g. in quantum opto- or electromechanics, as well as spin-mechanical applications, additionally incorporating color centers.

%
%
%

\begin{acknowledgments}
We gratefully acknowledge financial support from the Deutsche Forschungsgemeinschaft (DFG, German
Research Foundation) through Project-IDs No.
429529648-SFB/TRR 306 (project B03) and No. WE 4721/1-1, as well as under Germany’s Excellence
Strategy-EXC-2111-390814868.
The research is part of the Munich Quantum Valley, which is supported by the Bavarian state government with funds from the Hightech Agenda Bavaria.
We thank Sergey Flyax for his help in developing the peak classification algorithm required to produce the comprehensive dataset.
\end{acknowledgments}

\section*{Conflict of Interest Statement}

The authors have no conflicts to disclose.

\section*{Data Availability Statement}

The data that support the findings of this article
are openly available \cite{hochreiter_monolithic_4hsic_2025_supplemental_data}.

\appendix

\section{Fabrication and post-processing of samples}
\label{app:recipes}

Sample A has been fabricated and exposed to multiple thermal annealing steps before the mechanical characterization described in this work. Details on process and post-processing routines are described in Ref.~\onlinecite{hochreiter_electrochemical_etching_2023}.\\

Sample B is processed as follows:
\begin{enumerate}
    \item Electrochemical etching (ECE) of resonators, see Ref.~\onlinecite{hochreiter_electrochemical_etching_2023}.
    \item Initial sample characterization with AFM, SEM, LDV and interferometry.
    
    \item First atomic layer etching (ALE): 
    \begin{itemize}
        \item Number of ALE cycles: $98$ 
        \item RIE power: $10$\,W
        \item SiC thickness reduction: $20$\,nm
    \end{itemize} 
    Re-characterization of the sample with AFM, SEM, LDV and interferometry.

    \item Second atomic layer etching (ALE): 
        \begin{itemize}
        \item Number of ALE cycles: $340$
        \item RIE power: \SI{8}{W} (Note: The power was reduced to \SI{8}{W}, as this power resulted in the smoothest surfaces in previous measurements.)
        \item SiC thickness reduction: \SI{140}{\nm}.
    \end{itemize} 
    Re-characterization of the sample with AFM, SEM, LDV and interferometry.
    
    \item First thermal annealing (TA): 
    \begin{itemize}
        \item Temperature: \SI{1200}{\celsius} 
        \item Atmosphere: \SI{900}{\milli\bar} Argon
        \item Duration: \SI{30}{\min}
    \end{itemize}
    Re-characterization of the sample with AFM, SEM, LDV and interferometry.
    
    \item Second thermal annealing (TA): 
    \begin{itemize}
        \item Temperature sequence: \SI{1275}{\celsius} (\SI{15}{\min}), \SI{1350}{\celsius} (\SI{15}{min}), \SI{1400}{\celsius} (\SI{15}{min}), \SI{1550}{\celsius} (\SI{15}{\min}), \SI{1700}{\celsius} (\SI{15}{min})
        \item Atmosphere: \SI{900}{\milli\bar} Argon
        \item Total duration: \SI{75}{\min}
    \end{itemize}
    Re-characterization of the sample with AFM, SEM, LDV and interferometry.
\end{enumerate}
Sample B was annealed at exactly the same temperature sequence as sample A, the only difference being that sample A was removed after each annealing step and exposed to ambient air for surface characterization, see Ref.~\onlinecite{hochreiter_electrochemical_etching_2023}. Additionally, all annealing durations for sample B were standardized to 30 minutes.

\section{AFM roughness measurements}
\label{app:afm}

\begin{figure*}
  \captionsetup[subfigure]{position=top,
                           singlelinecheck=off,
                           justification=raggedright}
    \centering
    \subfloat[after ECE: rms $= 322 \pm 12$ pm \label{subfig:AFM_sampleB_aECE}]{
    \begin{tikzpicture}
        \node[anchor=south west,inner sep=0] (image) at (0,0) {\includegraphics[width=4cm]{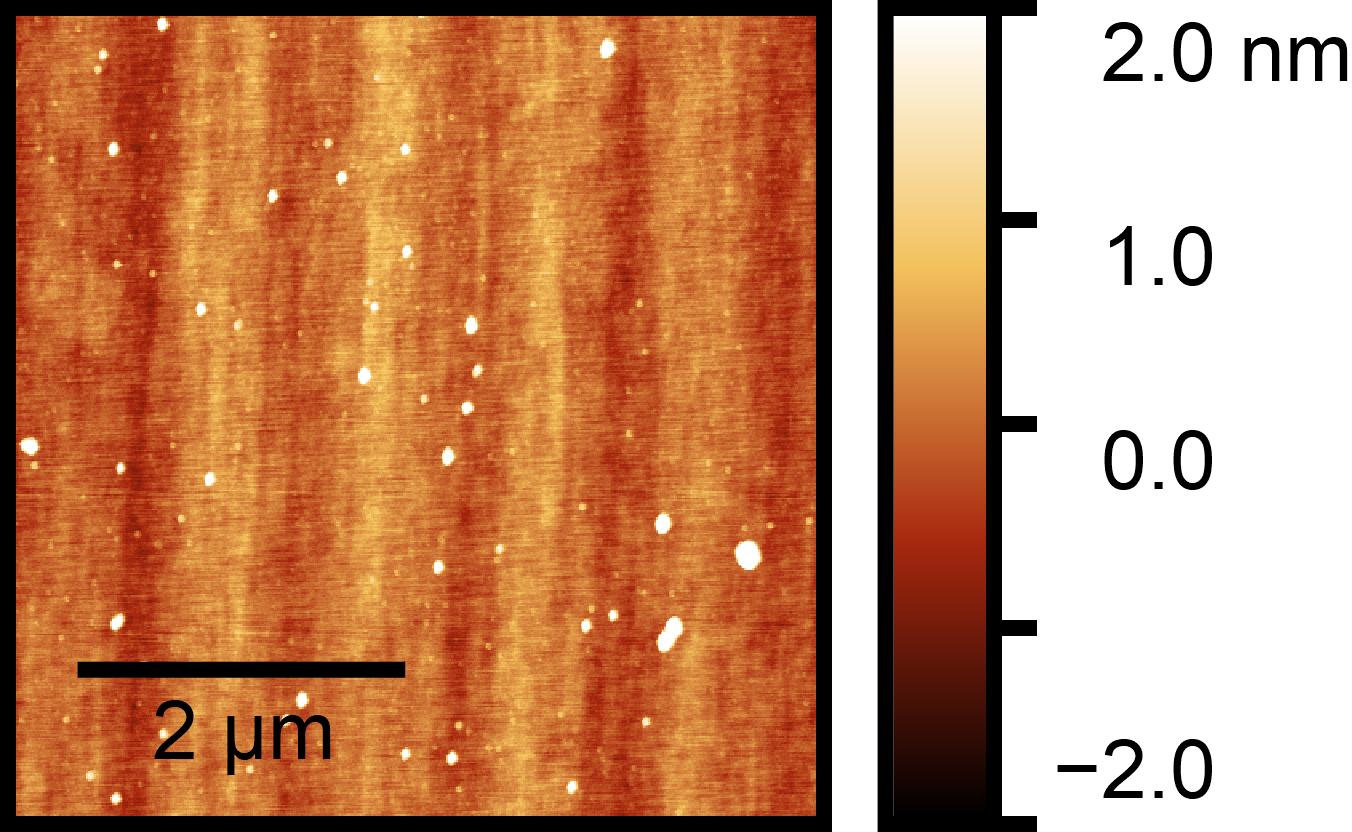}};
        \begin{scope}[x={(image.south east)},y={(image.north west)}]
            \draw [-{Triangle[length=3mm, width=2mm]}, line width = 0.65mm] (0.35,0.7) -- (0.05,0.7);
            \node[black, anchor = south] at (0.2, 0.73) {$[11\bar{2}0]$};
        \end{scope}
    \end{tikzpicture}
    }
    \quad
    \subfloat[after ALE1: rms $= 283 \pm 15$ pm\label{subfig:AFM_sampleB_aALE1}]{
    \begin{tikzpicture}
        \node[anchor=south west,inner sep=0] (image) at (0,0) {\includegraphics[width=4cm]{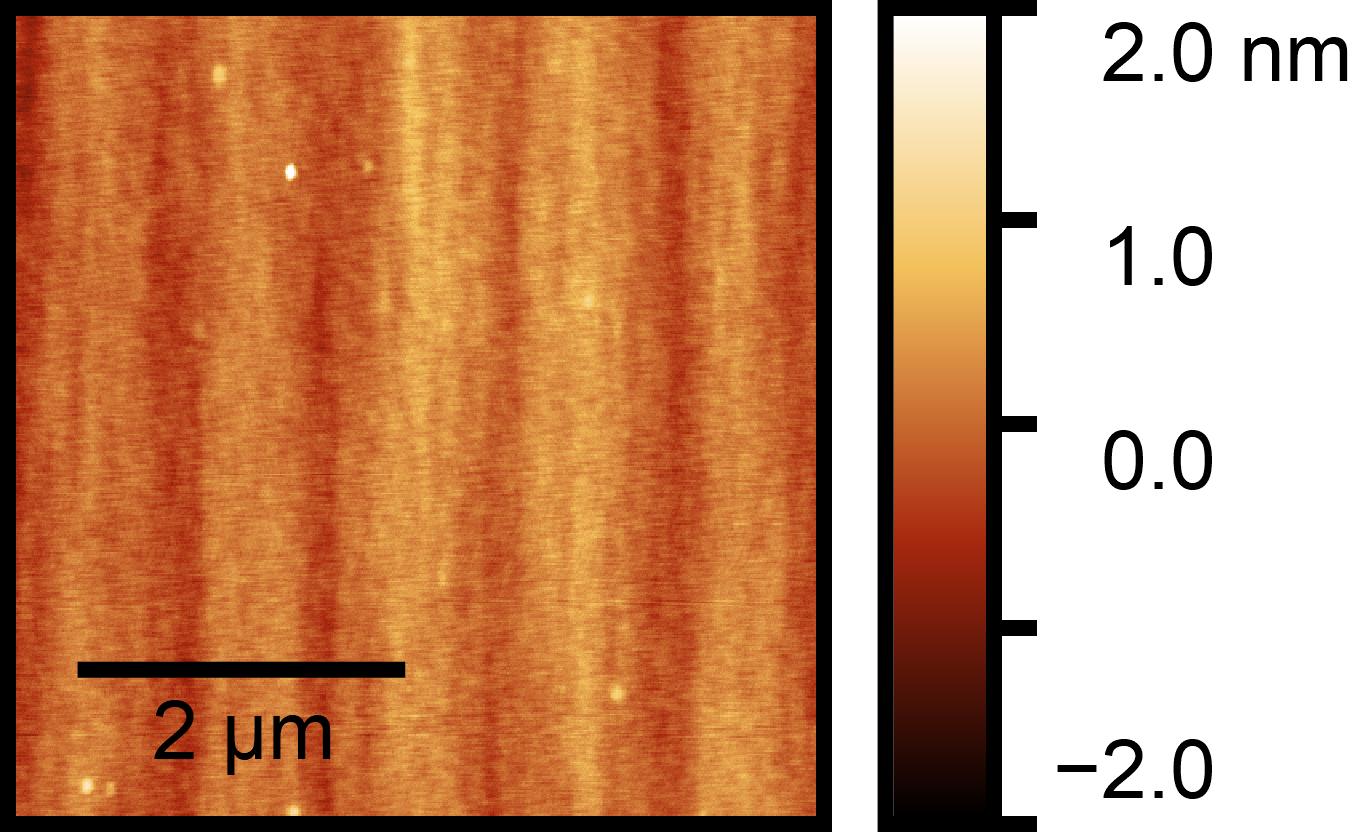}};
        \begin{scope}[x={(image.south east)},y={(image.north west)}]
            \draw [-{Triangle[length=3mm, width=2mm]}, line width = 0.65mm] (0.35,0.7) -- (0.05,0.7);
            \node[black, anchor = south] at (0.2, 0.73) {$[11\bar{2}0]$};
        \end{scope}
    \end{tikzpicture}
    }
    \quad
    \subfloat[after ALE2: rms $= 185 \pm 8$ pm\label{subfig:AFM_sampleB_aALE2}]{
    \begin{tikzpicture}
        \node[anchor=south west,inner sep=0] (image) at (0,0) {\includegraphics[width=4cm]{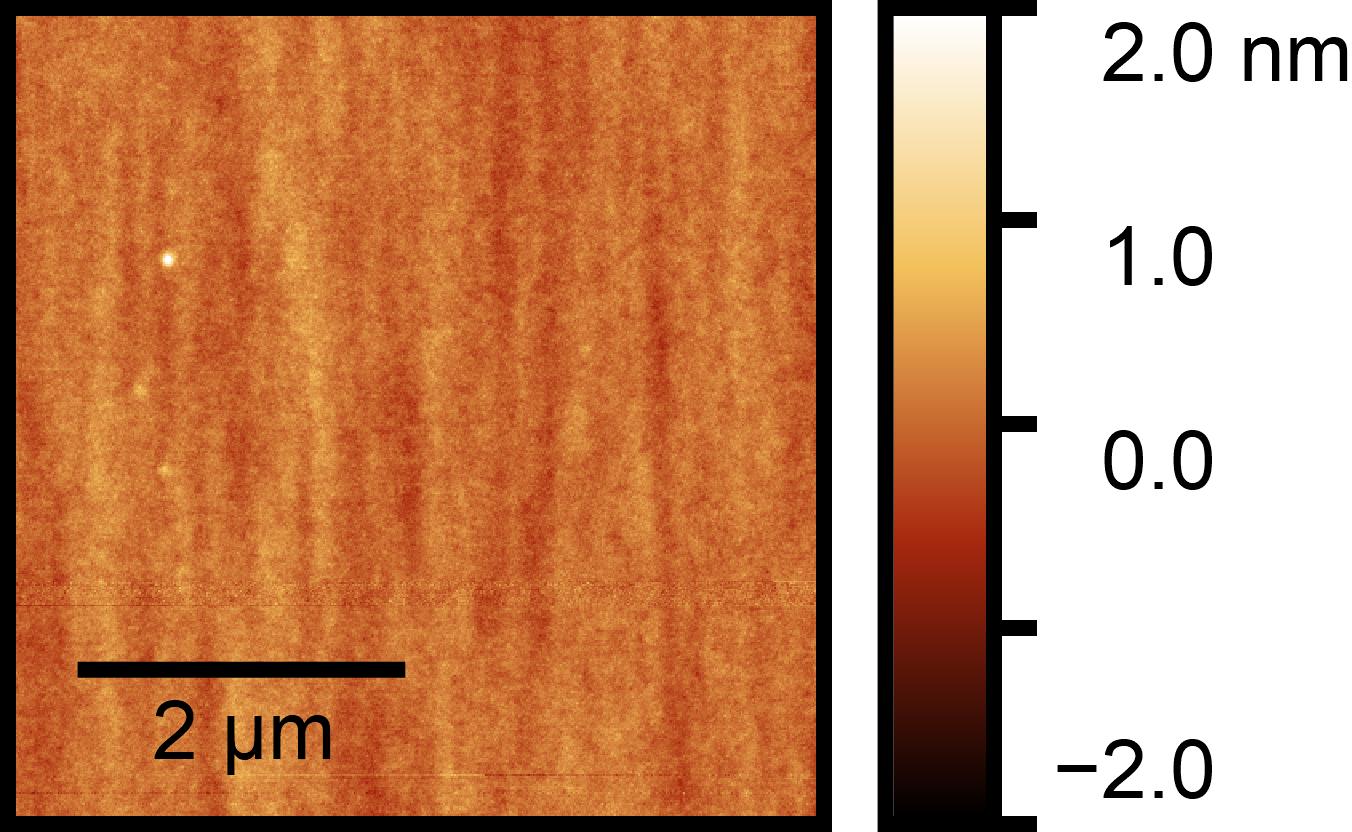}};
        \begin{scope}[x={(image.south east)},y={(image.north west)}]
            \draw [-{Triangle[length=3mm, width=2mm]}, line width = 0.65mm] (0.35,0.7) -- (0.05,0.7);
            \node[black, anchor = south] at (0.2, 0.73) {$[11\bar{2}0]$};
        \end{scope}
    \end{tikzpicture}
    }
    \newline
    \subfloat[after annealing (\SI{1200}{\celsius}): \newline rms $= 160 \pm 10$ pm \label{subfig:AFM_sampleB_a1200deg}]{
    \begin{tikzpicture}
        \node[anchor=south west,inner sep=0] (image) at (0,0) { \includegraphics[width=4cm]{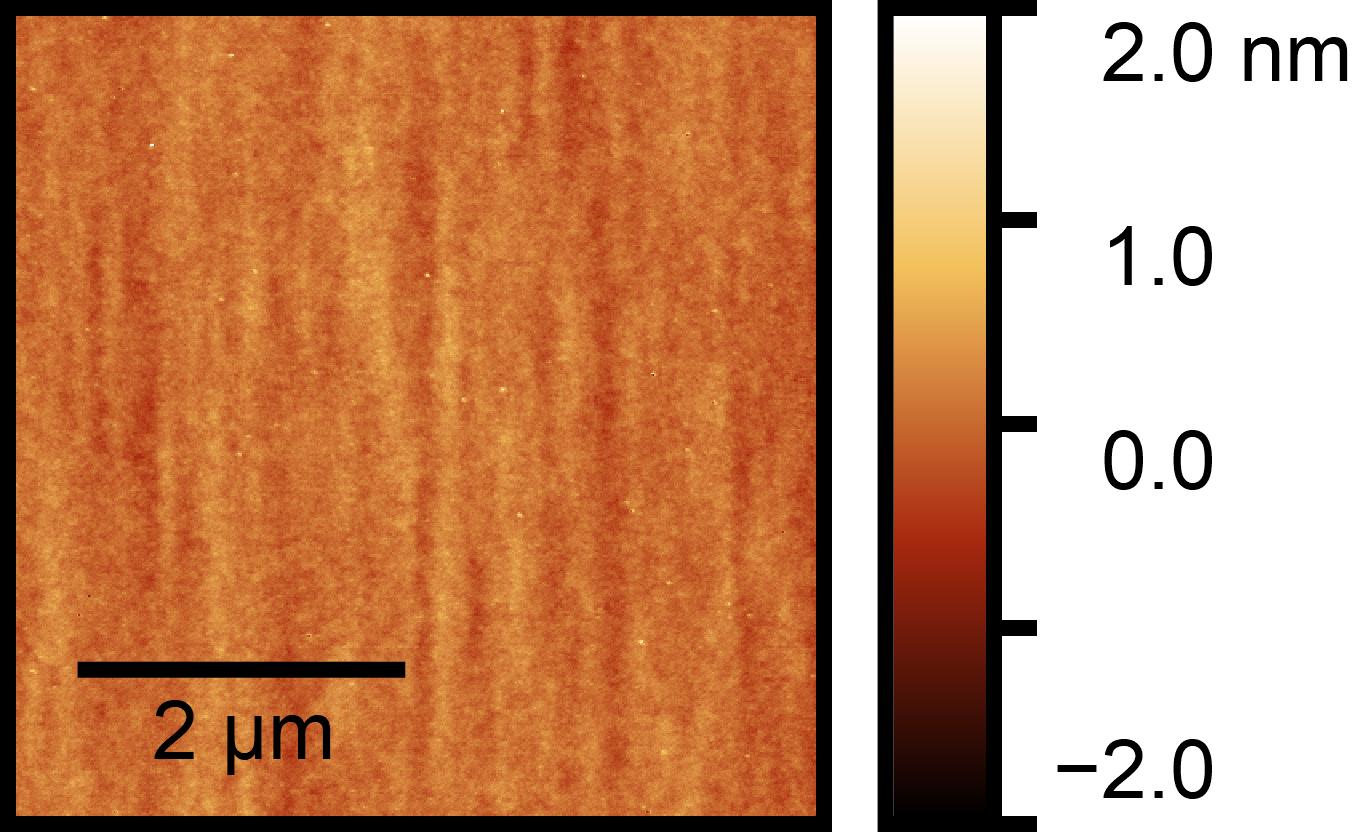}};
        \begin{scope}[x={(image.south east)},y={(image.north west)}]
            \draw [-{Triangle[length=3mm, width=2mm]}, line width = 0.65mm] (0.35,0.7) -- (0.05,0.7);
            \node[black, anchor = south] at (0.2, 0.73) {$[11\bar{2}0]$};
        \end{scope}
    \end{tikzpicture}
    }
    \quad
    \subfloat[after annealing (\SI{1700}{\celsius}): \newline rms = $(1.88\pm0.12)\cdot 10^3$ pm \label{subfig:AFM_sampleB_1700deg}]{
    \begin{tikzpicture}
        \node[anchor=south west,inner sep=0] (image) at (0,0) {\includegraphics[width=4cm]{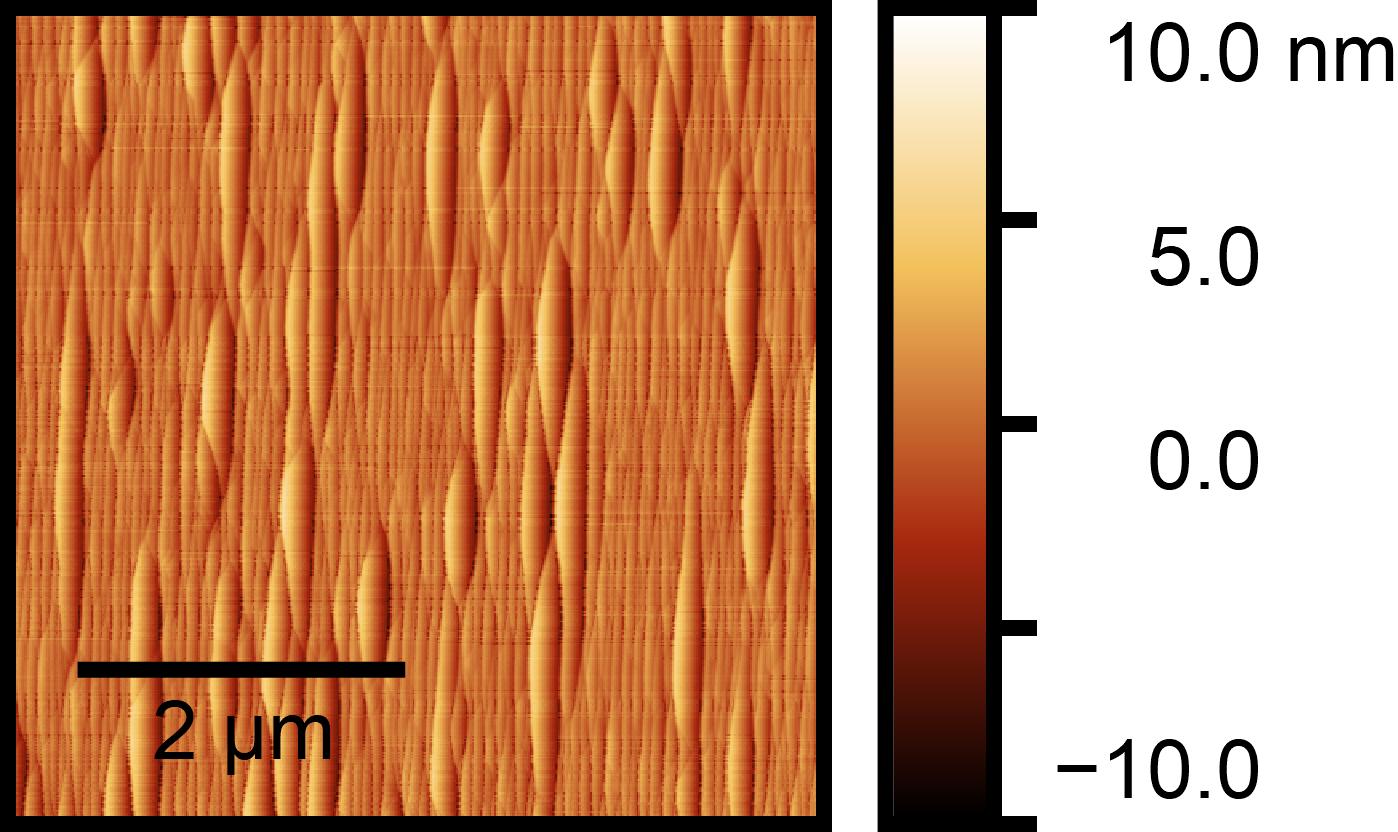}};
        \begin{scope}[x={(image.south east)},y={(image.north west)}]
            \draw [-{Triangle[length=3mm, width=2mm]}, line width = 0.65mm] (0.35,0.7) -- (0.05,0.7);
            \node[black, anchor = south] at (0.2, 0.73) {$[11\bar{2}0]$};
        \end{scope}
    \end{tikzpicture}
    }
    \quad
    \caption{AFM-scans in the $[11\bar{2}0]$ direction of sample B after each post-processing step. Scan size is roughly ${\SI{5}{\um}\times\SI{5}{\um}}$. Scale-bar: \SI{2}{\um}. \protect\subref*{subfig:AFM_sampleB_aECE} After ECE:
    point-like contaminations were masked in this particular processing protocol in order to facilitate the comparison of root-mean-square values. \protect\subref*{subfig:AFM_sampleB_aALE1} After ALE1 (\SI{20}{\nm} SiC removed). \protect\subref*{subfig:AFM_sampleB_aALE2} After ALE2 (additional \SI{140}{\nm} SiC removed). Smoother surface with low and homogeneously distributed terraces. \protect\subref*{subfig:AFM_sampleB_a1200deg} After first annealing at \SI{1200}{\celsius}. \protect\subref*{subfig:AFM_sampleB_1700deg} After stepwise annealing to \SI{1700}{\celsius}: Step-bunching and pronounced terrace formation.}
    
    \label{fig:AFM_scans_fabrication_processes}
\end{figure*}

AFM scans are conducted with a Park System NX10 at ambient pressure. Typical scan rates are 0.20 Hz with a lateral resolution of 10 nm per pixel. AFM scans were performed in the $[11\bar{2}0]$ direction. Steps and terraces naturally occur due to the 4°-off-axis n-type wafer. Gwyddion is used to analyze the AFM data. All scans are corrected by subtraction of a 2nd order polynomial to remove artifacts of the tip. Surface roughness is calculated for a 500 x 500 Pixel area, see Fig.~\ref{fig:AFM_scans_fabrication_processes}.

\section{Confirmation of validity of Euler Bernoulli assumptions}
\label{app:comsol}

\begin{figure}
  \begin{tikzpicture}
    \node[anchor=south west] at (0,0) {\includegraphics[trim={.97cm 0cm 0cm 0cm},clip,width=.95\linewidth]{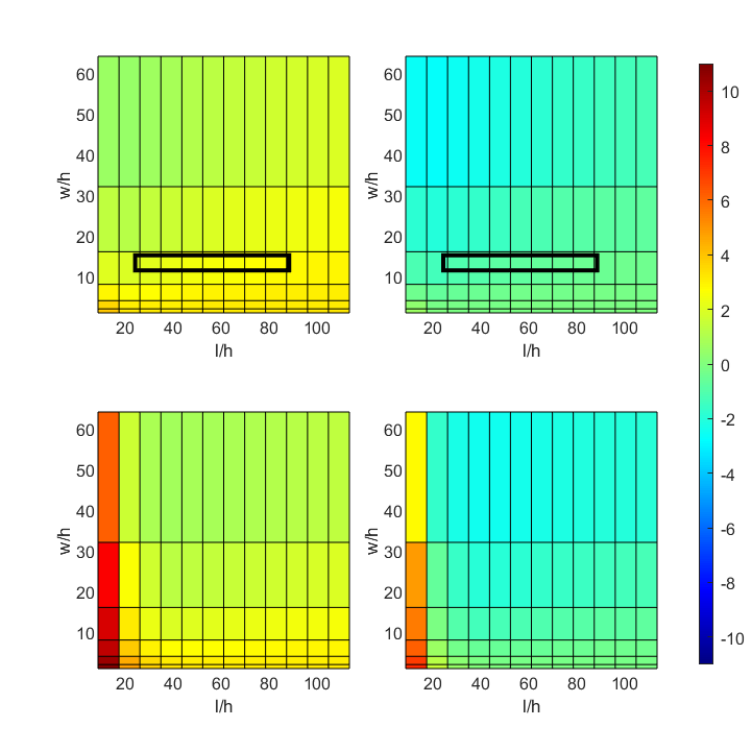}};
    %
    \begin{scope}[shift={(0,.5)}, x=1cm, y=1.04cm]
      \node at (.2,   7.8) {\subfloat[]{~\label{subfig:comsol-valid-singly-plate}}};
      \node at (3.8, 7.95) {\subfloat[]{~\label{subfig:comsol-valid-singly-beam}}};
      \node at (.2,   3.9) {\subfloat[]{~\label{subfig:comsol-valid-doubly-plate}}};
      \node at (3.8, 3.9) {\subfloat[]{~\label{subfig:comsol-valid-doubly-beam}}};
    \end{scope}
    %
    \node at (7.9, 8.4) {\scriptsize error (\unit{\percent})};
  \end{tikzpicture}
  \vspace{-.8cm}
  \caption{
    Relative deviation of predicted fundamental mode frequencies of Euler-Bernoulli and COMSOL models as a function of the length to thickness (${\length/\thick}$) and witdh to thickness ratios (${\width/\thick}$).
    Deviations are shown for
    \subref*{subfig:comsol-valid-singly-plate}~singly clamped plate with ${\poisson\neq0}$,
    \subref*{subfig:comsol-valid-singly-beam}~singly clamped beam with ${\poisson=0}$,
    \subref*{subfig:comsol-valid-doubly-plate}~doubly clamped plate with ${\poisson\neq0}$ and
    \subref*{subfig:comsol-valid-doubly-beam}~doubly clamped beam with ${\poisson=0}$.
    The black rectangles in \subref*{subfig:comsol-valid-singly-plate} and \subref*{subfig:comsol-valid-singly-beam} indicate the parameter range of cantilevers discussed here.
  }
  \label{subfig:comsol-valid-doubly-beam}
\end{figure}

Our stress and Young's modulus analysis is based on Euler-Bernoulli plate theory and therefore relies on the Euler-Bernoulli assumptions. Because of the intermediate width-to-thickness ratios ${\width/\thick}$ of our structures, it is not immediately clear whether the bending rididity of a slender beam, or that of a plate is adequate. Therefore, we compare the predictions for an Euler-Bernoulli beam and a plate to a COMSOL finite element model for a wide range of length and width to thickness ratios ${\length/\thick}$ and ${\width/\thick}$. \Figref{subfig:comsol-valid-doubly-beam} depicts the relative error, i.e. the difference of the calculated to the simulated eigenfrequency, normalized to the latter.
\Figref{subfig:comsol-valid-singly-plate} and \subref*{subfig:comsol-valid-singly-beam} show the relative deviation for a singly clamped cantilever geometry (c.f. \eqnref{eq:Eulerplate}). The frequency is slightly overestimated when the bending rigidity of a plate is used (see \subref*{subfig:comsol-valid-singly-plate}) and slightly underestimated assuming the bending rigidity of a beam (see \subref*{subfig:comsol-valid-singly-beam}).
\Figref{subfig:comsol-valid-doubly-plate} and \subref*{subfig:comsol-valid-doubly-beam} show the corresponding results for a doubly clamped geometry with clamped boundary conditions (but no prestress). Again frequencies are slightly overestimated in case of the plate assumption (see \subref*{subfig:comsol-valid-doubly-plate}) and slightly underestimated assuming a beam (see \subref*{subfig:comsol-valid-doubly-beam}). However, special care has to be taken for very short doubly clamped beams with ${\length/\thick<20}$, where errors up to \SI{10}{\percent} occur. Fortunately, this regime of no concern to us, since all devices discussed in this work feature ${\length/\thick>25}$.

Overall, we find the smallest errors (\SI{<2.4}{\percent}) for the discussed device geometries when using the bending rigidity of a plate, which we therefore apply in our analysis of stress and Young's modulus.

\section{Torsional admixture in flexural cantilever modes}

\begin{figure*}
  \centering
  \begin{tikzpicture}
    %
    \node[anchor=south west] at (0,.72) {\includegraphics[scale=.09]{
      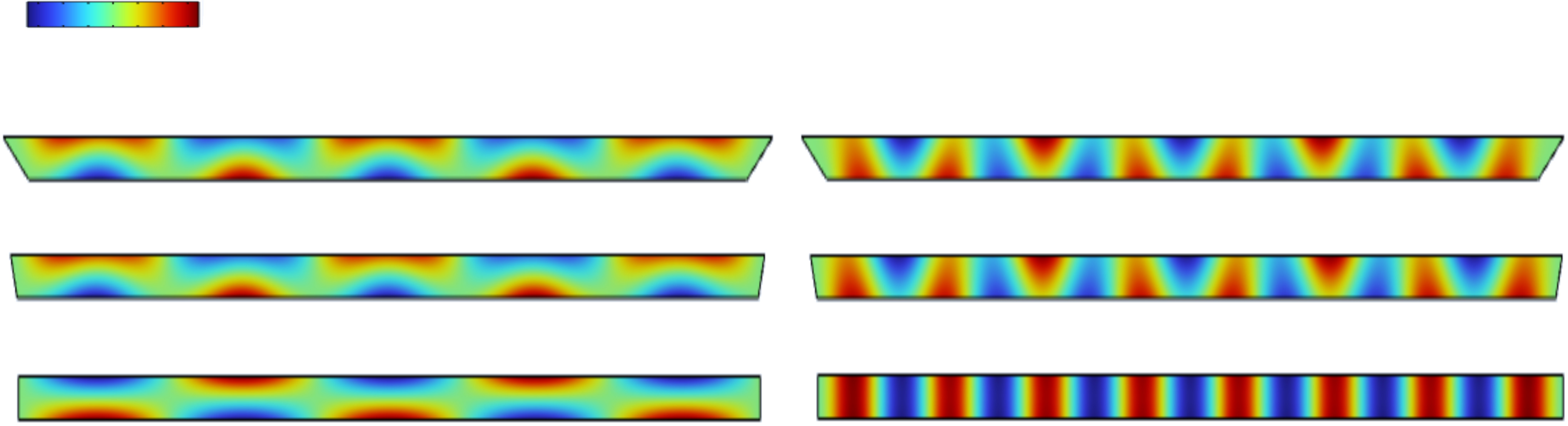}};
    %
    \node[anchor=south west] at (13,0) {\includegraphics[scale=\figscale]{
      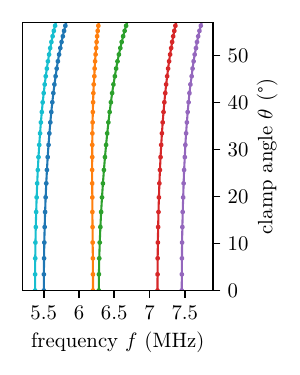}};
    %
    \begin{scope}[shift={(5.5,3.8)}, x={7mm}, y={4.5mm}]
      \fill[tabblue!30] (.3,0) -- (9.7,0) -- (10.3,1) -- (-.3,1) -- (.3,0);
      \node at (10.25,1.4) {\clampangle};
      \draw (9.7,0) -- (10.78,1.8);
      \draw (10,0) -- (10,1.9);
      \draw (0,0) -- (0,1.9);
      \draw (10,1.8) arc (90:60:1.3);
      \draw[<-] (0,1.5) -- (4.7,1.5);
      \node at (5,1.5) {\length};
      \draw[->] (5.4,1.5) -- (10,1.5);
    \end{scope}
    %
    \begin{scope}[shift={(0,0)}]
      \node at (5.1, 5.0) {\subfloat[]{~\label{subfig:clamp_angle_simu:legend}}};
      \node at (13.3, 5.3) {\subfloat[]{~\label{subfig:clamp_angle_simu:dispersion}}};
    \end{scope}
    \begin{scope}[shift={(-3.9,-.45)}, x={10.3mm}, y={10mm}]
      \node at (4.2, 4.1) {\subfloat[]{~\label{subfig:clamp_angle_simu:modes-l3}~}};
      \node[anchor=north west] at (4.4, 4.06) {\scriptsize ${\clampangle=\SI{50}{\degree}}$};
      \node at (4.2, 3.1) {\subfloat[]{~\label{subfig:clamp_angle_simu:modes-l2}~}};
      \node[anchor=north west] at (4.4, 3.06) {\scriptsize ${\clampangle=\SI{17}{\degree}}$};
      \node at (4.2, 2.1) {\subfloat[]{~\label{subfig:clamp_angle_simu:modes-l1}~}};
      \node[anchor=north west] at (4.4, 2.06) {\scriptsize ${\clampangle=0}$};
      \node at (10.5, 4.1) {\subfloat[]{~\label{subfig:clamp_angle_simu:modes-r3}~}};
      \node[anchor=north west] at (10.7, 4.06) {\scriptsize ${\clampangle=\SI{50}{\degree}}$};
      \node at (10.5, 3.1) {\subfloat[]{~\label{subfig:clamp_angle_simu:modes-r2}~}};
      \node[anchor=north west] at (10.7, 3.06) {\scriptsize ${\clampangle=\SI{17}{\degree}}$};
      \node at (10.5, 2.1) {\subfloat[]{~\label{subfig:clamp_angle_simu:modes-r1}~}};
      \node[anchor=north west] at (10.7, 2.06) {\scriptsize ${\clampangle=0}$};
    \end{scope}
    %
    \begin{scope}[shift={(0,0)}]
      \node at (1,4.5) {\scriptsize deflection (a.u.)};
    \end{scope}
  \end{tikzpicture}
  \vspace{-.8cm}
  \caption{
    Simulation results supporting the interpretation of mode shapes shown in \figref{fig:torsional_admixture} as hybridized torsional-flexural pairs, where hybridization is possible due the asymmetric clamp geometry. 
    \subref*{subfig:clamp_angle_simu:legend} Geometry of the COMSOL simulation. The two clamps of the bridge have a variable angle \clampangle, the average length \length of the beam is fixed to \SI{344}{\um}. All further parameters match the COMSOL simulation shown in \figref{fig:PN-14_LDV_spectrum}.
    \subref*{subfig:clamp_angle_simu:dispersion} Eigenfrequencies found by COMSOL in the vicinity of the measured modes shown in \figref{fig:torsional_admixture} for varying \clampangle. The angle ${\clampangle=0}$ corresponds to the symmetric case, increasing \clampangle corresponds to increasing asymmetry. The modes plotted in orange and green match the measured frequencies shown in \figref{fig:torsional_admixture}. For ${\clampangle=0}$, the orange branch corresponds to the 5${^\mathrm{th}}$ torsional mode and green branch corresponds to the 15${^\mathrm{th}}$ out-of-plane mode.
    (c,d,e) Mode maps of the orange branch of panel (b) for three different choices of \clampangle.
    (f,g,h) Corresponding mode maps of the green branch of panel (b).
    All mode maps are scaled for clarity without aspect ratio preservation, therefore the value of \clampangle does not match the angle expected from the image contours in (c-h).
  }
  \label{fig:clamp_angle_simu}
\end{figure*}

\begin{figure}
    \centering
    \includegraphics[width=8.5cm]{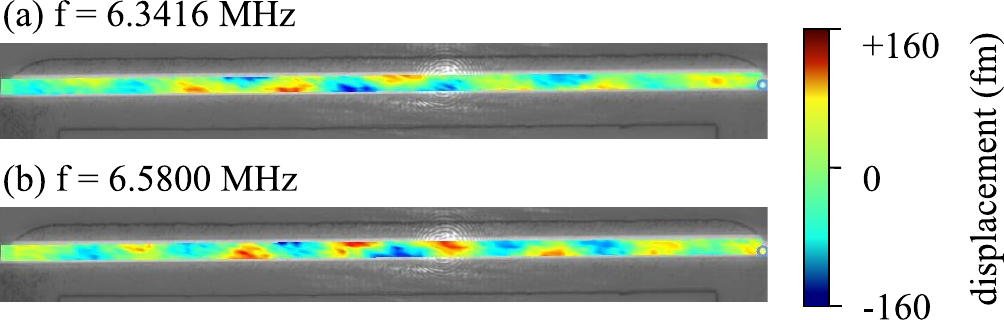}
    \caption{Measured mode-shapes of ${\SI{344}{\um}\times\SI{10}{\um}}$ beam obtained as Laser-Doppler vibrometry maps (red and blue indicate opposite directions of the relative amplitude with respect to a reference point; 594 scan points (6 along width, 99 along length) form a 2D surface representation with interpolated data). 15${^\mathrm{th}}$ flexural out-of-plane mode with pronounced torsional admixture. (a) Lower eigenfrequency and (b) higher eigenfrequency mode of the doublet.}
    \label{fig:torsional_admixture}
\end{figure}

We observe a notable phenomenon wherein, at eigenfrequencies exceeding the first torsional mode --- typically following the second or third torsional mode --- distinct out-of-plane modes exhibit significant torsional admixture. This aspect becomes apparent in the form of a doublet, with two mirror symmetric mode shapes, that are found in pairs. \Figref{fig:torsional_admixture} displays two modes close to the expected frequency for the 15th flexural out-of-plane mode, exhibiting a lower frequency contribution at \SI{6.3416}{\MHz}, and a higher frequency contribution at \SI{6.5800}{\MHz}.

The mixing of torsional and flexural modes is likely caused by the clamp point asymmetry in our devices. \Figref{fig:clamp_angle_simu} shows COMSOL simulations varying the clamp point angle \clampangle with all other parameters matching the device shown in \figref{fig:torsional_admixture}. 
For ${\clampangle=0}$ (symmetric case) we find the 5${^\mathrm{th}}$ torsional and the 15${^\mathrm{th}}$ out-of-plane mode in our simulation at frequencies matching for the two measured modes (\figref{fig:torsional_admixture}). As the clamp angle is gradually increased in the simulation, mixing of the two modes is observed.

Two important characteristics of the measured mode shapes are reproduced in the mixed mode shapes obtained in our simulation, namely the number of anti-nodes and the alternating positions of the anti-node's deflection maxima on the sides of the beam.
The simulated geometry does not intend to accurately model the clamp geometry of our fabricated devices. It however succeeds to grasp the effects of asymmetric clamps and to show how this may give rise to torsional-flexural mode mixing.

\section{Interferometric characterization setup}

\begin{figure*}
  \centering
  \includegraphics{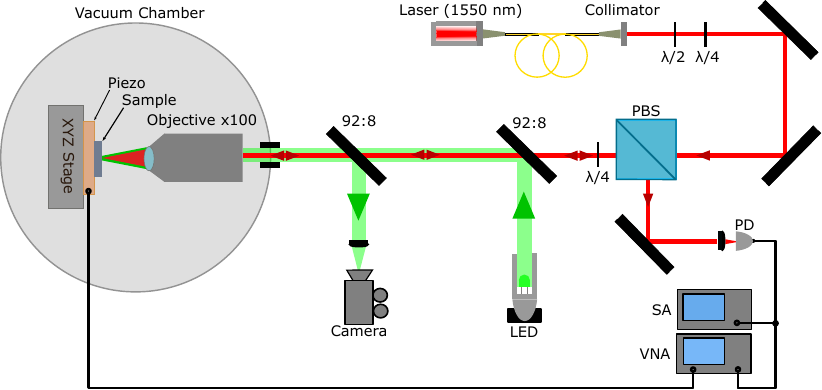}
  \caption{Interferometric measurement setup. The sample is placed in a vacuum chamber and actuated with a piezo glued to the chip. PBS - polarizing beam splitter, PD - photodetector, VNA - vector network analyzer, SA - spectrum analyzer.}
  \label{fig:Setup}
\end{figure*}

\Figref{fig:Setup} shows a sketch of the interferometric measurement setup. We focus \SI{1500}{\nm} laser light onto the structure under investigation on the sample surface and collect the reflected light with a fast photodetector.
Due to the interference of reflections from substrate and resonator surface, vibrations of the resonators are imprinted on the reflected light intensity, which in turn is demodulated by the vector network analyzer for frequency response measurements or by the spectrum analyzer to conduct ringdown measurements.
%
We use an infinity corrected objective with NA\,=\,0.7 and 100x magnification. The estimated depth of field is \SI{3}{\um}, which is larger than the vertical distance between resonator surface and substrate of \SI{\approx 1.5}{\um}, ensuring that light reflected from the resonator surface and the substrate is collected simultaneously.

We employ the following routine to select an optimal measurement position: We begin by placing the spot position on the edge of the resonator structure and set the focus in between the resonator surface and substrate. Subsequently, we fine tune $x$, $y$ and $z$ positions to maximize the mechanical response signal.

To navigate on the sample, an additional LED and a camera are coupled to the beam path with weakly reflecting 92:8 mirrors. The sample is moved by attocube positioners. The high frequency drive is applied by a piezo plate underneath the chip. During the measurement, the focus objective, the sample, and the xyz-positioner are held in a vacuum better than \SI{1e-3}{\milli\bar}.
 
Increasing the drive power applied to the piezo shaker increases the amplitude of the mechanical response and thus improves the signal to noise ratio. However, too large drive powers result in a nonlinear mechanical response, which is impractical for the purpose of determining eigenfrequency and linear damping rate. Therefore, for each resonance that we measure, we first determine the drive power at which nonlinear effects become visible and then record the response curve at drive powers well below the onset of nonlinearity. This strategy allows us to obtain the linear frequency response curve of each resonance with the optimal signal to noise ratio.

\begin{figure*}
  \centering
  \includegraphics[scale=\figscale]{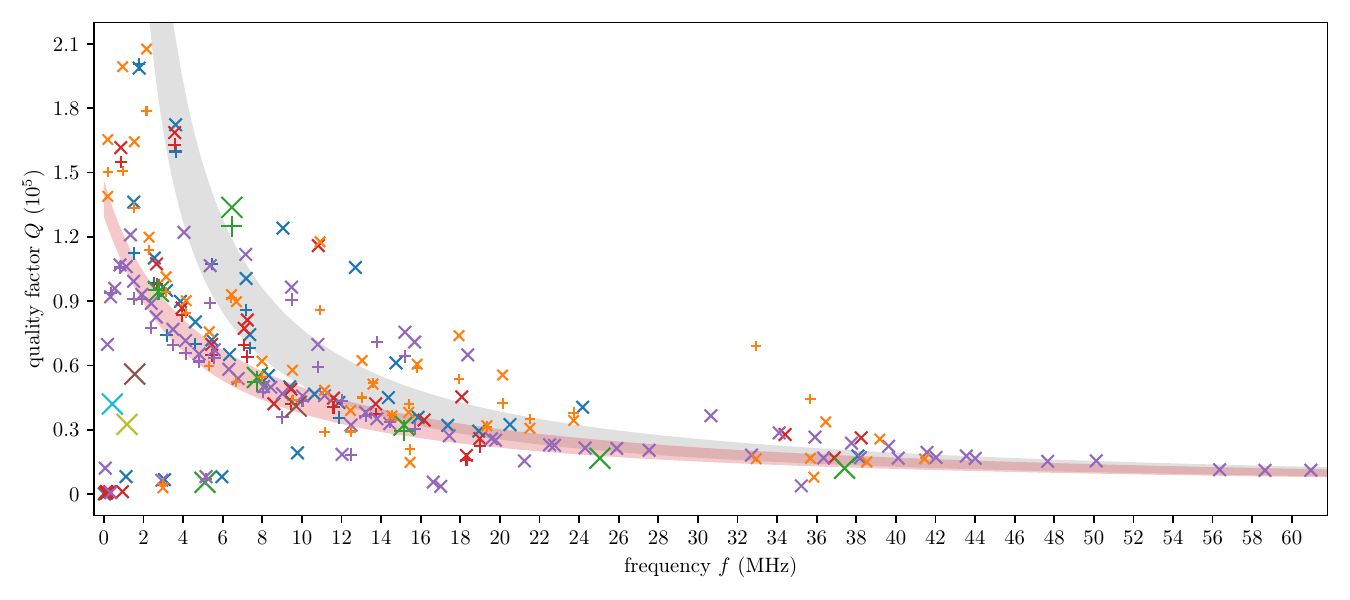}\\[-7cm]
  \raggedleft
  ~~\raisebox{.65cm}{sample A:}\includegraphics[scale=\figscale]{pn25-legend}~~~~\\[-.5cm]
  ~~\raisebox{.65cm}{sample B:}\includegraphics[scale=\figscale]{pn14-legend}~~~~\\[4cm]
\caption{Plot analog to \figref{fig:quality-factors} but with all measured \Q s on both samples. The areas shaded in gray and red trace \Qted and ${\qty(\Qted^{-1}+\Qconst^{-1})^{-1}}$ with ${\Qconst=\QconstVal}$. The width of the shaded are represents the propagated uncertainty of the Young's modulus ${\youngs=\youngsVal}$ that we obtained (\secref{sec:youngs}). The color code is identical to Figs.~\ref{fig:quality-factors} and \ref{fig:ted}.}
  \label{fig:full-q-vs-f-plot}
\end{figure*}

Because of the large size of our dataset it is difficult to distinguish the individual \Q{} data in the main text Figs.~\ref{fig:quality-factors} and \ref{fig:ted}. For more detailed insights, \figref{fig:full-q-vs-f-plot} displays all data shown in Figs.~\ref{fig:quality-factors} and \ref{fig:ted} without the space limitations of the main text. All data in hdf5 format is available under the repository described in the data availability statement.

\begin{figure*}
  \centering
    \begin{tikzpicture}
        \node[anchor=south west,inner sep=0] (image) at (0,0) {\includegraphics[scale=\figscale]{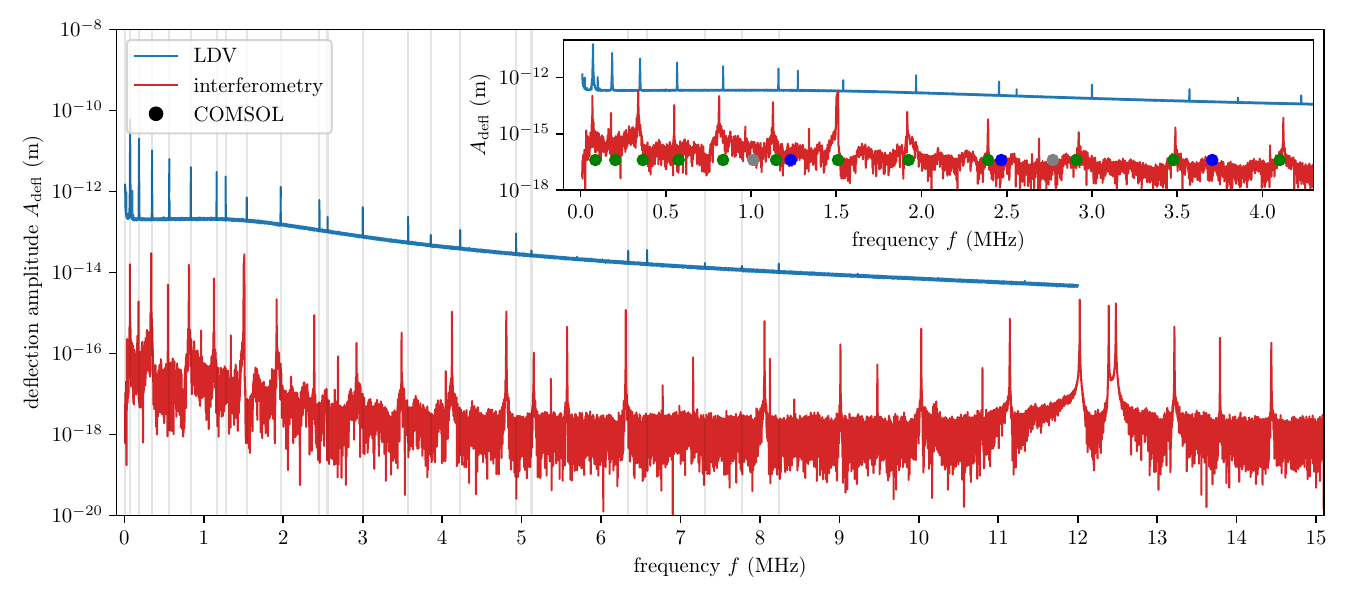}};
        \begin{scope}[x={(image.south east)},y={(image.north west)}]
            \node[anchor=south,inner sep=0] (image) at (0.5,.98) {\includegraphics[width=0.6\linewidth]{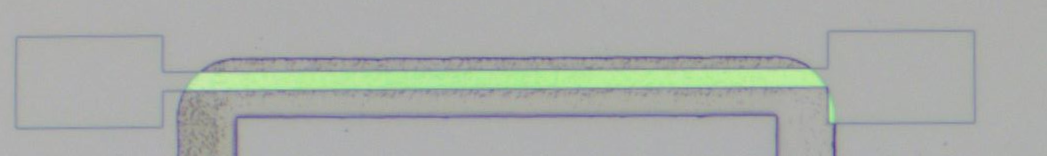}};
        \end{scope}
    \end{tikzpicture}
  \caption{Micrograph and spectrum of a \SI{344}{\um} long beam on sample B, taken with LDV (blue) and interferometry (red) before post-processing. The amplitude of the spectrum obtained by interferometry is expressed in arbitrary units scaled for clarity. The vibrational amplitude of the LDV data is calibrated and can be expressed in \si{m}. The spectrum corresponds to the LDV maps in Fig.~\ref{fig:Panel_SEM_optical_modes}. It is averaged over the $200$ spectra taken at each individual point of the measurement grid. The corresponding spectrum measured on the same resonator in the optical interferometry setup is shown in arbitrary units. Faint vertical gray lines are placed at the peaks of the LDV spectrum, highlighting the good correspondence of LDV peaks with interferometry peaks up to approx. \SI{5}{\MHz}. The insets shows a magnified view of the same data up to \SI{4}{\MHz}. In the zoom-in, circles indicate the frequencies of a COMSOL model using a tensile stress of \SI{5}{\MPa}. The circle colors indicate the mode family obtained from the COMSOL model, green: out-of-plane, gray: in-plane, blue: torsional.}
  \label{fig:PN-14_LDV_spectrum}
\end{figure*}

To demonstrate the agreement between optical interferometry and LDV measurements, \figref{fig:PN-14_LDV_spectrum} shows large span spectra of both techniques for a doubly clamped beam. In the frequency range up to \SI{5}{\MHz} (range of mode maps in \figref{fig:Panel_SEM_optical_modes}) we can match optical interferometry and LDV resonance peaks unambiguously. The small deviation of frequencies obtained by the two methods is not surprising, because the chamber pressure of the LDV (\SI{\approx10}{\milli\bar}) significantly exceeds the chamber pressure of the interferometric setup (\SI{<1e-3}{\milli\bar}). Furthermore the setups use different laser wavelengths, laser powers (\SI{100}{\uW}@\SI{1550}{\nm} for the interferometer, \SI{\leq2}{\mW}@\SI{532}{\nm} for the LDV) and objectives.
Frequencies and mode type found by a COMSOL simulation agree well with both LDV and interferometric data, supporting the results of our LDV mode shape analysis.

\section{Frequency response and ringdown fit functions}
\label{app:conventions}

The symbol conventions used in this paper are implied by the following equation of motion for the driven damped harmonic oscillator

\begin{align}
  \ddot{x} + \damping\dot{x} + \omeig^2 x = \ampdrive \cos (\omdrive\timevar),
\end{align}

where $x$ denotes the deflection, \damping is the damping rate (i.e.\ energy decay rate), ${\omeig=2\pi\freqeig}$ is the resonance frequency, \omdrive is the driving frequency, \ampdrive is the driving force normalized to effective mass, and \timevar denotes time. The fit functions used for the measured amplitude response curves and amplitude ringdown traces are given by

\begin{align}
  \amp_\mathrm{response}\qty(\omdrive) &= \frac{\damping\omeig\amp_0}{ \sqrt{\qty(\omeig^2-\omdrive^2)^2 + \damping^2\omdrive^2} } + \amp_\mathrm{noise},
  \label{eq:response}\\
  \amp_\mathrm{ringdown}\qty(\timevar) &= \amp_0 e^{-\timevar\damping/2} + \amp_\mathrm{noise},
\end{align}

with the signal amplitude ${\amp_0}$ and the experimental noise floor ${\amp_\mathrm{noise}}$. The parameters ${\amp_0}$, \omeig, \damping and ${\amp_\mathrm{noise}}$ are free fit parameters. All damping rates and amplitudes mentioned in the main text follow these conventions.

\section{Laser power dependence of eigenfrequencies and quality factors}
\label{app:laser-power}

For all interferometric measurements shown in this work we choose an optical power of \SI{100}{\uW} incident on the sample. This power is a trade-off between lower powers, which make detection of resonances cumbersome due to the diminishing signal-to-noise ratio, and higher powers at which thermal expansion induced by laser absorption and the resulting heating shifts the frequencies of doubly clamped structures.

To find the optimal laser power, we measure frequencies and quality factors of the first six resonances of a doubly clamped beam with length \SI{150}{\um} and a singly clamped cantilever with length \SI{30}{\um} different from the ones discussed in the main text. We measure the laser power incident on the sample using a power meter at the 92:8 beam splitter output opposing the LED (see \figref{fig:Setup}). The precise splitting ratio is determined by a priori calibrations.

\begin{figure*}[p]
  \centering
  \begin{tikzpicture}
    \node[anchor=south west] at (0,0) {\includegraphics[scale=\figscale]{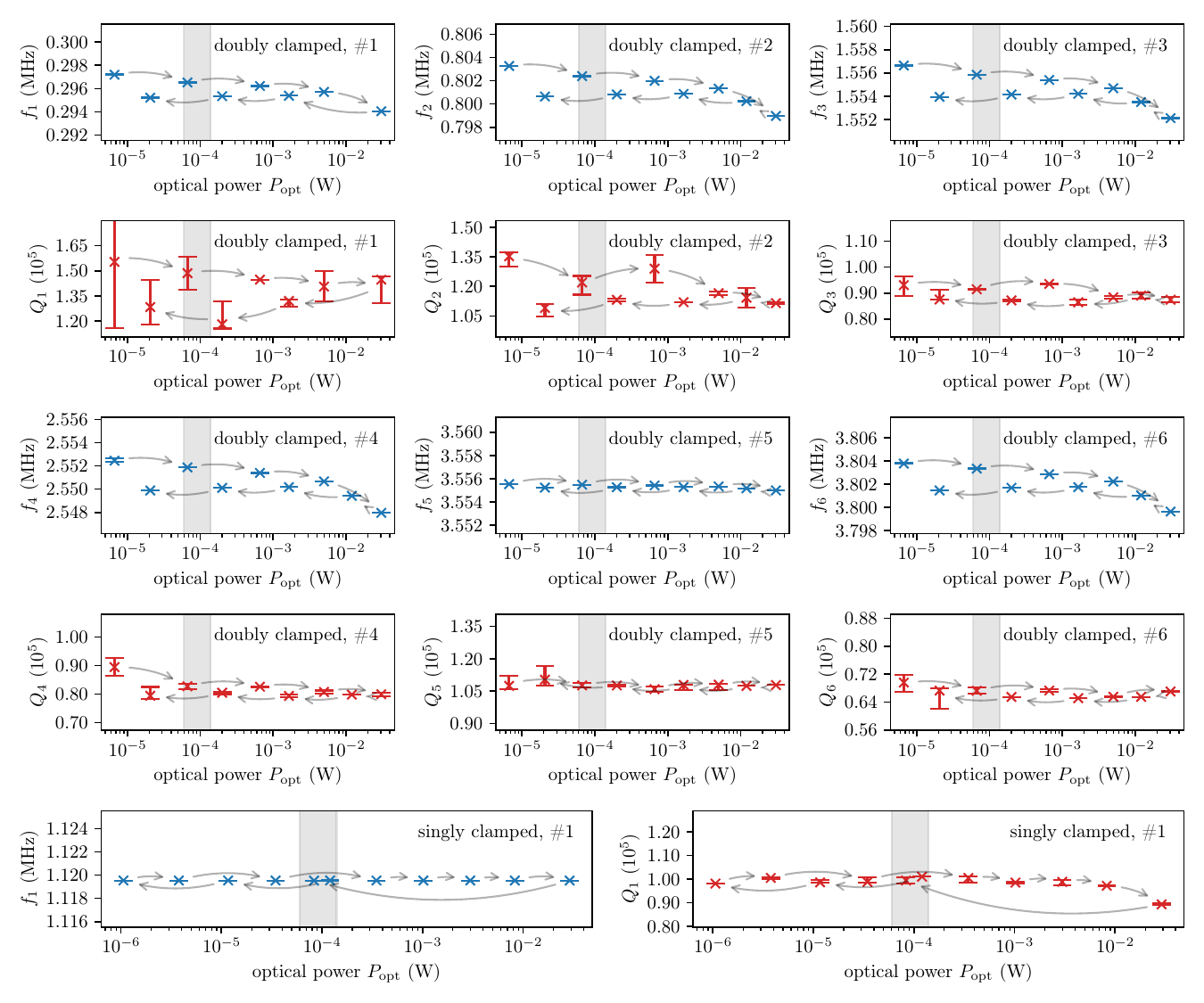}};
    %
    \begin{scope}[shift={(-.3,2.3)}, x=1cm, y=1.055cm]
      \node at (.7,    11.85) {\subfloat[]{~\label{subfig:laser-power-b1}}};
      \node at (6.5,    12) {\subfloat[]{~\label{subfig:laser-power-b2}}};
      \node at (12.4,   12) {\subfloat[]{~\label{subfig:laser-power-b3}}};
      \node at (.7,    9.25) {\subfloat[]{~\label{subfig:laser-power-b3}}};
      \node at (6.5,   9.25) {\subfloat[]{~\label{subfig:laser-power-b4}}};
      \node at (12.4,  9.25) {\subfloat[]{~\label{subfig:laser-power-b5}}};
      \node at (.7,    6.5) {\subfloat[]{~\label{subfig:laser-power-b6}}};
      \node at (6.5,   6.5) {\subfloat[]{~\label{subfig:laser-power-b7}}};
      \node at (12.4,  6.5) {\subfloat[]{~\label{subfig:laser-power-b8}}};
      \node at (.7,    3.75) {\subfloat[]{~\label{subfig:laser-power-b9}}};
      \node at (6.5,   3.75) {\subfloat[]{~\label{subfig:laser-power-b10}}};
      \node at (12.4,  3.75) {\subfloat[]{~\label{subfig:laser-power-b11}}};
      \node at (.7,    .9) {\subfloat[]{~\label{subfig:laser-power-c1}}};
      \node at (9.5,   .9) {\subfloat[]{~\label{subfig:laser-power-c2}}};
    \end{scope}
  \end{tikzpicture}
  \caption{Frequency and quality factor measurements of the first six resonances of a doubly clamped (a-l) and for the fundamental mode of a singly clamped (m-n) beam on sample A, for laser powers between \SI{1}{\uW} and \SI{30}{\mW}. Each frequency and quality factor is recorded multiple times at every laser power, error bars indicate minimum, maximum and median of the recorded values. All frequency axes are setup to span \SI{10}{\kHz}, all quality factor axes are setup to span \SI{\pm25}{\percent} to facilitate comparison. Faint gray arrows indicate the order in which measurements were taken. The data for the doubly clamped structure was recorded over a time of \SI{\approx6}{\hour}, the singly clamped data was recorded over a time of \SI{\approx2}{\hour}. \SI{5}{\min} of waiting time after each change of laser power were spent to allow for thermalization.
  }
  \label{fig:laser-power}
\end{figure*}

\Figref{fig:laser-power} shows frequencies and quality factors measured as a function of laser power for the two structures. All resonances except for \#5 of the doubly clamped structure display a shift to lower frequencies for laser powers above \SI{10}{\mW} and a superimposed slow overall drift towards lower frequency. We attribute the shift at \SI{>10}{\mW} laser power to laser absorption heating and the slow drift to the change of ambient temperature over the \SI{6}{\hour} of measurement. The frequency stability of resonance \#5 and the singly clamped cantilever are better than \SI{1}{\kHz}. The cantilever frequency has a vanishing susceptibility to thermal expansion because no stress can be established, which explains the stability. Resonance \#5 is likely of torsional character. The susceptibility of torsional modes to thermal expansion and the related stress change is much smaller than that of flexural modes.

The quality factors of the doubly clamped structure shown in \figref{fig:laser-power}(d-f,j-l) fluctuate by up to \SI{20}{\percent} for the first two resonances and by less than \SI{10}{\percent} for the remaining resonances. No significant correlation with the applied laser power is visible, confirming that detection laser absorption has no significant influence on the measured \Q. For the cantilever, a reduction of the measured \Q by \SI{\approx10}{\percent} at the highest investigated laser power is observed.

Overall, we find \SI{<1}{\percent} frequency shifts and no significant change of the measured quality factors for laser powers up to \SI{10}{\mW}, which is a factor of 100 above the power we employ in the interferometric measurements reported in the main text. We conclude that laser absorption heating and other laser power effects have are negligible for the results obtained by interferometric read-out reported in this work. For the results obtained by LDV, we did not evaluate the laser power dependence, because the validity of the LDV mode maps does not hinge on the minor frequency shifts caused by potential laser absorption heating.

\section{Estimation of extrinsic damping contributions}
\label{app:extrinsic}

Two extrinsic damping mechanisms have potential relevance for our experiments: gas damping and clamping loss. In the following, we estimate the corresponding extrinsic \Q and their scaling with parameters we varied experimentally. Comparison with our data shows that the extrinsic damping contributions in our measurements are negligible.

\emph{Gas damping:} The gas damping \Q scales with frequency, therefore the measured low frequency \Q are most susceptible to gas damping. The lowest frequency relevant in the manuscript is \SI{70}{\kHz}, given by the fundamental mode of the longest measured beam. Assuming residual gas with a molar mass of dry air and the worst-case pressure ${p=\SI{1e-3}{\milli\bar}}$ the predicted gas damping \Q \cite{funda-of-nanom-reson} at \SI{70}{\kHz} amounts to \num{7.5e5}. The measured \Q of these low frequency modes is \num{<7e4}, yielding a worst-case gas damping contribution of \SI{\approx10}{\percent}. The vast majority of data was measured at higher frequencies and thus has an even smaller gas damping contribution. We conclude that the effects of gas damping are negligible for the statements presented in our paper.

\emph{Clamping loss:} Exact formulas for the clamping loss of singly and doubly clamped beams have been derived \cite{photiadis_attachment_losses_2004,wilsonrae_intrinsic_dissipation_2008}. The resulting scaling behavior of the clamping loss with the geometry parameters of the nanomechanical resonator depends strongly on the modeling of the clamp point and substrate. For instance, assuming finite or infinite thickness substrate results in radically different ${\mathcal{O}(\length)}$ or ${\mathcal{O}(\length^5)}$ scaling, respectively \cite{photiadis_attachment_losses_2004}.

For the shortest singly clamped cantilever presented here with ${\length=\SI{15}{\um}}$, Refs.~\onlinecite{photiadis_attachment_losses_2004,wilsonrae_intrinsic_dissipation_2008} predict a fundamental flexural mode clamping loss ${\Q>10^6}$ for both for finite and for infinite thickness substrate. The clamping loss \Q increases for longer cantilevers according to both references. 

Reference~\onlinecite{wilsonrae_intrinsic_dissipation_2008} derives clamping loss expressions for arbitrary mode numbers ${n}$ of singly and doubly clamped beams, predicting the clamping loss scaling as ${\Q\sim n^{-4}}$ for higher order flexural modes. Our observation that the measured \Q for the various device lengths fall on the same ${1/f}$ trend is incompatible with the predicted length and mode number scaling of clamping loss. This leads us to the conclusion that clamping loss is not relevant for our analysis. We can, however, not ultimately exclude that some data we show have a measurable clamping loss contribution, because of the delicate dependence of clamping losses on the clamp point geometry and the fact the our clamp geometries deviate from the idealized geometries used in Refs.~\onlinecite{photiadis_attachment_losses_2004,wilsonrae_intrinsic_dissipation_2008}.

\section{Calculation of thermoelastic loss \Qted}
\label{app:ted}

We apply the thermoelastic damping model for flexural modes of thin beams derived by Zener \cite{zener_internal_friction_1937,lifshitz_thermoelastic_damping_2000,funda-of-nanom-reson}

\begin{align}
  \Delta &= \frac{\youngs T \alpha_{th}^{2}}{\dens c_{p}}, \\
  \tau &= \frac{\rho c_{p} h^{2}}{\pi^{2} \kappa}, \\
  \Qted^{-1}\qty(f) &= \Delta\frac{2\pi f\tau}{1+(2\pi f\tau)^2}
\end{align}

with the relaxation strength $\Delta$, relaxation time $\tau$, Young's modulus \youngs, temperature $T$, thermal expansion coefficient $\alpha_\mathrm{th}$, mass density $\dens$, specific heat capacity $c_p$, thickness in bending direction $h$, thermal conductivity $\kappa$, frequency $f$ and the thermoelastic quality factor \Qted. The chosen values for the material parameters and the respective references are listed in Table~\ref{tab:matparams}.

\begin{table}
  \begin{ruledtabular}\begin{tabular}{rrrll}
    mass density                  & $\dens=$    & \num{3.2e3} 
          & \si{\kg\per\cubic\m}  & \cite{mesquita_refinement_crystal_1967} \\
    Young's modulus               & $\youngs=$  & \youngsNumVal
          & \si{\GPa}             & Sec.~\ref{sec:youngs}  \\
    th. expansion coeff. & $\alpha=$   & \num{4.5e-6}
          & (for 6H-SiC)          & \cite{levinshtein_properties_2001} \\
    thermal conductivity          & $\kappa=$   & \num{370}     
          & \si{\W\per\m\per\K}   & \cite{levinshtein_properties_2001} \\
    specific heat capacity        & $c_v=$      & \num{690}     
          & \si{\J\per\kg\per\K}  & \cite{levinshtein_properties_2001} \\
    temperature                   & $T=$        & \num{300}
          & \si{\K}               & \\
  \end{tabular}\end{ruledtabular}
  \caption{Material parameters and literature values for the calculation of \Qted.}
  \label{tab:matparams}
\end{table}

\section{Determination of stress \stress}
\label{app:stress-fit}

The investigated singly-clamped cantilevers are assumed to be fully relaxed and stress-free; hence their dynamics is well described by Eq.~\ref{eq:Eulerplate}.

To extract the tensile stress of doubly clamped beams from the measured eigenmode spectra, the eigenfrequencies of a stressed Euler-Bernoulli plate have to be derived from its equation of motion \cite{funda-of-nanom-reson}
\begin{align}
  \youngs I\dv[4]{\deflvar(\posvar,\om)}{\posvar} 
  - \stress A\dv[2]{\deflvar(\posvar,\om)}{\posvar} 
  -\om^2 \dens A \deflvar(\posvar,\om) = 0,
  \label{eq:ebb-eom}
\end{align}
with the Youngs modulus \youngs, area moment of inertia $I$, the deflection \deflvar, the position along the beam \posvar, the angular frequency \om, the tensile stress \stress, the beam's cross sectional area $A$ and the mass density \dens.
Applying simply-supported boundary conditions 
\begin{align}
  \deflvar(\posvar=0) = \deflvar(\posvar=\length) 
  = \dv[2]{\deflvar}{\posvar}\bigg|_{\posvar=0}
  = \dv[2]{\deflvar}{\posvar}\bigg|_{\posvar=\length} = 0,
\end{align}
with the beam length \length yields the convenient analytic solution \cite{funda-of-nanom-reson}
\begin{align}
  \om_\modeno=\frac{\modeno^2\pi^2}{\length^2}\sqrt{\frac{\youngs\thick^2}{12(1-\poisson^2)\dens}}\sqrt{1+\frac{12(1-\poisson^2)\stress\length^2}{\modeno^2\pi^2\youngs\thick^2}}.
  \label{eq:ebb-analytic}
\end{align}
for the eigenfrequency ${\om_\modeno}$ of the \modeno{}-th mode. Note that we apply the area moment of inertia for a plate, because our devices feature a width to thickness ratio ${\width/\thick>5}$. In our measurements the only unknown in this equation is the stress \stress, which we obtain by performing a least-squares fit of the first six measured eigenfrequencies of a doubly clamped beam.

However, simply supported boundary conditions allow for arbitrary ${\dv{\deflvar}{\posvar}}$ values at the clamp points, which is not realistic for the devices under consideration. Applying the more accurate clamped boundary conditions
\begin{align}
  \deflvar(\posvar=0) = \deflvar(\posvar=\length) 
  = \dv{\deflvar}{\posvar}\bigg|_{\posvar=0} 
  = \dv{\deflvar}{\posvar}\bigg|_{\posvar=\length} = 0
  \label{eq:clamped-bc}
\end{align}
renders an analytical solution as in \eqnref{eq:ebb-analytic} impossible. To obtain a numerical solution, we start with the general solution of the 4th order differential equation \eqnref{eq:ebb-eom}. From the four integration constants of the general solution, three can be eliminated by inserting the first three boundary conditions from \eqnref{eq:clamped-bc}. The fourth constant can be arbitrarily chosen to,~e.g.~$1$,~because it sets the arbitrary eigenmode amplitude. This procedure is possible without any further assumptions, with either a great deal of patience or the use of computer algebra software. We do not show the resulting expression here because it fills many pages.

This exact expression for ${\deflvar(\posvar)}$ thus fulfills \eqnref{eq:ebb-eom} and the first three boundary conditions of \eqnref{eq:clamped-bc} with the last remaining unknown being the frequency \om. As mentioned before, an overall exact solution does not exist, therefore it is not surprising that inserting the expression into the fourth boundary condition does not yield an analytically solvable equation. However, we can efficiently solve the resulting equation numerically by using the eigenfrequencies of the simply-supported problem \eqnref{eq:ebb-analytic} as starting values, finally arriving at numerically exact values for the eigenfrequencies with clamped boundary conditions, which are discussed in Fig.~\ref{fig:Euler-Bernoulli-fit}(c) and (d).

\section{Calculating Young's modulus from the elastic constants ${c_{ij}}$}
\label{app:youngs-from-cij}

Not all references shown in \figref{subfig:youngs-literature} specify an explicit value for the Young's modulus of 4H-SiC, but instead present the elastic constants ${c_{ij}}$. The stiffness matrix $\mathbf{C}$ relates stress $\sigma$ and strain $\epsilon$. In order to calculate the Young's modulus along crystallographic directions, one has to invert the stiffness matrix $\mathbf{C}$ of hexagonal SiC \cite{chen_study_elastic_2019}:
\begin{align}
    \mathbf{C} = \begin{bmatrix}
    c_{11} & c_{12} & c_{13} & 0 & 0 & 0 \\
    c_{12} & c_{11} & c_{13} & 0 & 0 & 0 \\
    c_{13} & c_{13} & c_{33} & 0 & 0 & 0 \\
    0 & 0 & 0 & c_{44} & 0 & 0 \\
    0 & 0 & 0 & 0 & c_{44} & 0 \\
    0 & 0 & 0 & 0 & 0 & \frac{c_{11} - c_{12}}{2}
    \end{bmatrix},
\end{align}
which results in the compliance matrix $\mathbf{S}$:
\begin{align}
    \mathbf{S} = \begin{bmatrix}
    s_{11} & s_{12} & s_{13} & 0 & 0 & 0 \\
    s_{12} & s_{11} & s_{13} & 0 & 0 & 0 \\
    s_{13} & s_{13} & s_{33} & 0 & 0 & 0 \\
    0 & 0 & 0 & s_{44} & 0 & 0 \\
    0 & 0 & 0 & 0 & s_{44} & 0 \\
    0 & 0 & 0 & 0 & 0 & s_{66}
    \end{bmatrix},
\end{align}
with the relevant $s_{ij}$ for our calculations: 
\begin{align}
s_{11} = \frac{c_{11}c_{33}-c_{13}^2}{c_{11}^2c_{33}-2c_{11}c_{13}^2-c_{12}^2c_{33}+2c_{12}c_{13}^2}.
\end{align}
The Young's modulus \youngs for hexagonal SiC is isotropic in the basal plane (0001), resulting in $E_1 = E_2 = 1/s_{11}$:
\begin{align}
  \youngs = 
  \frac{c_{11}^{2} c_{33} - 2 c_{11} c_{13}^{2} - c_{12}^{2} c_{33} + 2 c_{12} c_{13}^{2}}{c_{11} c_{33} - c_{13}^{2}}.
\end{align}
Consequently, we exclusively use the Young's modulus ${\youngs = E_1 = E_2}$ within the basal plane, as all our devices are oriented along this plane, which is orthogonal to the wafer's c-axis.


\bibliography{main}

\begin{thebibliography}{56}%
\makeatletter
\providecommand \@ifxundefined [1]{%
 \@ifx{#1\undefined}
}%
\providecommand \@ifnum [1]{%
 \ifnum #1\expandafter \@firstoftwo
 \else \expandafter \@secondoftwo
 \fi
}%
\providecommand \@ifx [1]{%
 \ifx #1\expandafter \@firstoftwo
 \else \expandafter \@secondoftwo
 \fi
}%
\providecommand \natexlab [1]{#1}%
\providecommand \enquote  [1]{``#1''}%
\providecommand \bibnamefont  [1]{#1}%
\providecommand \bibfnamefont [1]{#1}%
\providecommand \citenamefont [1]{#1}%
\providecommand \href@noop [0]{\@secondoftwo}%
\providecommand \href [0]{\begingroup \@sanitize@url \@href}%
\providecommand \@href[1]{\@@startlink{#1}\@@href}%
\providecommand \@@href[1]{\endgroup#1\@@endlink}%
\providecommand \@sanitize@url [0]{\catcode `\\12\catcode `\$12\catcode `\&12\catcode `\#12\catcode `\^12\catcode `\_12\catcode `\%12\relax}%
\providecommand \@@startlink[1]{}%
\providecommand \@@endlink[0]{}%
\providecommand \url  [0]{\begingroup\@sanitize@url \@url }%
\providecommand \@url [1]{\endgroup\@href {#1}{\urlprefix }}%
\providecommand \urlprefix  [0]{URL }%
\providecommand \Eprint [0]{\href }%
\providecommand \doibase [0]{https://doi.org/}%
\providecommand \selectlanguage [0]{\@gobble}%
\providecommand \bibinfo  [0]{\@secondoftwo}%
\providecommand \bibfield  [0]{\@secondoftwo}%
\providecommand \translation [1]{[#1]}%
\providecommand \BibitemOpen [0]{}%
\providecommand \bibitemStop [0]{}%
\providecommand \bibitemNoStop [0]{.\EOS\space}%
\providecommand \EOS [0]{\spacefactor3000\relax}%
\providecommand \BibitemShut  [1]{\csname bibitem#1\endcsname}%
\let\auto@bib@innerbib\@empty
\bibitem [{\citenamefont {Lukin}\ \emph {et~al.}(2020{\natexlab{a}})\citenamefont {Lukin}, \citenamefont {Guidry},\ and\ \citenamefont {Vu{\v{c}}kovi{\'c}}}]{Lukin_Color_center_SiC_review_2020}%
  \BibitemOpen
  \bibfield  {author} {\bibinfo {author} {\bibfnamefont {D.~M.}\ \bibnamefont {Lukin}}, \bibinfo {author} {\bibfnamefont {M.~A.}\ \bibnamefont {Guidry}},\ and\ \bibinfo {author} {\bibfnamefont {J.}~\bibnamefont {Vu{\v{c}}kovi{\'c}}},\ }\bibfield  {title} {\bibinfo {title} {Integrated quantum photonics with silicon carbide: challenges and prospects},\ }\href {https://doi.org/10.1103/PRXQuantum.1.020102} {\bibfield  {journal} {\bibinfo  {journal} {PRX Quantum}\ }\textbf {\bibinfo {volume} {1}},\ \bibinfo {pages} {020102} (\bibinfo {year} {2020}{\natexlab{a}})}\BibitemShut {NoStop}%
\bibitem [{\citenamefont {Lukin}\ \emph {et~al.}(2020{\natexlab{b}})\citenamefont {Lukin}, \citenamefont {Dory}, \citenamefont {Guidry}, \citenamefont {Yang}, \citenamefont {Mishra}, \citenamefont {Trivedi}, \citenamefont {Radulaski}, \citenamefont {Sun}, \citenamefont {Vercruysse}, \citenamefont {Ahn} \emph {et~al.}}]{Lukin_Purcell_enhancement_SiC_SiCOI-2020}%
  \BibitemOpen
  \bibfield  {author} {\bibinfo {author} {\bibfnamefont {D.~M.}\ \bibnamefont {Lukin}}, \bibinfo {author} {\bibfnamefont {C.}~\bibnamefont {Dory}}, \bibinfo {author} {\bibfnamefont {M.~A.}\ \bibnamefont {Guidry}}, \bibinfo {author} {\bibfnamefont {K.~Y.}\ \bibnamefont {Yang}}, \bibinfo {author} {\bibfnamefont {S.~D.}\ \bibnamefont {Mishra}}, \bibinfo {author} {\bibfnamefont {R.}~\bibnamefont {Trivedi}}, \bibinfo {author} {\bibfnamefont {M.}~\bibnamefont {Radulaski}}, \bibinfo {author} {\bibfnamefont {S.}~\bibnamefont {Sun}}, \bibinfo {author} {\bibfnamefont {D.}~\bibnamefont {Vercruysse}}, \bibinfo {author} {\bibfnamefont {G.~H.}\ \bibnamefont {Ahn}}, \emph {et~al.},\ }\bibfield  {title} {\bibinfo {title} {{4H}-silicon-carbide-on-insulator for integrated quantum and nonlinear photonics},\ }\href {https://doi.org/10.1038/s41566-019-0556-6} {\bibfield  {journal} {\bibinfo  {journal} {Nat. Photonics}\ }\textbf {\bibinfo {volume} {14}},\ \bibinfo {pages} {330} (\bibinfo {year} {2020}{\natexlab{b}})}\BibitemShut {NoStop}%
\bibitem [{\citenamefont {Yang}\ \emph {et~al.}(2023)\citenamefont {Yang}, \citenamefont {Guidry}, \citenamefont {Lukin}, \citenamefont {Yang},\ and\ \citenamefont {Vu{\v{c}}kovi{\'c}}}]{Yang_SiCOI_photonics_2023}%
  \BibitemOpen
  \bibfield  {author} {\bibinfo {author} {\bibfnamefont {J.}~\bibnamefont {Yang}}, \bibinfo {author} {\bibfnamefont {M.~A.}\ \bibnamefont {Guidry}}, \bibinfo {author} {\bibfnamefont {D.~M.}\ \bibnamefont {Lukin}}, \bibinfo {author} {\bibfnamefont {K.}~\bibnamefont {Yang}},\ and\ \bibinfo {author} {\bibfnamefont {J.}~\bibnamefont {Vu{\v{c}}kovi{\'c}}},\ }\bibfield  {title} {\bibinfo {title} {Inverse-designed silicon carbide quantum and nonlinear photonics},\ }\href {https://doi.org/doi.org/10.1038/s41377-023-01253-9} {\bibfield  {journal} {\bibinfo  {journal} {Light: Sci. Appl.}\ }\textbf {\bibinfo {volume} {12}},\ \bibinfo {pages} {201} (\bibinfo {year} {2023})}\BibitemShut {NoStop}%
\bibitem [{\citenamefont {Sato}\ \emph {et~al.}(2009)\citenamefont {Sato}, \citenamefont {Abe}, \citenamefont {Shoji}, \citenamefont {Suda},\ and\ \citenamefont {Kondo}}]{Sato_Chi2_2009}%
  \BibitemOpen
  \bibfield  {author} {\bibinfo {author} {\bibfnamefont {H.}~\bibnamefont {Sato}}, \bibinfo {author} {\bibfnamefont {M.}~\bibnamefont {Abe}}, \bibinfo {author} {\bibfnamefont {I.}~\bibnamefont {Shoji}}, \bibinfo {author} {\bibfnamefont {J.}~\bibnamefont {Suda}},\ and\ \bibinfo {author} {\bibfnamefont {T.}~\bibnamefont {Kondo}},\ }\bibfield  {title} {\bibinfo {title} {Accurate measurements of second-order nonlinear optical coefficients of {6H} and {4H} silicon carbide},\ }\href {https://doi.org/10.1364/JOSAB.26.001892} {\bibfield  {journal} {\bibinfo  {journal} {J. Opt. Soc. Am. B}\ }\textbf {\bibinfo {volume} {26}},\ \bibinfo {pages} {1892} (\bibinfo {year} {2009})}\BibitemShut {NoStop}%
\bibitem [{\citenamefont {De~Leonardis}\ \emph {et~al.}(2017)\citenamefont {De~Leonardis}, \citenamefont {Soref},\ and\ \citenamefont {Passaro}}]{DeLeonardis_Chi3_2017}%
  \BibitemOpen
  \bibfield  {author} {\bibinfo {author} {\bibfnamefont {F.}~\bibnamefont {De~Leonardis}}, \bibinfo {author} {\bibfnamefont {R.~A.}\ \bibnamefont {Soref}},\ and\ \bibinfo {author} {\bibfnamefont {V.}~\bibnamefont {Passaro}},\ }\bibfield  {title} {\bibinfo {title} {Dispersion of nonresonant third-order nonlinearities in silicon carbide},\ }\href {https://doi.org/10.1038/srep40924} {\bibfield  {journal} {\bibinfo  {journal} {Sci. Rep.}\ }\textbf {\bibinfo {volume} {7}},\ \bibinfo {pages} {1} (\bibinfo {year} {2017})}\BibitemShut {NoStop}%
\bibitem [{\citenamefont {Dietz}\ \emph {et~al.}(2023)\citenamefont {Dietz}, \citenamefont {Jiang}, \citenamefont {Day}, \citenamefont {Bhave},\ and\ \citenamefont {Hu}}]{Dietz_V_Si_spin-acoustic_coupling-2023}%
  \BibitemOpen
  \bibfield  {author} {\bibinfo {author} {\bibfnamefont {J.~R.}\ \bibnamefont {Dietz}}, \bibinfo {author} {\bibfnamefont {B.}~\bibnamefont {Jiang}}, \bibinfo {author} {\bibfnamefont {A.~M.}\ \bibnamefont {Day}}, \bibinfo {author} {\bibfnamefont {S.~A.}\ \bibnamefont {Bhave}},\ and\ \bibinfo {author} {\bibfnamefont {E.~L.}\ \bibnamefont {Hu}},\ }\bibfield  {title} {\bibinfo {title} {Spin-acoustic control of silicon vacancies in {4H} silicon carbide},\ }\href {https://doi.org/10.1038/s41928-023-01029-4} {\bibfield  {journal} {\bibinfo  {journal} {Nat. Electron.}\ }\textbf {\bibinfo {volume} {6}},\ \bibinfo {pages} {739} (\bibinfo {year} {2023})}\BibitemShut {NoStop}%
\bibitem [{\citenamefont {Nagy}\ \emph {et~al.}(2019)\citenamefont {Nagy}, \citenamefont {Niethammer}, \citenamefont {Widmann}, \citenamefont {Chen}, \citenamefont {Udvarhelyi}, \citenamefont {Bonato}, \citenamefont {Hassan}, \citenamefont {Karhu}, \citenamefont {Ivanov}, \citenamefont {Son} \emph {et~al.}}]{Nagy_V_Si-2019}%
  \BibitemOpen
  \bibfield  {author} {\bibinfo {author} {\bibfnamefont {R.}~\bibnamefont {Nagy}}, \bibinfo {author} {\bibfnamefont {M.}~\bibnamefont {Niethammer}}, \bibinfo {author} {\bibfnamefont {M.}~\bibnamefont {Widmann}}, \bibinfo {author} {\bibfnamefont {Y.-C.}\ \bibnamefont {Chen}}, \bibinfo {author} {\bibfnamefont {P.}~\bibnamefont {Udvarhelyi}}, \bibinfo {author} {\bibfnamefont {C.}~\bibnamefont {Bonato}}, \bibinfo {author} {\bibfnamefont {J.~U.}\ \bibnamefont {Hassan}}, \bibinfo {author} {\bibfnamefont {R.}~\bibnamefont {Karhu}}, \bibinfo {author} {\bibfnamefont {I.~G.}\ \bibnamefont {Ivanov}}, \bibinfo {author} {\bibfnamefont {N.~T.}\ \bibnamefont {Son}}, \emph {et~al.},\ }\bibfield  {title} {\bibinfo {title} {High-fidelity spin and optical control of single silicon-vacancy centres in silicon carbide},\ }\href {https://doi.org/10.1038/s41467-019-09873-9} {\bibfield  {journal} {\bibinfo  {journal} {Nat. Commun.}\ }\textbf {\bibinfo {volume} {10}},\ \bibinfo {pages} {1} (\bibinfo {year} {2019})}\BibitemShut {NoStop}%
\bibitem [{\citenamefont {Udvarhelyi}\ \emph {et~al.}(2020)\citenamefont {Udvarhelyi}, \citenamefont {Thiering}, \citenamefont {Morioka}, \citenamefont {Babin}, \citenamefont {Kaiser}, \citenamefont {Lukin}, \citenamefont {Ohshima}, \citenamefont {Ul-Hassan}, \citenamefont {Son}, \citenamefont {Vu{\v{c}}kovi{\'c}} \emph {et~al.}}]{Udvarhelyi_strain_V_Si-2020}%
  \BibitemOpen
  \bibfield  {author} {\bibinfo {author} {\bibfnamefont {P.}~\bibnamefont {Udvarhelyi}}, \bibinfo {author} {\bibfnamefont {G.}~\bibnamefont {Thiering}}, \bibinfo {author} {\bibfnamefont {N.}~\bibnamefont {Morioka}}, \bibinfo {author} {\bibfnamefont {C.}~\bibnamefont {Babin}}, \bibinfo {author} {\bibfnamefont {F.}~\bibnamefont {Kaiser}}, \bibinfo {author} {\bibfnamefont {D.}~\bibnamefont {Lukin}}, \bibinfo {author} {\bibfnamefont {T.}~\bibnamefont {Ohshima}}, \bibinfo {author} {\bibfnamefont {J.}~\bibnamefont {Ul-Hassan}}, \bibinfo {author} {\bibfnamefont {N.~T.}\ \bibnamefont {Son}}, \bibinfo {author} {\bibfnamefont {J.}~\bibnamefont {Vu{\v{c}}kovi{\'c}}}, \emph {et~al.},\ }\bibfield  {title} {\bibinfo {title} {Vibronic states and their effect on the temperature and strain dependence of silicon-vacancy qubits in {4H-SiC}},\ }\href {https://doi.org/10.1103/PhysRevApplied.13.054017} {\bibfield  {journal} {\bibinfo  {journal} {Phys. Rev. Appl.}\ }\textbf {\bibinfo {volume} {13}},\ \bibinfo {pages} {054017} (\bibinfo {year} {2020})}\BibitemShut {NoStop}%
\bibitem [{\citenamefont {{Sementilli}}\ \emph {et~al.}(2021)\citenamefont {{Sementilli}}, \citenamefont {Romero},\ and\ \citenamefont {Bowen}}]{sementilli_review_2021}%
  \BibitemOpen
  \bibfield  {author} {\bibinfo {author} {\bibfnamefont {L.}~\bibnamefont {{Sementilli}}}, \bibinfo {author} {\bibfnamefont {E.}~\bibnamefont {Romero}},\ and\ \bibinfo {author} {\bibfnamefont {W.~P.}\ \bibnamefont {Bowen}},\ }\bibfield  {title} {\bibinfo {title} {Nanomechanical dissipation and strain engineering},\ }\href {https://doi.org/10.1002/adfm.202105247} {\bibfield  {journal} {\bibinfo  {journal} {Adv. Funct. Mater.}\ }\textbf {\bibinfo {volume} {32}},\ \bibinfo {pages} {2105247} (\bibinfo {year} {2021})}\BibitemShut {NoStop}%
\bibitem [{\citenamefont {Engelsen}\ \emph {et~al.}(2024)\citenamefont {Engelsen}, \citenamefont {Beccari},\ and\ \citenamefont {Kippenberg}}]{engelsen_review_2024}%
  \BibitemOpen
  \bibfield  {author} {\bibinfo {author} {\bibfnamefont {N.~J.}\ \bibnamefont {Engelsen}}, \bibinfo {author} {\bibfnamefont {A.}~\bibnamefont {Beccari}},\ and\ \bibinfo {author} {\bibfnamefont {T.~J.}\ \bibnamefont {Kippenberg}},\ }\bibfield  {title} {\bibinfo {title} {Ultrahigh-quality-factor micro- and nanomechanical resonators using dissipation dilution},\ }\href {https://doi.org/10.1038/s41565-023-01597-8} {\bibfield  {journal} {\bibinfo  {journal} {Nat. Nanotechnol.}\ }\textbf {\bibinfo {volume} {19}},\ \bibinfo {pages} {725} (\bibinfo {year} {2024})}\BibitemShut {NoStop}%
\bibitem [{\citenamefont {Ghaffari}\ \emph {et~al.}(2013)\citenamefont {Ghaffari}, \citenamefont {Chandorkar}, \citenamefont {Wang}, \citenamefont {Ng}, \citenamefont {Ahn}, \citenamefont {Hong}, \citenamefont {Yang},\ and\ \citenamefont {Kenny}}]{Ghaffari_Akhiezer_limit_damping-2013}%
  \BibitemOpen
  \bibfield  {author} {\bibinfo {author} {\bibfnamefont {S.}~\bibnamefont {Ghaffari}}, \bibinfo {author} {\bibfnamefont {S.~A.}\ \bibnamefont {Chandorkar}}, \bibinfo {author} {\bibfnamefont {S.}~\bibnamefont {Wang}}, \bibinfo {author} {\bibfnamefont {E.~J.}\ \bibnamefont {Ng}}, \bibinfo {author} {\bibfnamefont {C.~H.}\ \bibnamefont {Ahn}}, \bibinfo {author} {\bibfnamefont {V.}~\bibnamefont {Hong}}, \bibinfo {author} {\bibfnamefont {Y.}~\bibnamefont {Yang}},\ and\ \bibinfo {author} {\bibfnamefont {T.~W.}\ \bibnamefont {Kenny}},\ }\bibfield  {title} {\bibinfo {title} {Quantum limit of quality factor in silicon micro and nano mechanical resonators},\ }\href {https://doi.org/10.1038/srep03244} {\bibfield  {journal} {\bibinfo  {journal} {Sci. Rep.}\ }\textbf {\bibinfo {volume} {3}},\ \bibinfo {pages} {1} (\bibinfo {year} {2013})}\BibitemShut {NoStop}%
\bibitem [{\citenamefont {Hamelin}\ \emph {et~al.}(2019)\citenamefont {Hamelin}, \citenamefont {Yang}, \citenamefont {Daruwalla}, \citenamefont {Wen},\ and\ \citenamefont {Ayazi}}]{Hamelin_Akhiezer_SiC_limit-2019}%
  \BibitemOpen
  \bibfield  {author} {\bibinfo {author} {\bibfnamefont {B.}~\bibnamefont {Hamelin}}, \bibinfo {author} {\bibfnamefont {J.}~\bibnamefont {Yang}}, \bibinfo {author} {\bibfnamefont {A.}~\bibnamefont {Daruwalla}}, \bibinfo {author} {\bibfnamefont {H.}~\bibnamefont {Wen}},\ and\ \bibinfo {author} {\bibfnamefont {F.}~\bibnamefont {Ayazi}},\ }\bibfield  {title} {\bibinfo {title} {Monocrystalline silicon carbide disk resonators on phononic crystals with ultra-low dissipation bulk acoustic wave modes},\ }\href {https://doi.org/10.1038/s41598-019-54278-9} {\bibfield  {journal} {\bibinfo  {journal} {Sci. Rep.}\ }\textbf {\bibinfo {volume} {9}},\ \bibinfo {pages} {1} (\bibinfo {year} {2019})}\BibitemShut {NoStop}%
\bibitem [{\citenamefont {Imboden}\ and\ \citenamefont {Mohanty}(2014)}]{imboden_review_dissipation_2014}%
  \BibitemOpen
  \bibfield  {author} {\bibinfo {author} {\bibfnamefont {M.}~\bibnamefont {Imboden}}\ and\ \bibinfo {author} {\bibfnamefont {P.}~\bibnamefont {Mohanty}},\ }\bibfield  {title} {\bibinfo {title} {Dissipation in nanoelectromechanical systems},\ }\href {https://doi.org/http://dx.doi.org/10.1016/j.physrep.2013.09.003} {\bibfield  {journal} {\bibinfo  {journal} {Phys. Rep.}\ }\textbf {\bibinfo {volume} {534}},\ \bibinfo {pages} {89} (\bibinfo {year} {2014})}\BibitemShut {NoStop}%
\bibitem [{\citenamefont {Hochreiter}\ \emph {et~al.}(2023)\citenamefont {Hochreiter}, \citenamefont {Groß}, \citenamefont {Möller}, \citenamefont {Krieger},\ and\ \citenamefont {Weber}}]{hochreiter_electrochemical_etching_2023}%
  \BibitemOpen
  \bibfield  {author} {\bibinfo {author} {\bibfnamefont {A.}~\bibnamefont {Hochreiter}}, \bibinfo {author} {\bibfnamefont {F.}~\bibnamefont {Groß}}, \bibinfo {author} {\bibfnamefont {M.-N.}\ \bibnamefont {Möller}}, \bibinfo {author} {\bibfnamefont {M.}~\bibnamefont {Krieger}},\ and\ \bibinfo {author} {\bibfnamefont {H.~B.}\ \bibnamefont {Weber}},\ }\bibfield  {title} {\bibinfo {title} {Electrochemical etching strategy for shaping monolithic {3D} structures from {4H-SiC} wafers},\ }\href {https://doi.org/10.1038/s41598-023-46110-2} {\bibfield  {journal} {\bibinfo  {journal} {Sci. Rep.}\ }\textbf {\bibinfo {volume} {13}} (\bibinfo {year} {2023})}\BibitemShut {NoStop}%
\bibitem [{\citenamefont {Rühl}\ \emph {et~al.}(2018)\citenamefont {Rühl}, \citenamefont {Ott}, \citenamefont {Götzinger}, \citenamefont {Krieger},\ and\ \citenamefont {Weber}}]{Ruehl_controlled_color_center_generation_2018}%
  \BibitemOpen
  \bibfield  {author} {\bibinfo {author} {\bibfnamefont {M.}~\bibnamefont {Rühl}}, \bibinfo {author} {\bibfnamefont {C.}~\bibnamefont {Ott}}, \bibinfo {author} {\bibfnamefont {S.}~\bibnamefont {Götzinger}}, \bibinfo {author} {\bibfnamefont {M.}~\bibnamefont {Krieger}},\ and\ \bibinfo {author} {\bibfnamefont {H.~B.}\ \bibnamefont {Weber}},\ }\bibfield  {title} {\bibinfo {title} {Controlled generation of intrinsic near-infrared color centers in {4H-SiC} via proton irradiation and annealing},\ }\href {https://doi.org/10.1063/1.5045859} {\bibfield  {journal} {\bibinfo  {journal} {Appl. Phys. Lett.}\ }\textbf {\bibinfo {volume} {113}},\ \bibinfo {pages} {122102} (\bibinfo {year} {2018})}\BibitemShut {NoStop}%
\bibitem [{\citenamefont {Emtsev}\ \emph {et~al.}(2009)\citenamefont {Emtsev}, \citenamefont {Bostwick}, \citenamefont {Horn}, \citenamefont {Jobst}, \citenamefont {Kellogg}, \citenamefont {Ley}, \citenamefont {McChesney}, \citenamefont {Ohta}, \citenamefont {Reshanov}, \citenamefont {R{\"o}hrl} \emph {et~al.}}]{emtsev_graphene_growth_Argon_6H-SiC_2009}%
  \BibitemOpen
  \bibfield  {author} {\bibinfo {author} {\bibfnamefont {K.~V.}\ \bibnamefont {Emtsev}}, \bibinfo {author} {\bibfnamefont {A.}~\bibnamefont {Bostwick}}, \bibinfo {author} {\bibfnamefont {K.}~\bibnamefont {Horn}}, \bibinfo {author} {\bibfnamefont {J.}~\bibnamefont {Jobst}}, \bibinfo {author} {\bibfnamefont {G.~L.}\ \bibnamefont {Kellogg}}, \bibinfo {author} {\bibfnamefont {L.}~\bibnamefont {Ley}}, \bibinfo {author} {\bibfnamefont {J.~L.}\ \bibnamefont {McChesney}}, \bibinfo {author} {\bibfnamefont {T.}~\bibnamefont {Ohta}}, \bibinfo {author} {\bibfnamefont {S.~A.}\ \bibnamefont {Reshanov}}, \bibinfo {author} {\bibfnamefont {J.}~\bibnamefont {R{\"o}hrl}}, \emph {et~al.},\ }\bibfield  {title} {\bibinfo {title} {Towards wafer-size graphene layers by atmospheric pressure graphitization of silicon carbide},\ }\href {https://doi.org/10.1038/nmat2382} {\bibfield  {journal} {\bibinfo  {journal} {Nat. Mater.}\ }\textbf {\bibinfo {volume} {8}},\ \bibinfo {pages} {203} (\bibinfo {year} {2009})}\BibitemShut {NoStop}%
\bibitem [{\citenamefont {Villanueva}\ and\ \citenamefont {Schmid}(2014)}]{villanueva_evidence_surface_2014}%
  \BibitemOpen
  \bibfield  {author} {\bibinfo {author} {\bibfnamefont {L.~G.}\ \bibnamefont {Villanueva}}\ and\ \bibinfo {author} {\bibfnamefont {S.}~\bibnamefont {Schmid}},\ }\bibfield  {title} {\bibinfo {title} {Evidence of surface loss as ubiquitous limiting damping mechanism in {SiN} micro- and nanomechanical resonators},\ }\href {https://doi.org/10.1103/PhysRevLett.113.227201} {\bibfield  {journal} {\bibinfo  {journal} {Phys. Rev. Lett.}\ }\textbf {\bibinfo {volume} {113}},\ \bibinfo {pages} {227201} (\bibinfo {year} {2014})}\BibitemShut {NoStop}%
\bibitem [{\citenamefont {Romero}\ \emph {et~al.}(2020)\citenamefont {Romero}, \citenamefont {Valenzuela}, \citenamefont {Kermany}, \citenamefont {Sementilli}, \citenamefont {Iacopi},\ and\ \citenamefont {Bowen}}]{romero_engineering_dissipation_2020}%
  \BibitemOpen
  \bibfield  {author} {\bibinfo {author} {\bibfnamefont {E.}~\bibnamefont {Romero}}, \bibinfo {author} {\bibfnamefont {V.~M.}\ \bibnamefont {Valenzuela}}, \bibinfo {author} {\bibfnamefont {A.~R.}\ \bibnamefont {Kermany}}, \bibinfo {author} {\bibfnamefont {L.}~\bibnamefont {Sementilli}}, \bibinfo {author} {\bibfnamefont {F.}~\bibnamefont {Iacopi}},\ and\ \bibinfo {author} {\bibfnamefont {W.~P.}\ \bibnamefont {Bowen}},\ }\bibfield  {title} {\bibinfo {title} {Engineering the dissipation of crystalline micromechanical resonators},\ }\href {https://doi.org/10.1103/PhysRevApplied.13.044007} {\bibfield  {journal} {\bibinfo  {journal} {Phys. Rev. Appl.}\ }\textbf {\bibinfo {volume} {13}},\ \bibinfo {pages} {044007} (\bibinfo {year} {2020})}\BibitemShut {NoStop}%
\bibitem [{\citenamefont {Xu}\ \emph {et~al.}(2024)\citenamefont {Xu}, \citenamefont {Shin}, \citenamefont {Sberna}, \citenamefont {{van der Kolk}}, \citenamefont {Cupertino}, \citenamefont {Bessa},\ and\ \citenamefont {Norte}}]{xu_highstrength_amorphous_2024}%
  \BibitemOpen
  \bibfield  {author} {\bibinfo {author} {\bibfnamefont {M.}~\bibnamefont {Xu}}, \bibinfo {author} {\bibfnamefont {D.}~\bibnamefont {Shin}}, \bibinfo {author} {\bibfnamefont {P.~M.}\ \bibnamefont {Sberna}}, \bibinfo {author} {\bibfnamefont {R.}~\bibnamefont {{van der Kolk}}}, \bibinfo {author} {\bibfnamefont {A.}~\bibnamefont {Cupertino}}, \bibinfo {author} {\bibfnamefont {M.~A.}\ \bibnamefont {Bessa}},\ and\ \bibinfo {author} {\bibfnamefont {R.~A.}\ \bibnamefont {Norte}},\ }\bibfield  {title} {\bibinfo {title} {"high‑strength amorphous silicon carbide for nanomechanics"},\ }\href {https://doi.org/10.1002/adma.202306513} {\bibfield  {journal} {\bibinfo  {journal} {Adv. Mater.}\ }\textbf {\bibinfo {volume} {36}},\ \bibinfo {pages} {2306513} (\bibinfo {year} {2024})}\BibitemShut {NoStop}%
\bibitem [{\citenamefont {Manjeshwar}\ \emph {et~al.}(2023)\citenamefont {Manjeshwar}, \citenamefont {Ciers}, \citenamefont {Hellman}, \citenamefont {Bläsing}, \citenamefont {Strittmatter},\ and\ \citenamefont {Wieczorek}}]{manjeshwar_highq_trampoline_2023}%
  \BibitemOpen
  \bibfield  {author} {\bibinfo {author} {\bibfnamefont {S.~K.}\ \bibnamefont {Manjeshwar}}, \bibinfo {author} {\bibfnamefont {A.}~\bibnamefont {Ciers}}, \bibinfo {author} {\bibfnamefont {F.}~\bibnamefont {Hellman}}, \bibinfo {author} {\bibfnamefont {J.}~\bibnamefont {Bläsing}}, \bibinfo {author} {\bibfnamefont {A.}~\bibnamefont {Strittmatter}},\ and\ \bibinfo {author} {\bibfnamefont {W.}~\bibnamefont {Wieczorek}},\ }\bibfield  {title} {\bibinfo {title} {High-q trampoline resonators from strained crystalline ingap for integrated free-space optomechanics},\ }\href {https://doi.org/10.1021/acs.nanolett.3c00996} {\bibfield  {journal} {\bibinfo  {journal} {Nano Lett.}\ }\textbf {\bibinfo {volume} {23}},\ \bibinfo {pages} {5076} (\bibinfo {year} {2023})}\BibitemShut {NoStop}%
\bibitem [{\citenamefont {Ciers}\ \emph {et~al.}(2024)\citenamefont {Ciers}, \citenamefont {Jung}, \citenamefont {Ciers}, \citenamefont {Nindito}, \citenamefont {Pfeifer}, \citenamefont {Dadgar}, \citenamefont {Strittmatter},\ and\ \citenamefont {Wieczorek}}]{ciers_AlN_2024}%
  \BibitemOpen
  \bibfield  {author} {\bibinfo {author} {\bibfnamefont {A.}~\bibnamefont {Ciers}}, \bibinfo {author} {\bibfnamefont {A.}~\bibnamefont {Jung}}, \bibinfo {author} {\bibfnamefont {J.}~\bibnamefont {Ciers}}, \bibinfo {author} {\bibfnamefont {L.~R.}\ \bibnamefont {Nindito}}, \bibinfo {author} {\bibfnamefont {H.}~\bibnamefont {Pfeifer}}, \bibinfo {author} {\bibfnamefont {A.}~\bibnamefont {Dadgar}}, \bibinfo {author} {\bibfnamefont {A.}~\bibnamefont {Strittmatter}},\ and\ \bibinfo {author} {\bibfnamefont {W.}~\bibnamefont {Wieczorek}},\ }\bibfield  {title} {\bibinfo {title} {Nanomechanical crystalline {AlN} resonators with high quality factors for quantum optoelectromechanics},\ }\bibfield  {journal} {\bibinfo  {journal} {Adv. Mater.}\ }\href {https://doi.org/10.1002/adma.202403155} {10.1002/adma.202403155} (\bibinfo {year} {2024})\BibitemShut {NoStop}%
\bibitem [{\citenamefont {Cadeddu}\ \emph {et~al.}(2016)\citenamefont {Cadeddu}, \citenamefont {Braakman}, \citenamefont {Tütüncüoglu}, \citenamefont {Matteini}, \citenamefont {Rüffer}, \citenamefont {Fontcuberta~i Morral},\ and\ \citenamefont {Poggio}}]{cadeddu_timeresolved_nonlinear_2016}%
  \BibitemOpen
  \bibfield  {author} {\bibinfo {author} {\bibfnamefont {D.}~\bibnamefont {Cadeddu}}, \bibinfo {author} {\bibfnamefont {F.~R.}\ \bibnamefont {Braakman}}, \bibinfo {author} {\bibfnamefont {G.}~\bibnamefont {Tütüncüoglu}}, \bibinfo {author} {\bibfnamefont {F.}~\bibnamefont {Matteini}}, \bibinfo {author} {\bibfnamefont {D.}~\bibnamefont {Rüffer}}, \bibinfo {author} {\bibfnamefont {A.}~\bibnamefont {Fontcuberta~i Morral}},\ and\ \bibinfo {author} {\bibfnamefont {M.}~\bibnamefont {Poggio}},\ }\bibfield  {title} {\bibinfo {title} {Time-resolved nonlinear coupling between orthogonal flexural modes of a pristine gaas nanowire},\ }\href {https://doi.org/10.1021/acs.nanolett.5b03822} {\bibfield  {journal} {\bibinfo  {journal} {Nano Lett.}\ }\textbf {\bibinfo {volume} {16}},\ \bibinfo {pages} {926} (\bibinfo {year} {2016})}\BibitemShut {NoStop}%
\bibitem [{\citenamefont {Tao}\ \emph {et~al.}(2014)\citenamefont {Tao}, \citenamefont {Boss}, \citenamefont {Moores},\ and\ \citenamefont {Degen}}]{tao_singlecrystal_diamond_2014}%
  \BibitemOpen
  \bibfield  {author} {\bibinfo {author} {\bibfnamefont {Y.}~\bibnamefont {Tao}}, \bibinfo {author} {\bibfnamefont {J.~M.}\ \bibnamefont {Boss}}, \bibinfo {author} {\bibfnamefont {B.~A.}\ \bibnamefont {Moores}},\ and\ \bibinfo {author} {\bibfnamefont {C.~L.}\ \bibnamefont {Degen}},\ }\bibfield  {title} {\bibinfo {title} {Single-crystal diamond nanomechanical resonators with quality factors exceeding one million},\ }\href {https://doi.org/10.1038/ncomms4638} {\bibfield  {journal} {\bibinfo  {journal} {Nat. Commun.}\ }\textbf {\bibinfo {volume} {5}} (\bibinfo {year} {2014})}\BibitemShut {NoStop}%
\bibitem [{\citenamefont {Metcalf}\ \emph {et~al.}(2009)\citenamefont {Metcalf}, \citenamefont {Pate}, \citenamefont {Photiadis},\ and\ \citenamefont {Houston}}]{metcalf_thermoelastic_damping_2009}%
  \BibitemOpen
  \bibfield  {author} {\bibinfo {author} {\bibfnamefont {T.~H.}\ \bibnamefont {Metcalf}}, \bibinfo {author} {\bibfnamefont {B.~B.}\ \bibnamefont {Pate}}, \bibinfo {author} {\bibfnamefont {D.~M.}\ \bibnamefont {Photiadis}},\ and\ \bibinfo {author} {\bibfnamefont {B.~H.}\ \bibnamefont {Houston}},\ }\bibfield  {title} {\bibinfo {title} {Thermoelastic damping in micromechanical resonators},\ }\href {https://doi.org/10.1063/1.3190509} {\bibfield  {journal} {\bibinfo  {journal} {Appl. Phys. Lett.}\ }\textbf {\bibinfo {volume} {95}},\ \bibinfo {pages} {061903} (\bibinfo {year} {2009})}\BibitemShut {NoStop}%
\bibitem [{\citenamefont {Adachi}\ \emph {et~al.}(2013)\citenamefont {Adachi}, \citenamefont {Watanabe}, \citenamefont {Okamoto}, \citenamefont {Yamaguchi}, \citenamefont {Kimoto},\ and\ \citenamefont {Suda}}]{adachi_singlecrystalline_4hsic_2013}%
  \BibitemOpen
  \bibfield  {author} {\bibinfo {author} {\bibfnamefont {K.}~\bibnamefont {Adachi}}, \bibinfo {author} {\bibfnamefont {N.}~\bibnamefont {Watanabe}}, \bibinfo {author} {\bibfnamefont {H.}~\bibnamefont {Okamoto}}, \bibinfo {author} {\bibfnamefont {H.}~\bibnamefont {Yamaguchi}}, \bibinfo {author} {\bibfnamefont {T.}~\bibnamefont {Kimoto}},\ and\ \bibinfo {author} {\bibfnamefont {J.}~\bibnamefont {Suda}},\ }\bibfield  {title} {\bibinfo {title} {Single-crystalline {4H-SiC} micro cantilevers with a high quality factor},\ }\href {https://doi.org/10.1016/j.sna.2013.04.014} {\bibfield  {journal} {\bibinfo  {journal} {Sens. Actuators, A}\ }\textbf {\bibinfo {volume} {197}},\ \bibinfo {pages} {122} (\bibinfo {year} {2013})}\BibitemShut {NoStop}%
\bibitem [{\citenamefont {Sementilli}\ \emph {et~al.}(2025)\citenamefont {Sementilli}, \citenamefont {Lukin}, \citenamefont {Lee}, \citenamefont {Yang}, \citenamefont {Romero}, \citenamefont {Vučković},\ and\ \citenamefont {Bowen}}]{sementilli_lowdissipation_nanomechanical_2025}%
  \BibitemOpen
  \bibfield  {author} {\bibinfo {author} {\bibfnamefont {L.}~\bibnamefont {Sementilli}}, \bibinfo {author} {\bibfnamefont {D.~M.}\ \bibnamefont {Lukin}}, \bibinfo {author} {\bibfnamefont {H.}~\bibnamefont {Lee}}, \bibinfo {author} {\bibfnamefont {J.}~\bibnamefont {Yang}}, \bibinfo {author} {\bibfnamefont {E.}~\bibnamefont {Romero}}, \bibinfo {author} {\bibfnamefont {J.}~\bibnamefont {Vučković}},\ and\ \bibinfo {author} {\bibfnamefont {W.~P.}\ \bibnamefont {Bowen}},\ }\bibfield  {title} {\bibinfo {title} {Low-dissipation nanomechanical devices from monocrystalline silicon carbide},\ }\href {https://doi.org/10.1021/acs.nanolett.4c06475} {\bibfield  {journal} {\bibinfo  {journal} {Nano Lett.}\ }\textbf {\bibinfo {volume} {25}},\ \bibinfo {pages} {6069} (\bibinfo {year} {2025})}\BibitemShut {NoStop}%
\bibitem [{\citenamefont {Sato}\ \emph {et~al.}(2014)\citenamefont {Sato}, \citenamefont {Adachi}, \citenamefont {Okamoto}, \citenamefont {Yamaguchi}, \citenamefont {Kimoto},\ and\ \citenamefont {Suda}}]{sato_fabrication_electrostatically_2014}%
  \BibitemOpen
  \bibfield  {author} {\bibinfo {author} {\bibfnamefont {K.}~\bibnamefont {Sato}}, \bibinfo {author} {\bibfnamefont {K.}~\bibnamefont {Adachi}}, \bibinfo {author} {\bibfnamefont {H.}~\bibnamefont {Okamoto}}, \bibinfo {author} {\bibfnamefont {H.}~\bibnamefont {Yamaguchi}}, \bibinfo {author} {\bibfnamefont {T.}~\bibnamefont {Kimoto}},\ and\ \bibinfo {author} {\bibfnamefont {J.}~\bibnamefont {Suda}},\ }\bibfield  {title} {\bibinfo {title} {Fabrication of electrostatically actuated {4H-SiC} microcantilever resonators by using n/p/n epitaxial structures and doping-selective electrochemical etching},\ }in\ \href {https://doi.org/10.4028/www.scientific.net/MSF.778-780.780} {\emph {\bibinfo {booktitle} {Silicon Carbide and Related Materials 2013}}},\ \bibinfo {series} {Materials Science Forum}, Vol.\ \bibinfo {volume} {778}\ (\bibinfo  {publisher} {Trans Tech Publications Ltd},\ \bibinfo {year} {2014})\ pp.\ \bibinfo {pages} {780--783}\BibitemShut {NoStop}%
\bibitem [{\citenamefont {Locke}\ \emph {et~al.}(2000)\citenamefont {Locke}, \citenamefont {Tobar},\ and\ \citenamefont {Ivanov}}]{locke_sapphire_2000}%
  \BibitemOpen
  \bibfield  {author} {\bibinfo {author} {\bibfnamefont {C.~R.}\ \bibnamefont {Locke}}, \bibinfo {author} {\bibfnamefont {M.~E.}\ \bibnamefont {Tobar}},\ and\ \bibinfo {author} {\bibfnamefont {E.~N.}\ \bibnamefont {Ivanov}},\ }\bibfield  {title} {\bibinfo {title} {Monolithic sapphire parametric transducer operation at cryogenic temperatures},\ }\href {https://doi.org/10.1063/1.1150684} {\bibfield  {journal} {\bibinfo  {journal} {Rev. Sci. Instrum.}\ }\textbf {\bibinfo {volume} {71}},\ \bibinfo {pages} {2737} (\bibinfo {year} {2000})}\BibitemShut {NoStop}%
\bibitem [{\citenamefont {Galliou}\ \emph {et~al.}(2011)\citenamefont {Galliou}, \citenamefont {Imbaud}, \citenamefont {Goryachev}, \citenamefont {Bourquin},\ and\ \citenamefont {Abbé}}]{galliou_quartz_2011}%
  \BibitemOpen
  \bibfield  {author} {\bibinfo {author} {\bibfnamefont {S.}~\bibnamefont {Galliou}}, \bibinfo {author} {\bibfnamefont {J.}~\bibnamefont {Imbaud}}, \bibinfo {author} {\bibfnamefont {M.}~\bibnamefont {Goryachev}}, \bibinfo {author} {\bibfnamefont {R.}~\bibnamefont {Bourquin}},\ and\ \bibinfo {author} {\bibfnamefont {P.}~\bibnamefont {Abbé}},\ }\bibfield  {title} {\bibinfo {title} {Losses in high quality quartz crystal resonators at cryogenic temperatures},\ }\href {https://doi.org/10.1063/1.3559611} {\bibfield  {journal} {\bibinfo  {journal} {Appl. Phys. Lett.}\ }\textbf {\bibinfo {volume} {98}},\ \bibinfo {pages} {091911} (\bibinfo {year} {2011})}\BibitemShut {NoStop}%
\bibitem [{\citenamefont {González}\ and\ \citenamefont {Saulson}(1994)}]{gonzalez_brownian_1994}%
  \BibitemOpen
  \bibfield  {author} {\bibinfo {author} {\bibfnamefont {G.~I.}\ \bibnamefont {González}}\ and\ \bibinfo {author} {\bibfnamefont {P.~R.}\ \bibnamefont {Saulson}},\ }\bibfield  {title} {\bibinfo {title} {Brownian motion of a mass suspended by an anelastic wire},\ }\href {https://doi.org/http://dx.doi.org/10.1121/1.410467} {\bibfield  {journal} {\bibinfo  {journal} {J. Acoust. Soc. Am.}\ }\textbf {\bibinfo {volume} {96}},\ \bibinfo {pages} {207} (\bibinfo {year} {1994})}\BibitemShut {NoStop}%
\bibitem [{\citenamefont {Kajima}\ \emph {et~al.}(1999)\citenamefont {Kajima}, \citenamefont {Kusumi}, \citenamefont {Moriwaki},\ and\ \citenamefont {Mio}}]{kajima_1999}%
  \BibitemOpen
  \bibfield  {author} {\bibinfo {author} {\bibfnamefont {M.}~\bibnamefont {Kajima}}, \bibinfo {author} {\bibfnamefont {N.}~\bibnamefont {Kusumi}}, \bibinfo {author} {\bibfnamefont {S.}~\bibnamefont {Moriwaki}},\ and\ \bibinfo {author} {\bibfnamefont {N.}~\bibnamefont {Mio}},\ }\bibfield  {title} {\bibinfo {title} {Wide-band measurement of mechanical thermal noise using a laser interferometer},\ }\href {https://doi.org/10.1016/s0375-9601(99)00828-2} {\bibfield  {journal} {\bibinfo  {journal} {Phys. Lett. A}\ }\textbf {\bibinfo {volume} {264}},\ \bibinfo {pages} {251} (\bibinfo {year} {1999})}\BibitemShut {NoStop}%
\bibitem [{\citenamefont {Bernardini}\ \emph {et~al.}(1999)\citenamefont {Bernardini}, \citenamefont {Majorana}, \citenamefont {Ogawa}, \citenamefont {Puppo}, \citenamefont {Rapagnani}, \citenamefont {Ricci},\ and\ \citenamefont {Testi}}]{bernardini_1999}%
  \BibitemOpen
  \bibfield  {author} {\bibinfo {author} {\bibfnamefont {A.}~\bibnamefont {Bernardini}}, \bibinfo {author} {\bibfnamefont {E.}~\bibnamefont {Majorana}}, \bibinfo {author} {\bibfnamefont {Y.}~\bibnamefont {Ogawa}}, \bibinfo {author} {\bibfnamefont {P.}~\bibnamefont {Puppo}}, \bibinfo {author} {\bibfnamefont {P.}~\bibnamefont {Rapagnani}}, \bibinfo {author} {\bibfnamefont {F.}~\bibnamefont {Ricci}},\ and\ \bibinfo {author} {\bibfnamefont {G.}~\bibnamefont {Testi}},\ }\bibfield  {title} {\bibinfo {title} {Characterization of mechanical dissipation spectral behavior using a gravitomagnetic pendulum},\ }\href {https://doi.org/10.1016/s0375-9601(99)00146-2} {\bibfield  {journal} {\bibinfo  {journal} {Phys. Lett. A}\ }\textbf {\bibinfo {volume} {255}},\ \bibinfo {pages} {142} (\bibinfo {year} {1999})}\BibitemShut {NoStop}%
\bibitem [{\citenamefont {Fedorov}\ \emph {et~al.}(2017)\citenamefont {Fedorov}, \citenamefont {Sudhir}, \citenamefont {Schilling}, \citenamefont {Schütz}, \citenamefont {Wilson},\ and\ \citenamefont {Kippenberg}}]{fedorov_structural-damping_2017}%
  \BibitemOpen
  \bibfield  {author} {\bibinfo {author} {\bibfnamefont {S.}~\bibnamefont {Fedorov}}, \bibinfo {author} {\bibfnamefont {V.}~\bibnamefont {Sudhir}}, \bibinfo {author} {\bibfnamefont {R.}~\bibnamefont {Schilling}}, \bibinfo {author} {\bibfnamefont {H.}~\bibnamefont {Schütz}}, \bibinfo {author} {\bibfnamefont {D.}~\bibnamefont {Wilson}},\ and\ \bibinfo {author} {\bibfnamefont {T.}~\bibnamefont {Kippenberg}},\ }\bibfield  {title} {\bibinfo {title} {Evidence for structural damping in a high-stress silicon nitride nanobeam and its implications for quantum optomechanics},\ }\href {https://doi.org/10.1016/j.physleta.2017.05.046} {\bibfield  {journal} {\bibinfo  {journal} {Phys. Lett. A}\ }\textbf {\bibinfo {volume} {382}},\ \bibinfo {pages} {2251} (\bibinfo {year} {2017})}\BibitemShut {NoStop}%
\bibitem [{\citenamefont {Kla\ss{}}\ \emph {et~al.}()\citenamefont {Kla\ss{}}, \citenamefont {Wilson-Rae},\ and\ \citenamefont {Weig}}]{klass_unpublished}%
  \BibitemOpen
  \bibfield  {author} {\bibinfo {author} {\bibfnamefont {Y.}~\bibnamefont {Kla\ss{}}}, \bibinfo {author} {\bibfnamefont {I.}~\bibnamefont {Wilson-Rae}},\ and\ \bibinfo {author} {\bibfnamefont {E.}~\bibnamefont {Weig}},\ }\bibfield  {title} {\bibinfo {title} {Constancy of the undiluted inverse {Q} from the stress-diluted dissipation of crystalline nanomechanical resonators},\ }\href@noop {} {\bibinfo  {journal} {in preparation}\ }\BibitemShut {NoStop}%
\bibitem [{\citenamefont {Chen}\ \emph {et~al.}(2019)\citenamefont {Chen}, \citenamefont {Fahim}, \citenamefont {Suhling},\ and\ \citenamefont {Jaeger}}]{chen_study_elastic_2019}%
  \BibitemOpen
\bibfield  {journal} {  }\bibfield  {author} {\bibinfo {author} {\bibfnamefont {J.}~\bibnamefont {Chen}}, \bibinfo {author} {\bibfnamefont {A.}~\bibnamefont {Fahim}}, \bibinfo {author} {\bibfnamefont {J.~C.}\ \bibnamefont {Suhling}},\ and\ \bibinfo {author} {\bibfnamefont {R.~C.}\ \bibnamefont {Jaeger}},\ }\bibfield  {title} {\bibinfo {title} {A study of the elastic constants of {4H} silicon carbide ({4H-SiC})},\ }in\ \href {https://doi.org/10.1109/ITHERM.2019.8757291} {\emph {\bibinfo {booktitle} {2019 18th {IEEE} Intersociety Conference on Thermal and Thermomechanical Phenomena in Electronic Systems (ITherm)}}}\ (\bibinfo {year} {2019})\ pp.\ \bibinfo {pages} {835--840}\BibitemShut {NoStop}%
\bibitem [{\citenamefont {Schmid}\ \emph {et~al.}(2016)\citenamefont {Schmid}, \citenamefont {Villanueva},\ and\ \citenamefont {Roukes}}]{funda-of-nanom-reson}%
  \BibitemOpen
  \bibfield  {author} {\bibinfo {author} {\bibfnamefont {S.}~\bibnamefont {Schmid}}, \bibinfo {author} {\bibfnamefont {L.~G.}\ \bibnamefont {Villanueva}},\ and\ \bibinfo {author} {\bibfnamefont {M.~L.}\ \bibnamefont {Roukes}},\ }\href@noop {} {\emph {\bibinfo {title} {Fundamentals of nanomechanical resonators}}},\ Vol.~\bibinfo {volume} {49}\ (\bibinfo  {publisher} {Springer},\ \bibinfo {year} {2016})\BibitemShut {NoStop}%
\bibitem [{\citenamefont {Ben~Messaoud}\ \emph {et~al.}(2019)\citenamefont {Ben~Messaoud}, \citenamefont {Michaud}, \citenamefont {Certon}, \citenamefont {Camarda}, \citenamefont {Piluso}, \citenamefont {Colin}, \citenamefont {Barcella},\ and\ \citenamefont {Alquier}}]{messaoud_investigation_youngs_2019}%
  \BibitemOpen
  \bibfield  {author} {\bibinfo {author} {\bibfnamefont {J.}~\bibnamefont {Ben~Messaoud}}, \bibinfo {author} {\bibfnamefont {J.-F.}\ \bibnamefont {Michaud}}, \bibinfo {author} {\bibfnamefont {D.}~\bibnamefont {Certon}}, \bibinfo {author} {\bibfnamefont {M.}~\bibnamefont {Camarda}}, \bibinfo {author} {\bibfnamefont {N.}~\bibnamefont {Piluso}}, \bibinfo {author} {\bibfnamefont {L.}~\bibnamefont {Colin}}, \bibinfo {author} {\bibfnamefont {F.}~\bibnamefont {Barcella}},\ and\ \bibinfo {author} {\bibfnamefont {D.}~\bibnamefont {Alquier}},\ }\bibfield  {title} {\bibinfo {title} {Investigation of the {Young's} modulus and the residual stress of {4H-SiC} circular membranes on {4H-SiC} substrates},\ }\href {https://doi.org/10.3390/mi10120801} {\bibfield  {journal} {\bibinfo  {journal} {Micromachines}\ }\textbf {\bibinfo {volume} {10}},\ \bibinfo {pages} {801} (\bibinfo {year} {2019})}\BibitemShut {NoStop}%
\bibitem [{\citenamefont {Xu}\ \emph {et~al.}(2018)\citenamefont {Xu}, \citenamefont {Xia}, \citenamefont {Chen}, \citenamefont {Wu}, \citenamefont {Gang},\ and\ \citenamefont {Huang}}]{xu_hightemperature_mechanical_2018}%
  \BibitemOpen
  \bibfield  {author} {\bibinfo {author} {\bibfnamefont {W.-W.}\ \bibnamefont {Xu}}, \bibinfo {author} {\bibfnamefont {F.}~\bibnamefont {Xia}}, \bibinfo {author} {\bibfnamefont {L.}~\bibnamefont {Chen}}, \bibinfo {author} {\bibfnamefont {M.}~\bibnamefont {Wu}}, \bibinfo {author} {\bibfnamefont {T.}~\bibnamefont {Gang}},\ and\ \bibinfo {author} {\bibfnamefont {Y.}~\bibnamefont {Huang}},\ }\bibfield  {title} {\bibinfo {title} {High-temperature mechanical and thermodynamic properties of silicon carbide polytypes},\ }\href {https://doi.org/10.1016/j.jallcom.2018.07.299} {\bibfield  {journal} {\bibinfo  {journal} {J. Alloy. Compd.}\ }\textbf {\bibinfo {volume} {768}},\ \bibinfo {pages} {722} (\bibinfo {year} {2018})}\BibitemShut {NoStop}%
\bibitem [{\citenamefont {Islam}\ \emph {et~al.}(2012)\citenamefont {Islam}, \citenamefont {Huang},\ and\ \citenamefont {Zhao}}]{islam_singlecrystal_sic_2012}%
  \BibitemOpen
  \bibfield  {author} {\bibinfo {author} {\bibfnamefont {M.~M.}\ \bibnamefont {Islam}}, \bibinfo {author} {\bibfnamefont {C.~F.}\ \bibnamefont {Huang}},\ and\ \bibinfo {author} {\bibfnamefont {F.}~\bibnamefont {Zhao}},\ }\bibfield  {title} {\bibinfo {title} {Single-crystal {SiC} resonators by photoelectrochemical etching},\ }in\ \href {https://doi.org/10.4028/www.scientific.net/MSF.717-720.529} {\emph {\bibinfo {booktitle} {Silicon Carbide and Related Materials 2011}}},\ \bibinfo {series} {Materials Science Forum}, Vol.\ \bibinfo {volume} {717}\ (\bibinfo  {publisher} {Trans Tech Publications Ltd},\ \bibinfo {year} {2012})\ pp.\ \bibinfo {pages} {529--532}\BibitemShut {NoStop}%
\bibitem [{\citenamefont {Karmann}\ \emph {et~al.}(1989)\citenamefont {Karmann}, \citenamefont {Helbig},\ and\ \citenamefont {Stein}}]{karmann_piezoelectric_properties_1989}%
  \BibitemOpen
  \bibfield  {author} {\bibinfo {author} {\bibfnamefont {S.}~\bibnamefont {Karmann}}, \bibinfo {author} {\bibfnamefont {R.}~\bibnamefont {Helbig}},\ and\ \bibinfo {author} {\bibfnamefont {R.~A.}\ \bibnamefont {Stein}},\ }\bibfield  {title} {\bibinfo {title} {Piezoelectric properties and elastic constants of {4H} and {6H} {SiC} at temperatures 4-320 k},\ }\href {https://doi.org/10.1063/1.344477} {\bibfield  {journal} {\bibinfo  {journal} {J. Appl. Phys.}\ }\textbf {\bibinfo {volume} {66}},\ \bibinfo {pages} {3922} (\bibinfo {year} {1989})}\BibitemShut {NoStop}%
\bibitem [{\citenamefont {Pizzagalli}(2014)}]{pizzagalli_stability_mobility_2014}%
  \BibitemOpen
  \bibfield  {author} {\bibinfo {author} {\bibfnamefont {L.}~\bibnamefont {Pizzagalli}},\ }\bibfield  {title} {\bibinfo {title} {Stability and mobility of screw dislocations in {4H}, {2H} and {3C} silicon carbide},\ }\href {https://doi.org/10.1016/j.actamat.2014.06.053} {\bibfield  {journal} {\bibinfo  {journal} {Acta Mater.}\ }\textbf {\bibinfo {volume} {78}},\ \bibinfo {pages} {236} (\bibinfo {year} {2014})}\BibitemShut {NoStop}%
\bibitem [{\citenamefont {Yahagi}\ \emph {et~al.}()\citenamefont {Yahagi}, \citenamefont {Ohji}, \citenamefont {Yamaguchi}, \citenamefont {Takahashi}, \citenamefont {Nakano}, \citenamefont {Iijima},\ and\ \citenamefont {Tatami}}]{yahagi_deformation_behavior}%
  \BibitemOpen
  \bibfield  {author} {\bibinfo {author} {\bibfnamefont {T.}~\bibnamefont {Yahagi}}, \bibinfo {author} {\bibfnamefont {T.}~\bibnamefont {Ohji}}, \bibinfo {author} {\bibfnamefont {H.}~\bibnamefont {Yamaguchi}}, \bibinfo {author} {\bibfnamefont {T.}~\bibnamefont {Takahashi}}, \bibinfo {author} {\bibfnamefont {H.}~\bibnamefont {Nakano}}, \bibinfo {author} {\bibfnamefont {M.}~\bibnamefont {Iijima}},\ and\ \bibinfo {author} {\bibfnamefont {J.}~\bibnamefont {Tatami}},\ }\bibfield  {title} {\bibinfo {title} {Deformation behavior and fracture strength of single-crystal {4H-SiC} determined by microcantilever bending tests},\ }\href {https://doi.org/10.1002/adem.202400095} {\bibfield  {journal} {\bibinfo  {journal} {Adv. Eng. Mater.}\ }\textbf {\bibinfo {volume} {26}},\ \bibinfo {pages} {2400095}}\BibitemShut {NoStop}%
\bibitem [{\citenamefont {Nuruzzaman}\ \emph {et~al.}(2015)\citenamefont {Nuruzzaman}, \citenamefont {Islam}, \citenamefont {Alam}, \citenamefont {Shah},\ and\ \citenamefont {Karim}}]{nuruzzaman_structural_elastic_2015}%
  \BibitemOpen
  \bibfield  {author} {\bibinfo {author} {\bibfnamefont {M.}~\bibnamefont {Nuruzzaman}}, \bibinfo {author} {\bibfnamefont {M.~A.}\ \bibnamefont {Islam}}, \bibinfo {author} {\bibfnamefont {M.~A.}\ \bibnamefont {Alam}}, \bibinfo {author} {\bibfnamefont {M.~H.}\ \bibnamefont {Shah}},\ and\ \bibinfo {author} {\bibfnamefont {A.}~\bibnamefont {Karim}},\ }\bibfield  {title} {\bibinfo {title} {Structural, elastic and electronic properties of 2h-and {4H-SiC}},\ }\href@noop {} {\bibfield  {journal} {\bibinfo  {journal} {Int. J. Eng. Res. Appl.}\ }\textbf {\bibinfo {volume} {5}},\ \bibinfo {pages} {48} (\bibinfo {year} {2015})}\BibitemShut {NoStop}%
\bibitem [{\citenamefont {Yang}\ \emph {et~al.}(2020)\citenamefont {Yang}, \citenamefont {Hamelin},\ and\ \citenamefont {Ayazi}}]{yang_investigating_elastic_2020}%
  \BibitemOpen
  \bibfield  {author} {\bibinfo {author} {\bibfnamefont {J.}~\bibnamefont {Yang}}, \bibinfo {author} {\bibfnamefont {B.}~\bibnamefont {Hamelin}},\ and\ \bibinfo {author} {\bibfnamefont {F.}~\bibnamefont {Ayazi}},\ }\bibfield  {title} {\bibinfo {title} {Investigating elastic anisotropy of {4H-SiC} using ultra-high q bulk acoustic wave resonators},\ }\href@noop {} {\bibfield  {journal} {\bibinfo  {journal} {J Microelectromech Syst}\ }\textbf {\bibinfo {volume} {29}},\ \bibinfo {pages} {1473} (\bibinfo {year} {2020})}\BibitemShut {NoStop}%
\bibitem [{\citenamefont {Sakakima}\ \emph {et~al.}(2018)\citenamefont {Sakakima}, \citenamefont {Takamoto}, \citenamefont {Murakami}, \citenamefont {Hatano}, \citenamefont {Goryu}, \citenamefont {Hirohata},\ and\ \citenamefont {Izumi}}]{sakakima_development_method_2018}%
  \BibitemOpen
  \bibfield  {author} {\bibinfo {author} {\bibfnamefont {H.}~\bibnamefont {Sakakima}}, \bibinfo {author} {\bibfnamefont {S.}~\bibnamefont {Takamoto}}, \bibinfo {author} {\bibfnamefont {Y.}~\bibnamefont {Murakami}}, \bibinfo {author} {\bibfnamefont {A.}~\bibnamefont {Hatano}}, \bibinfo {author} {\bibfnamefont {A.}~\bibnamefont {Goryu}}, \bibinfo {author} {\bibfnamefont {K.}~\bibnamefont {Hirohata}},\ and\ \bibinfo {author} {\bibfnamefont {S.}~\bibnamefont {Izumi}},\ }\bibfield  {title} {\bibinfo {title} {Development of a method to evaluate the stress distribution in {4H-SiC} power devices},\ }\href {https://doi.org/10.7567/JJAP.57.106602} {\bibfield  {journal} {\bibinfo  {journal} {Jpn. J. Appl. Phys.}\ }\textbf {\bibinfo {volume} {57}},\ \bibinfo {pages} {106602} (\bibinfo {year} {2018})}\BibitemShut {NoStop}%
\bibitem [{\citenamefont {Kamitani}\ \emph {et~al.}(1997)\citenamefont {Kamitani}, \citenamefont {Grimsditch}, \citenamefont {Nipko}, \citenamefont {Loong}, \citenamefont {Okada},\ and\ \citenamefont {Kimura}}]{kamitani_elastic_constants_1997}%
  \BibitemOpen
  \bibfield  {author} {\bibinfo {author} {\bibfnamefont {K.}~\bibnamefont {Kamitani}}, \bibinfo {author} {\bibfnamefont {M.}~\bibnamefont {Grimsditch}}, \bibinfo {author} {\bibfnamefont {J.~C.}\ \bibnamefont {Nipko}}, \bibinfo {author} {\bibfnamefont {C.-K.}\ \bibnamefont {Loong}}, \bibinfo {author} {\bibfnamefont {M.}~\bibnamefont {Okada}},\ and\ \bibinfo {author} {\bibfnamefont {I.}~\bibnamefont {Kimura}},\ }\bibfield  {title} {\bibinfo {title} {The elastic constants of silicon carbide: A {Brillouin}-scattering study of {4H} and {6H} {SiC} single crystals},\ }\href {https://doi.org/10.1063/1.366100} {\bibfield  {journal} {\bibinfo  {journal} {J. Appl. Phys.}\ }\textbf {\bibinfo {volume} {82}},\ \bibinfo {pages} {3152} (\bibinfo {year} {1997})}\BibitemShut {NoStop}%
\bibitem [{\citenamefont {Iuga}\ \emph {et~al.}(2007)\citenamefont {Iuga}, \citenamefont {Steinle-Neumann},\ and\ \citenamefont {Meinhardt}}]{iuga_abinitio_simulation_2007}%
  \BibitemOpen
  \bibfield  {author} {\bibinfo {author} {\bibfnamefont {M.}~\bibnamefont {Iuga}}, \bibinfo {author} {\bibfnamefont {G.}~\bibnamefont {Steinle-Neumann}},\ and\ \bibinfo {author} {\bibfnamefont {J.}~\bibnamefont {Meinhardt}},\ }\bibfield  {title} {\bibinfo {title} {Ab-initio simulation of elastic constants for some ceramic materials},\ }\href {https://doi.org/10.1140/epjb/e2007-00209-1} {\bibfield  {journal} {\bibinfo  {journal} {Eur. Phys. J. B}\ }\textbf {\bibinfo {volume} {58}},\ \bibinfo {pages} {127} (\bibinfo {year} {2007})}\BibitemShut {NoStop}%
\bibitem [{\citenamefont {Khan}\ \emph {et~al.}(2002)\citenamefont {Khan}, \citenamefont {Zhou}, \citenamefont {Kumar},\ and\ \citenamefont {Adesida}}]{Khan-RIE_ICP_damage_Cl_F_chemistry-2002}%
  \BibitemOpen
  \bibfield  {author} {\bibinfo {author} {\bibfnamefont {F.}~\bibnamefont {Khan}}, \bibinfo {author} {\bibfnamefont {L.}~\bibnamefont {Zhou}}, \bibinfo {author} {\bibfnamefont {V.}~\bibnamefont {Kumar}},\ and\ \bibinfo {author} {\bibfnamefont {I.}~\bibnamefont {Adesida}},\ }\bibfield  {title} {\bibinfo {title} {Low-damage etching of silicon carbide in cl2-based plasmas},\ }\href {https://doi.org/10.1149/1.1482059} {\bibfield  {journal} {\bibinfo  {journal} {J. Electrochem. Soc.}\ }\textbf {\bibinfo {volume} {149}},\ \bibinfo {pages} {G420} (\bibinfo {year} {2002})}\BibitemShut {NoStop}%
\bibitem [{\citenamefont {Michaels}\ \emph {et~al.}(2023)\citenamefont {Michaels}, \citenamefont {Delegan}, \citenamefont {Tsaturyan}, \citenamefont {Renzas}, \citenamefont {Awschalom}, \citenamefont {Eden},\ and\ \citenamefont {Heremans}}]{Michaels_SiC_ALE-2023}%
  \BibitemOpen
  \bibfield  {author} {\bibinfo {author} {\bibfnamefont {J.}~\bibnamefont {Michaels}}, \bibinfo {author} {\bibfnamefont {N.}~\bibnamefont {Delegan}}, \bibinfo {author} {\bibfnamefont {Y.}~\bibnamefont {Tsaturyan}}, \bibinfo {author} {\bibfnamefont {J.}~\bibnamefont {Renzas}}, \bibinfo {author} {\bibfnamefont {D.}~\bibnamefont {Awschalom}}, \bibinfo {author} {\bibfnamefont {J.}~\bibnamefont {Eden}},\ and\ \bibinfo {author} {\bibfnamefont {F.}~\bibnamefont {Heremans}},\ }\bibfield  {title} {\bibinfo {title} {Bias-pulsed atomic layer etching of 4h-silicon carbide producing subangstrom surface roughness},\ }\href {https://doi.org/10.1116/6.0002447} {\bibfield  {journal} {\bibinfo  {journal} {J. Vac. Sci. Technol.}\ }\textbf {\bibinfo {volume} {41}},\ \bibinfo {pages} {032607} (\bibinfo {year} {2023})}\BibitemShut {NoStop}%
\bibitem [{\citenamefont {Hochreiter}\ \emph {et~al.}(2025)\citenamefont {Hochreiter}, \citenamefont {Bredol}, \citenamefont {Demiralp}, \citenamefont {David}, \citenamefont {Weber},\ and\ \citenamefont {Weig}}]{hochreiter_monolithic_4hsic_2025_supplemental_data}%
  \BibitemOpen
  \bibfield  {author} {\bibinfo {author} {\bibfnamefont {A.}~\bibnamefont {Hochreiter}}, \bibinfo {author} {\bibfnamefont {P.}~\bibnamefont {Bredol}}, \bibinfo {author} {\bibfnamefont {B.}~\bibnamefont {Demiralp}}, \bibinfo {author} {\bibfnamefont {F.}~\bibnamefont {David}}, \bibinfo {author} {\bibfnamefont {H.~B.}\ \bibnamefont {Weber}},\ and\ \bibinfo {author} {\bibfnamefont {E.~M.}\ \bibnamefont {Weig}},\ }\bibfield  {title} {\bibinfo {title} {Data for publication: Monolithic {4H-SiC} nanomechanical resonators with high intrinsic quality factors},\ }\href {https://doi.org/10.5281/zenodo.14765535} {10.5281/zenodo.14765535} (\bibinfo {year} {2025})\BibitemShut {NoStop}%
\bibitem [{\citenamefont {Photiadis}\ and\ \citenamefont {Judge}(2004)}]{photiadis_attachment_losses_2004}%
  \BibitemOpen
  \bibfield  {author} {\bibinfo {author} {\bibfnamefont {D.~M.}\ \bibnamefont {Photiadis}}\ and\ \bibinfo {author} {\bibfnamefont {J.~A.}\ \bibnamefont {Judge}},\ }\bibfield  {title} {\bibinfo {title} {Attachment losses of high {Q} oscillators},\ }\href {https://doi.org/10.1063/1.1773928} {\bibfield  {journal} {\bibinfo  {journal} {Appl. Phys. Lett.}\ }\textbf {\bibinfo {volume} {85}},\ \bibinfo {pages} {482} (\bibinfo {year} {2004})}\BibitemShut {NoStop}%
\bibitem [{\citenamefont {Wilson-Rae}(2008)}]{wilsonrae_intrinsic_dissipation_2008}%
  \BibitemOpen
  \bibfield  {author} {\bibinfo {author} {\bibfnamefont {I.}~\bibnamefont {Wilson-Rae}},\ }\bibfield  {title} {\bibinfo {title} {Intrinsic dissipation in nanomechanical resonators due to phonon tunneling},\ }\href {https://doi.org/10.1103/PhysRevB.77.245418} {\bibfield  {journal} {\bibinfo  {journal} {Phys. Rev. B}\ }\textbf {\bibinfo {volume} {77}},\ \bibinfo {pages} {245418} (\bibinfo {year} {2008})}\BibitemShut {NoStop}%
\bibitem [{\citenamefont {Zener}(1937)}]{zener_internal_friction_1937}%
  \BibitemOpen
  \bibfield  {author} {\bibinfo {author} {\bibfnamefont {C.}~\bibnamefont {Zener}},\ }\bibfield  {title} {\bibinfo {title} {Internal friction in solids. {I.} {Theory} of internal friction in reeds},\ }\href {https://doi.org/10.1103/PhysRev.52.230} {\bibfield  {journal} {\bibinfo  {journal} {Phys. Rev.}\ }\textbf {\bibinfo {volume} {52}},\ \bibinfo {pages} {230} (\bibinfo {year} {1937})}\BibitemShut {NoStop}%
\bibitem [{\citenamefont {Lifshitz}\ and\ \citenamefont {Roukes}(2000)}]{lifshitz_thermoelastic_damping_2000}%
  \BibitemOpen
  \bibfield  {author} {\bibinfo {author} {\bibfnamefont {R.}~\bibnamefont {Lifshitz}}\ and\ \bibinfo {author} {\bibfnamefont {M.~L.}\ \bibnamefont {Roukes}},\ }\bibfield  {title} {\bibinfo {title} {Thermoelastic damping in micro- and nanomechanical systems},\ }\href {https://doi.org/10.1103/PhysRevB.61.5600} {\bibfield  {journal} {\bibinfo  {journal} {Phys. Rev. B}\ }\textbf {\bibinfo {volume} {61}},\ \bibinfo {pages} {5600} (\bibinfo {year} {2000})}\BibitemShut {NoStop}%
\bibitem [{\citenamefont {Gomes~de Mesquita}(1967)}]{mesquita_refinement_crystal_1967}%
  \BibitemOpen
  \bibfield  {author} {\bibinfo {author} {\bibfnamefont {A.~H.}\ \bibnamefont {Gomes~de Mesquita}},\ }\bibfield  {title} {\bibinfo {title} {Refinement of the crystal structure of {SiC} type {6H}},\ }\href {https://doi.org/10.1107/S0365110X67003275} {\bibfield  {journal} {\bibinfo  {journal} {Acta. Cryst.}\ }\textbf {\bibinfo {volume} {23}},\ \bibinfo {pages} {610} (\bibinfo {year} {1967})}\BibitemShut {NoStop}%
\bibitem [{\citenamefont {Levinshtein}\ \emph {et~al.}(2001)\citenamefont {Levinshtein}, \citenamefont {Rumyantsev},\ and\ \citenamefont {Shur}}]{levinshtein_properties_2001}%
  \BibitemOpen
  \bibfield  {author} {\bibinfo {author} {\bibfnamefont {M.~E.}\ \bibnamefont {Levinshtein}}, \bibinfo {author} {\bibfnamefont {S.~L.}\ \bibnamefont {Rumyantsev}},\ and\ \bibinfo {author} {\bibfnamefont {M.~S.}\ \bibnamefont {Shur}},\ }\href@noop {} {\emph {\bibinfo {title} {Properties of Advanced Semiconductor Materials: {GaN}, {AIN}, {InN}, {BN}, {SiC}, {SiGe}}}}\ (\bibinfo  {publisher} {John Wiley \& Sons},\ \bibinfo {year} {2001})\BibitemShut {NoStop}%
\end{thebibliography}%

\end{document}